\documentclass[a4paper,12pt]{article}
\usepackage{graphicx}
\usepackage{dcolumn}
\usepackage{color}
\usepackage{bm}
\usepackage{psfig}
\usepackage{epsfig}
\usepackage{graphicx}
\usepackage{graphics}
\usepackage{color}

\usepackage{amsthm}
\usepackage{amssymb}
\usepackage{amsmath}

\usepackage{psfrag}

\RequirePackage{palette}
\getcolor{red}

\theoremstyle{plain} 
\theoremstyle{plain} 
\theoremstyle{plain} 

\title{An Asymptotic Preserving Scheme for the Euler equations in a
  strong magnetic field}  

\author{P. Degond$^{(1)}$, F. Deluzet$^{(1)}$, A. Sangam$^{(1)}$,
  M-H. Vignal$^{(1)}$}  
\date{July 23, 2008}

\begin{document}

\maketitle

\vspace{0.5 cm}

\begin{center}
(1)\, Institut de Math\'ematiques de Toulouse \\
UMR 5219 (CNRS-UPS-INSA-UT1-UT2)\\
Equipe MIP (Math\'ematiques pour l'Industrie et la Physique) \\
Universit\'e Paul Sabatier\\
118, route de Narbonne,  31062 TOULOUSE cedex, 
FRANCE\\
email: pierre.degond@math.univ-toulouse.fr, 
fabrice.deluzet@math.univ-toulouse.fr, 
afeintou.sangam@math.univ-toulouse.fr, 
marie-helene.vignal@math.univ-toulouse.fr
\end{center}

\maketitle

\begin{abstract} 
This paper is concerned with the numerical approximation of the  
isothermal Euler equations for charged particles subject to the
Lorentz force (the 'Euler-Lorentz' system). 
When the magnetic field is large, or equivalently, 
when the parameter $\varepsilon$ representing the non-dimensional ion
cyclotron frequency tends to zero, the so-called drift-fluid (or 
gyro-fluid) approximation is obtained. In this limit, 
the parallel motion relative to the magnetic field 
direction splits from perpendicular motion and is given 
implicitly by the constraint of zero total force along the
magnetic  field lines. In this paper, we provide a well-posed
elliptic equation for the parallel velocity which in turn allows
us to construct an Asymptotic-Preserving (AP) scheme for the Euler-Lorentz
system. This scheme gives rise to both a consistent approximation
of the Euler-Lorentz model when $\varepsilon$ is finite and a consistent
approximation of the drift limit when $\varepsilon \to 0$. Above all,
it does not require any constraint on the space and time steps related
to the small value of $\varepsilon$. Numerical results are presented,
which confirm the AP character of the scheme and its Asymptotic
Stability. 
\end{abstract}


\medskip
{\bf Acknowledgements:} The authors wish to express their gratitude
to G.~Falchetto, X.~Garbet and M.~Ottaviani from the 
CEA-Cadarache for fruitul 
discussions and encouragements. This work has been partially supported 
by the Marie Curie Actions of the European 
Commission in the frame of the DEASE project (MEST-CT-2005-021122), by
the CNRS and the Association Euratom-CEA in the framework of the 
contract 'Gyrostab' and by the CEA-Saclay in the framework of the 
contrat 'Astre' \# SAV 34160.

%%%%%%%%%%%%%%%%%%%%%%%%%%%%%%%%%%%%%%%%%%%%%%%%%%%%%%%%%%%%%%%%
%%%%%%%%%%%%%%%%%%%%%%%%%%%%%%%%%%%%%%%%%%%%%%%%%%%%%%%%%%%%%%%%
%%%%%%%%%%%%%%%%%%%%%%%%%%%%%%%%%%%%%%%%%%%%%%%%%%%%%%%%%%%%%%%%
%%%%%%%%%%%%%%%%%%%%%%%%%%%%%%%%%%%%%%%%%%%%%%%%%%%%%%%%%%%%%%%%
%%%%%%%%%%%%%%%%%%%%%%%%%%%%%%%%%%%%%%%%%%%%%%%%%%%%%%%%%%%%%%%%
\setcounter{equation}{0}
\section{Introduction}
\label{DA1}

This paper is concerned with the construction of a numerical scheme
for the system of isothermal Euler equations for charged particles
subject to the Lorentz force
(which we will refer to as the \textit{Euler-Lorentz} system). More
precisely, we are interested in the regime where the inertia terms
which  balance the pressure and Lorentz forces in the momentum balance
equation are scaled by a small parameter~$\varepsilon$. The
parameter~$\varepsilon$ represents the inverse of the ion
gyrofrequency around the magnetic field axis scaled by a
characteristic time of the experiment. When $\varepsilon$ tends to
zero, the so-called \textit{drift-fluid} or \textit{gyro-fluid} 
regime is reached~\cite{Hazeltine_Meiss,Lifshitz}. 

In the drift-fluid approximation, particles are confined along the
magnetic field lines. As a consequence, the dynamics along the
magnetic field lines is much quicker than across it. In the limit
$\varepsilon \to 0$, the parallel motion assumes an instantaneous
equilibrium in which the pressure force equilibrates the electric
force. This equilibrium is attained through acoustic waves propagating
at infinite velocity in a similar fashion as what happens in a low
Mach number fluid. These acoustic waves adjust the parallel velocity
instantaneously in such a way that this equilibrium is satisfied at
all times. In this way, the parallel velocity plays the role of a
Lagrange multiplier of the constraint of zero total aligned force. The
first goal of this paper is to give an equivalent formulation of the
drift-fluid approximation that enables us to calculate this parallel
velocity (contrary to what is sometimes written in the literature
\cite{Hazeltine_Meiss}, it is possible to find such an equation). 

The second goal is to design a valid scheme for both regimes 
$\varepsilon \sim 1$ and $\varepsilon \to 0$. This scheme gives rise
to both a consistent approximation of the Euler-Lorentz model when
$\varepsilon$ is finite and a consistent approximation of the
drift-fluid limit when $\varepsilon \to 0$. Above all, it does not
require any constraint on the space and time steps related to the
small value of $\varepsilon$. This type of schemes is usually referred
to as \textit{Asymptotic Preserving schemes (AP)}.

Asymptotic Preserving schemes have been proposed in a variety of
contexts, such as hydrodynamic or diffusion limits of kinetic model 
 \cite{CJR, JPT1, JPT2, Jin, Pareschi_Russo,
   Buet_Cordier_Lucquin_Mancini, Gosse_Toscani}, relaxation limits of
 hyperbolic models \cite{Jin-JCP, JL, Gosse_Toscani_2}, relaxation
 limits of Complex-Ginzburg-Landau equations \cite{DJT}, low-Mach
 number limits of compressible fluid models \cite{DJL}. In the plasma
 physics context, these schemes have appeared in relation with the
 quasi-neutral limit of the Euler-Poisson system \cite{CDV_CRAS_05,
   CDV, DLV} or Vlasov-Poisson system \cite{DDN_CRAS_06,BCDS}.

Such schemes are of great potential interest to the simulation of
strongly magnetized plasmas such as those encountered in space plasmas
or in Tokamak devices like ITER. First of all, there are several
advantages to using the original Euler-Lorentz model instead of the
limit drift-fluid model. Indeed, the drift-fluid model is a
mathematically complex model: the constraint of zero total force makes
it a mixed-type model, with certain characteristics of an elliptic
problem, like the incompressible Navier-Stokes equation. Dealing with
this constraint is numerically challenging, and is at least as
difficult as dealing with the incompressibility constraint in the
Navier-Stokes equation. In the literature, various drift-fluid models
have been proposed on physical
grounds~\cite{Beer_CowleyHammett,Beer_Hammett,Dorland_Hammett,
Falchetto_Ottaviani_PRL04,Garbet_PhPl01,Hallatschek_PhPl00,
Hammett_Dorland,Naulin_PhPl03,Naulin_PhPl05,Ottaviani_Manfredi_PhPl99,
Scott_PlPhCF97,Scott_PhPl05,Xu_PhPl00}. However, their relations to
the drift-fluid model which can be derived from the formal asymptotic
analysis developped below are unclear. This is why we find preferable
to rely on the original Euler-Lorentz model, in which the momentum 
conservation equation directly follows from first physical principles.  

Another advantage  of AP schemes is their ability to deal equally well
with the asymptotic regime $\varepsilon \to 0$ and the 'normal'
situation $\varepsilon =O(1)$. This is potentially very interesting
for situations in which part of the simulation domain reaches the
asymptotic regime and part of it does not. The usual approach for
dealing with such occurences is through multiphysics domain
decomposition: the full Euler-Lorentz model is used in the region
where $\varepsilon =O(1)$ and the drift-fluid limit model is used
where $\varepsilon \ll 1$ (we assume the dimensionless parameter
$\varepsilon$ is computed using local estimates of the magnetic field
strength and can be considered as a function of the space and time
coordinates). There are several drawbacks in using this approach. The
first one is the choice of the position of the interface (or
cross-talk region), which can influence the outcome of the
simulation. If the interface evolves in time, an algorithm for
interface motion has to be devised and some adaptive remeshing has to
be implemented, which requires heavy code developments and can be
quite CPU time consuming. Determining the right coupling strategy
between the two models can also be quite challenging and the outcome
of the numerical simulations may also depend on that choice. Because
these questions do not have a straigthforward answer, multiphysics
domain decomposition strategies often lack robustness and
reliability. Here, using the original model with an AP discretization
method everywhere prevents from having to introduce questionable
physical artefacts in the model and permits to use the same code
everywhere in both regimes.  

The potential application of these methods is the simulation of fluid
turbulence in Tokamak plasmas. There has been considerable literature
on this
problem \cite{Beer_CowleyHammett,Beer_Hammett,Dimits,Dorland_Hammett,
Hallatschek_PhPl00,Hammett_Dorland,Xu_PhPl00}. 
The present work is far from being at a comparable development stage,
since the present numerical tests are restricted to given uniform
magnetic and electric fields in a two-dimensional
setting. Nonetheless, this is an unavoidable intermediate step to  
check the performances of the method. The numerical tests being
successfull, this approach will soon be extended to an arbitrary given
magnetic field and coupled to the dynamics of the electron fluid
through quasineutrality assumptions.   

The assumption that the ion fluid is isothermal is only made for
simplicity. An energy equation for the ion fluid can be considered
instead. The approach extends easily to this case. We will report on
it in future work.  

This paper is then organized as follows. The isothermal Euler-Lorentz
model in the drift-fluid scaling and the drift-fluid limit 
are presented in section~\ref{DA2}. 
An Asymptotic Preserving time discretization of the
isothermal Euler-Lorentz model in the drift-fluid scaling and a full
discretization of this scheme in a reduced 2-dimensional setting 
are proposed in section~\ref{DA5}. Numerical results are given in
section~\ref{DA6} and finally, conclusions are given in
section~\ref{DA7}.

%%%%%%%%%%%%%%%%%%%%%%%%%%%%%%%%%%%%%%%%%%%%%%%%%%%%%%%%%%%%%%%%
%%%%%%%%%%%%%%%%%%%%%%%%%%%%%%%%%%%%%%%%%%%%%%%%%%%%%%%%%%%%%%%%
%%%%%%%%%%%%%%%%%%%%%%%%%%%%%%%%%%%%%%%%%%%%%%%%%%%%%%%%%%%%%%%%
%%%%%%%%%%%%%%%%%%%%%%%%%%%%%%%%%%%%%%%%%%%%%%%%%%%%%%%%%%%%%%%%
%%%%%%%%%%%%%%%%%%%%%%%%%%%%%%%%%%%%%%%%%%%%%%%%%%%%%%%%%%%%%%%%
\setcounter{equation}{0}
\section{The isothermal Euler-Lorentz model and the drift-fluid
  limit}
  \label{DA2}

%%%%%%%%%%%%%%%%%%%%%%%%%%%%%%%%%%%%%%%%%%%%%%%%%%%%%%%%%%%%%%%%
\subsection{The Euler-Lorentz model}
\label{ss_DA2}

We are concerned with the Euler-Lorentz model describing the
isothermal flow of positive ions in a tokamak. In this model, we
neglect the electrons and suppose that the electric and magnetic
fields are given. In the  \textit{drift-fluid asymptotics}, we
let~$\varepsilon$ be a typical scaled value of the gyro-period of 
the particles, i.e. the period of their rotation motion about the
magnetic field axis. The scaled isothermal Euler-Lorentz model takes
the form: 
\begin{eqnarray}
  &&\partial_t{n_{\varepsilon}} + \bm \nabla \cdot
  \left(n_{\varepsilon} \bm u_{\varepsilon}\right) = 0\,,\label{ec}\\
  &&\varepsilon \Big[\partial_t{\left(n_{\varepsilon}\bm
      u_{\varepsilon}\right)} + \bm \nabla \cdot\left(n_{\varepsilon} 
    \bm u_{\varepsilon} \otimes \bm u_{\varepsilon}\right)\Big] +
  T\,\bm \nabla n_{\varepsilon} = n_{\varepsilon} \left(\bm E + \bm
  u_{\varepsilon} \times \bm B \right)\,,\label{em}
\end{eqnarray} 
where $n_{\varepsilon}$, $\bm u_{\varepsilon}$ are the density and
the velocity of ions, respectively. The quantity $T$ is the ion
temperature. Here, the electric field $\bm E$ and the magnetic field
$\bm B$ are assumed to be given functions. The symbol $\bm \nabla$ is
the gradient operator while $\bm \nabla \cdot$ denotes the divergent
operator.

This scaled model is obtained from the unscaled Euler-Lorentz model by
introducing characteristic scales for length $x_0$, time $t_0$,
velocity $u_0$, density $n_0$, temperature $T_0$, electric field $E_0$
and magnetic field $B_0$. As usual, we set $x_0=u_0t_0$ and the
characteristic electric and magnetic fields are assumed to follow the
relation $E_0 = u_0 B_0$, so that the gyro-frequency of the ions is
given by $\omega = qB_0/m = qE_0/(mu_0)$ (where $q$ is the
  ion electric charge). In doing so, two dimensionless parameters
appear, the Mach number $M=u_0/c_s$ where $c_s = (T_0/m)^{1/2}$ is the
sound speed (and $m$ is the ion mass) on the one hand, and the scaled
gyro-period $\varepsilon = m/(q B_0 t_0)$. In the drift-fluid
asymptotics, we assume that the Mach number and the gyro-period are
linked by $M = \sqrt \varepsilon$, which leads to the scaled
problem~(\ref{ec}), (\ref{em}).    

The following notations will be useful: the director of the magnetic
field is denoted by $\bm b = \bm B/B$ where $B$ is the Euclidean norm
of $\bm B$.   

Any vector quantity~$\bm v$ can be split into its
\textit{parallel}~($\parallel$) and \textit{perpendicular}~($\perp$)
parts as follows:      
\begin{equation*}\label{DA2e2} 
  \bm v = \bm v_{\parallel} + \bm v_{\perp} = v_{\parallel} \bm b +
  \bm v_{\perp}\,, \quad v_{\parallel} = \bm v \cdot \bm b \,, \quad
  \bm v_{\perp} = \bm b \times \left(\bm v \times \bm b\right)\,. 
\end{equation*}
Next, we introduce the \textit{parallel gradient} $\nabla_{\parallel}
\phi = \bm b \cdot \bm \nabla \phi $ for any scalar
function~$\phi$. The quantity $\nabla_{\parallel} \phi $ is a
scalar. In the same way, we also introduce the \textit{parallel
  divergence}, given for any vector field $ \bm v$ by
$\nabla_{\parallel} \cdot (v_{\parallel}) = \bm \nabla \cdot
\left(v_{\parallel} \bm b \right)$. This operator can be related to
the parallel gradient by following equalities,
\begin{equation}\label{DA2e7}
  \nabla_{\parallel} \cdot v_{\parallel} = \bm \nabla
    \cdot \Big( \frac{v_{\parallel}}{B} \bm B \Big) =
    B\,\nabla_{\parallel} \Big(\frac{v_{\parallel}}{B}\Big)\,,  
\end{equation}
since the magnetic field is a divergence-free vector. We
can write $\bm \nabla \cdot \bm v = \bm \nabla \cdot \bm v_{\perp} +
\nabla_{\parallel} \cdot \left(v_{\parallel} \right)$. Note that
$\nabla_{\parallel} \cdot \left(v_{\parallel}\right)$ is also a
scalar. More generally, we introduce the \textit{parallel divergence}
$\nabla_{\parallel} \cdot\, \phi$, acting on a scalar~$\phi$, by
$\nabla_{\parallel}\cdot \phi  = \bm \nabla \cdot \left(\phi \bm
b\right)$. The operators $\nabla_{\parallel}$ and
$\nabla_{\parallel}\cdot$ are formal adjoints. Let us consider two
scalar-valued fonctions $\phi$ and $\psi$ defined on a regular
domain~$\Omega$, and let us assume also that $\phi$ and $\psi$ vanish
on the boundary $\partial{\Omega}$, for simplicity. We have,        
\begin{equation*}\label{DA2e10}
  \begin{aligned}  
    \int_{\Omega}{\Big[\nabla_{\parallel} \Big(\nabla_{\parallel}
	\cdot \phi \Big)\Big] \psi \,d\bm x}&=\int_{\Omega}{\Big[\bm
	b \cdot \bm \nabla \Big(\bm \nabla \cdot \left(\phi \bm
	b\right)\Big)\Big] \psi \,d\bm x}\\ 
	&=-\int_{\Omega}{\Big(\bm \nabla \cdot \left(\phi \bm
	b\right)\Big)\Big(\bm \nabla \cdot \left(\psi \bm
	b\right)\Big)\,d\bm x}\\ 
	&=-\int_{\Omega}{\Big(\nabla_{\parallel} \cdot  \phi\Big)
	  \Big(\nabla_{\parallel} \cdot  \psi  \Big)\,d\bm x}\,.
  \end{aligned}
\end{equation*}

%%%%%%%%%%%%%%%%%%%%%%%%%%%%%%%%%%%%%%%%%%%%%%%%%%%%%%%%%%%%%%%%
\subsection{The Drift-fluid Limit}
\label{DA3}

The formal limit $\varepsilon \to 0$ in the
isothermal Euler-Lorentz model \eqref{ec},
\eqref{em}, leads to the so-called isothermal drift-fluid model:
\begin{eqnarray}
  &&\partial_t{n} + \bm \nabla \cdot \left(n \bm u \right) =
  0\,,\label{DA3e1} \\ 
  &&T\,\bm \nabla n = n \left(\bm E + \bm u \times \bm B
  \right)\,.\label{DA3e2} 
\end{eqnarray}   
The constraint~\eqref{DA3e2} completely determines the velocity $\bm
u$. Indeed, taking the parallel and perpendicular components
of~(\ref{DA3e2}) leads to   
\begin{eqnarray}\label{DA3e5}
  & & n \bm u_{\perp} = \frac1{B} \bm b \times
  \left(T\,\bm \nabla n - n \bm E\right)\,, \\
  & & \label{DA3e6} T\, \nabla_\parallel n - n E_\parallel = 0
  \,.
\end{eqnarray} 
After dividing by $n$, we find that the first term at the right-hand
side of~(\ref{DA3e5}) is the diamagnetic drift velocity while the
second one is the $\bm E \times \bm B$ drift velocity.  

Eq.~(\ref{DA3e6}) can be recast in the form of an elliptic equation
for $u_\parallel$. Indeed, (\ref{DA3e1}) can be written: 
\begin{eqnarray}
  &&\partial_t{n} + \bm \nabla_\bot \cdot \left(n \bm u_\bot \right)
  + \nabla_\parallel \cdot \left(n  u_\parallel \right)  =
  0\,,
  \label{DA3e1_2} 
\end{eqnarray}   
Applying $\nabla_\parallel$ to (\ref{DA3e1_2}), noting that
$[\nabla_\parallel, \partial_t] = - \partial_t b \cdot \nabla$ (where
$[\cdot,\cdot]$ denotes the commutator) and inserting (\ref{DA3e6})
leads to:  
\begin{eqnarray}
  && \hspace{-0.5cm} - \nabla_\parallel( \nabla_\parallel \cdot 
  \left(n  u_\parallel \right) ) = 
  \partial_t \left( \frac{n E_\parallel}{T} \right) -  \partial_t b
  \cdot \nabla n + \nabla_\parallel ( \bm \nabla_\bot \cdot \left(n
  \bm u_\bot \right) )\,. 
  \label{DA3e13} 
\end{eqnarray}   
This is a one-dimensional elliptic equation for $u_\parallel$ along
the magnetic field lines, which is well-posed through the above
mentioned duality between the parallel gradient and parallel
divergence operators. Therefore the parallel component $u_{\parallel}$
can be computed explicitly through the resolution of this elliptic
equation, provided that adequate boundary conditions are given. The
boundary conditions depend on the specific test case under
consideration. They will be discussed in the numerical section
below.  
 
The drift-fluid model consists of equations~(\ref{DA3e1}),
(\ref{DA3e5}) and (\ref{DA3e13}).

%%%%%%%%%%%%%%%%%%%%%%%%%%%%%%%%%%%%%%%%%%%%%%%%%%%%%%%%%%%%%%%%
\subsection{A reformulation of the iso\-ther\-mal Euler-\-Lorentz
  model}
\label{ss_DA4}

The scaled Euler-Lorentz model in the drift-fluid asymptotics is a
singularly perturbed problem: in the drift-fluid limit, the type of
certain equations changes. Indeed, in the Euler-Lorentz model the
velocity is given by a time evolution equation of hyperbolic type
while  in the drift-fluid limit, the perpendicular velocity is given
by an algebraic equation, while the parallel component is found
through solving an elliptic type equation. To find an AP scheme, it is
essential to 'regularize' the perturbation, i.e. to reformulate the
Euler-Lorentz model in such a way that the limit equations for the
velocity appear explicitly in the system of equations. The goal of
this section is to find such a reformulation.  

For the perpendicular component of the momentum, we take the
cross-product of (\ref{em}) with $\bm b$, which leads to: 
\begin{equation}\label{DA4e2}
  \begin{aligned}
    & B\,\left(n_\varepsilon \bm u_\varepsilon \right)_{\perp} 
    - \varepsilon \partial_t{\Big(\bm b \times \left(n_\varepsilon
    \bm u_\varepsilon  \right)_{\perp}\Big)} =    - \bm b \times
    \Big[-T\,\bm \nabla n_\varepsilon  + n_\varepsilon  \bm E\Big] +  
    \\  
    &\hspace{3.0cm}+
    \varepsilon \Big[ - \Big(\partial_t{\bm b}\Big) \times
    \left(n_\varepsilon \bm u_\varepsilon \right) + 
    \bm b \times \Big(\bm \nabla \cdot \left(n_\varepsilon 
    \bm u_\varepsilon  \otimes \bm u_\varepsilon \right)\Big)\Big]
    \,. 
  \end{aligned}
\end{equation}
Formally, when $\varepsilon\; \to 0$ in eq.~\eqref{DA4e2}, we recover
the equation for the perpendicular component of the momentum in the
drift-fluid limit model~\eqref{DA3e5}.

We now take the scalar product of \eqref{em} with $\bm b$:
\begin{equation}\label{DA4e4}
  \begin{aligned}
    &\varepsilon \Big[\partial_t{\Big(\left(n_\varepsilon 
	u_\varepsilon \right)_{\parallel}\Big)} - \Big(\partial_t{\bm
	b}\Big) \cdot \left(n_\varepsilon \bm u_\varepsilon  \right) 
      + \bm b \cdot \Big(\bm \nabla \cdot \left(n_\varepsilon 
      \bm u_\varepsilon  \otimes \bm u_\varepsilon \right)\Big)\Big] =
    \\   
    &\hspace{6.0cm}
    = \bm b \cdot \Big[-T\,\bm \nabla n_\varepsilon  + n_\varepsilon
      \bm E\Big]\,. 
  \end{aligned}
\end{equation}
Since $u_\parallel$ cannot be computed explicitly from  this equation
in the limit $\varepsilon\; \to 0$, we are led to reformulate the
equation~\eqref{DA4e4}. We first take the time derivative of
(\ref{DA4e4}) and get 
\begin{eqnarray}
  && \varepsilon \Big[ \partial^2_t \Big( \left( n_\varepsilon 
    u_\varepsilon \right)_{\parallel} \Big) 
    - \partial_t \left( \Big(\partial_t{\bm b}\Big) \cdot
    \left(n_\varepsilon \bm u_\varepsilon  \right) \right) 
    + \partial_t \left( \bm b \cdot \Big( \bm \nabla \cdot \left(
    n_\varepsilon \bm u_\varepsilon  \otimes \bm u_\varepsilon \right)
    \Big) \right) \Big] = \nonumber \\   
  && \hspace{4.5cm} 
  =  \partial_t ( n_\varepsilon  \bm E_\parallel) - T \partial_t \bm b
  \cdot \bm \nabla n_\varepsilon - T\, \nabla_\parallel  \partial_t
  n_\varepsilon \,. 
  \label{DA4e4_2}
\end{eqnarray}
Now, applying $\nabla_\parallel$ to \eqref{ec} (rewritten in the same
fashion as \eqref{DA3e1_2}) leads to  
\begin{eqnarray}
  && \hspace{-1.2cm} \varepsilon  \partial^2_t \Big( \left(
  n_\varepsilon  u_\varepsilon \right)_{\parallel} \Big) 
  - T \nabla_\parallel (\nabla_\parallel \cdot (n_\varepsilon
  u_\varepsilon)_\parallel) = \nonumber \\ 
  && \hspace{-0.2cm} = \varepsilon \partial_t \left(
  \Big(\partial_t{\bm b}\Big) \cdot 
  \left(n_\varepsilon \bm u_\varepsilon  \right) \right) 
  - \varepsilon \partial_t \left( \bm b \cdot \Big( \bm \nabla \cdot
  \left( n_\varepsilon \bm u_\varepsilon  \otimes \bm u_\varepsilon
  \right) \Big) \right) + \nonumber \\   
  && \hspace{0.7cm} + \partial_t ( n_\varepsilon  \bm
  E_\parallel ) - T  \partial_t \bm b \cdot \bm \nabla n_\varepsilon   
  + T \nabla_\parallel (\nabla_\bot \cdot (n_\varepsilon \bm
  u_\varepsilon)_\bot)\,. 
  \label{DA4e5}
\end{eqnarray}

We notice that eq.~\eqref{DA3e13} is the formal limit of
eq.~\eqref{DA4e5} when $\varepsilon \to 0$. Eq.~\eqref{DA4e5} is a 
wave equation for $u_\parallel$ associated with the
elliptic operator $\nabla_\parallel \cdot (\nabla_\parallel )$, which
is well-posed provided that suitable boundary conditions are 
given. This wave equation describes the propagation of disturbances
along the magnetic field lines, which propagate at a velocity of order
$O(\varepsilon^{-1/2})$. In the limit $\varepsilon \to 0$, an
equilibrium described by (\ref{DA3e13}) is instantaneously reached 
through waves propagating at infinite speed. Eq. (\ref{DA4e5})
provides an equivalent formulation to (\ref{em}) for $u_\parallel$,
but which does not become singular when $\varepsilon \to 0$.   

Therefore, the reformulation of the Euler-Lorentz model consists of
eqs. (\ref{ec}), (\ref{DA4e2}), (\ref{DA4e5}).

%%%%%%%%%%%%%%%%%%%%%%%%%%%%%%%%%%%%%%%%%%%%%%%%%%%%%%%%%%%%%%%%
%%%%%%%%%%%%%%%%%%%%%%%%%%%%%%%%%%%%%%%%%%%%%%%%%%%%%%%%%%%%%%%%
%%%%%%%%%%%%%%%%%%%%%%%%%%%%%%%%%%%%%%%%%%%%%%%%%%%%%%%%%%%%%%%%
%%%%%%%%%%%%%%%%%%%%%%%%%%%%%%%%%%%%%%%%%%%%%%%%%%%%%%%%%%%%%%%%
%%%%%%%%%%%%%%%%%%%%%%%%%%%%%%%%%%%%%%%%%%%%%%%%%%%%%%%%%%%%%%%%
\setcounter{equation}{0}
\section{An Asymptotic Preserving scheme for the iso\-thermal
  Euler-Lorentz model in the drift-fluid approximation}
  \label{DA5}

%%%%%%%%%%%%%%%%%%%%%%%%%%%%%%%%%%%%%%%%%%%%%%%%%%%%%%%%%%%%%%%%
\subsection{Time semi-discrete scheme}
\label{DA5_1}

The purpose of this section is to build an AP scheme for the
Euler-Lorentz model, i.e. a scheme which is consistent with the
Euler-Lorentz model when $\varepsilon = O(1)$ and with the drift-fluid
limit model when $\varepsilon \ll 1$. The AP property mostly relies on
an appropriate time-discretization. We will investigate this point
first. Of course, we have in mind that time semi-discrete schemes of
hyperbolic problems are unstable unless some diffusion is added. In
this section, we assume that the gradient operators are
actually approximate operators which encompass the requested numerical
diffusion. The space discretization is discussed in detail in section
\ref{DA5_2}.

Our AP time semi-discrete scheme relies on use of the reformulated
equations (\ref{ec}), (\ref{DA4e2}), (\ref{DA4e5}). However, rather
than looking for a discretization of them, it is more efficient to
start from a discretization of the original formulation (\ref{ec}),
(\ref{em}) and find a scheme which allows the same reformulation as
the continuous problem and the derivation of the discrete equivalent
to (\ref{ec}), (\ref{DA4e2}), (\ref{DA4e5}). In this way, we are
guaranteed to find a suitable discretization also in the regime
$\varepsilon = O(1)$ which we could miss otherwise.

We first introduce some notations. Let $\bm B^m$ be the magnetic field
at time~$t^m$, $B^m$ its magnitude and $\bm b^m= \bm B^m/B^m$ its
director. For a given vector field $\bm v$, we denote by
$(v)_\parallel^m$ and $(\bm v)_\bot^m$ its parallel and perpendicular 
components with respect to $\bm b^m$. Similarly, we denote by
$\nabla_\parallel^m$ and $\nabla_\parallel^m \cdot$\,  the parallel
gradient and divergence operators respective to this field.

To calculate the solution of the isothermal Euler-Lorentz model in the
drift-fluid approximation~\eqref{ec}, \eqref{em}, we propose the following
time semi-discrete scheme,        
\begin{eqnarray}
  &&\frac{n_\varepsilon^{m+1} - n_\varepsilon^m}{\Delta t} + \bm
  \nabla \cdot \left(n_\varepsilon \bm u_\varepsilon \right)^{m+1} =
  0\,,\label{DA5e1}\\ 
  &&\varepsilon \Big[\frac{\left(n_\varepsilon \bm u_\varepsilon
  \right)^{m+1}-\left(n_\varepsilon \bm u_\varepsilon \right)^m}
    {\Delta t} + \bm \nabla \cdot \left(n_\varepsilon \bm
  u_\varepsilon \otimes \bm u_\varepsilon \right)^m\Big] + T \left(\bm
  \nabla n_\varepsilon^{\#}\right)^{m+1} 
  \notag\\
  &&\hspace{2.0cm}= n_\varepsilon^m \bm E^{m+1} + \left(n_\varepsilon
  \bm u_\varepsilon \right)^{m+1} \times \bm B^{m+1}\,.
  \label{DA5e2}
\end{eqnarray} 
Here, the quantity $\left(\bm \nabla n_\varepsilon^{\#}\right)^{m+1}$
is given by, 
\begin{equation}\label{DA5e3}
  \left(\bm \nabla n_\varepsilon^{\#}\right)^{m+1} = \left(\bm \nabla
  n_\varepsilon^m\right)_\perp^{m+1} + \left(\bm \nabla
  n_\varepsilon^{m+1}\right)_{\parallel}^{m+1} \,\bm b^{m+1}\,.
\end{equation}
In this scheme, the mass flux, the parallel component of the pressure
force and the Lorentz force are evaluated implicitly while the
perpendicular component of the pressure force  is evaluated
explicitly. We show that these choices permit a reformulation of the
scheme into discrete equivalents to eqs. (\ref{ec}), (\ref{DA4e2}),
(\ref{DA4e5}).  

We first investigate the transverse component and take
the cross-product of \eqref{DA5e2} with $\bm b^{m+1}$. This leads to,
\begin{equation}\label{DA5e8}
  \begin{aligned}
    &\left(n_\varepsilon \bm u_\varepsilon\right)^{m+1}_{\perp} -
    \frac{\varepsilon}{\Delta t} \frac1{B^{m+1}} \bm b^{m+1} \times
    \left(n_\varepsilon \bm u_\varepsilon \right)^{m+1}_{\perp} \\   
    &\hspace{0.5cm}= - \frac1{B^{m+1}} \bm b^{m+1} \times 
    \Big[\frac{\varepsilon}{\Delta t}\left(n_\varepsilon \bm
    u_\varepsilon \right)^m - \varepsilon \bm \nabla \cdot 
    \left(n_\varepsilon \bm u_\varepsilon \otimes 
    \bm u_\varepsilon \right)^m - T\,\bm \nabla n_\varepsilon^m 
    + n_\varepsilon^m \bm E^{m+1}\Big]\,,   
  \end{aligned}
\end{equation}
which is a discretization of eq.~\eqref{DA4e2},
where $(\partial_t{\bm b})
\times \left(n_\varepsilon \bm u_\varepsilon \right) \approx 
((\bm b^{m+1} - \bm b^m)/\Delta t) \times \left(n_\varepsilon \bm
u_\varepsilon \right)^m$.

We now compute the scalar product of \eqref{DA5e2} with $\bm
b^{m+1}$. We get:  
\begin{equation}\label{DA5e10}
  \begin{aligned}
    &\frac{\varepsilon}{\Delta t}\left( (n_\varepsilon \bm
    u_\varepsilon)^{m+1} \right)^{m+1}_{\parallel}  + 
    T\, \nabla_\parallel^{m+1} n_\varepsilon^{m+1} \\ 
    &\hspace{1.0cm}=\bm b^{m+1} \cdot \Big[\frac{\varepsilon}{\Delta
    t}\left(n_\varepsilon \bm u_\varepsilon\right)^m - \varepsilon
    \Big(\bm \nabla \cdot \left(n_\varepsilon \bm u_\varepsilon
    \otimes \bm u_\varepsilon\right)^m \Big) + n_\varepsilon^m \bm
    E^{m+1} \Big]\,,  
\end{aligned}
\end{equation}
which is a discrete version of eq.~\eqref{DA4e4}. Differentiation of
the discrete mass conservation equation (\ref{DA5e1}) in the parallel
direction gives 
\begin{equation}\label{DA5e13}
  \begin{aligned}
    &\nabla_\parallel^{m+1} n_\varepsilon^{m+1} =
    \nabla_\parallel^{m+1} n_\varepsilon^m - \Delta t \,
    \nabla_{\parallel}^{m+1}\Big(\bm \nabla \cdot \left(
    (n_\varepsilon \bm u_\varepsilon)^{m+1} \right)^{m+1}_{\perp}
    \Big)\\ 
    &\hspace{4.5cm}-\Delta t \,\nabla_{\parallel}^{m+1}
    \Big(\nabla^{m+1}_{\parallel} \cdot \left( (n_\varepsilon
    u_\varepsilon)^{m+1} \right)^{m+1}_{\parallel} \Big)\,,  
  \end{aligned}    
\end{equation}
which can be used to eliminate $n_\varepsilon^{m+1}$ in favor of
$(n_\varepsilon u_\varepsilon)^{m+1}_{\parallel}$
in~(\ref{DA5e10}). This leads to:   
\begin{equation}\label{DA5e14}
  \begin{aligned}
    &\frac{\varepsilon}{\Delta t} \left( (n_\varepsilon 
    \bm u_\varepsilon)^{m+1} \right)^{m+1}_{\parallel} 
    - T\,\Delta t \,\nabla_{\parallel}^{m+1}
    \Big(\nabla^{m+1}_{\parallel} \cdot \left( (n_\varepsilon
    u_\varepsilon)^{m+1} \right)^{m+1}_{\parallel} \Big) \\   
    &\hspace{1.0cm}=T\,\Delta t \,\nabla_{\parallel}^{m+1}
    \Big(\bm \nabla \cdot \left( (n_\varepsilon \bm
    u_\varepsilon)^{m+1} \right)^{m+1}_{\perp} \Big) 
    -T\,\nabla_{\parallel}^{m+1} n_\varepsilon^m\\  
    &\hspace{1.1cm}+\Big[\frac{\varepsilon}{\Delta
    t}\left(n_\varepsilon \bm u_\varepsilon \right)^m - \varepsilon
    \Big(\bm \nabla \cdot \left(n_\varepsilon \bm u_\varepsilon
    \otimes \bm u_\varepsilon \right)^m \Big) + n_\varepsilon^m \bm
    E^{m+1} \Big]_\parallel^{m+1} \,.  
\end{aligned}
\end{equation}

This equation is a one dimensional elliptic
equation (along the magnetic field lines) for the quantity
$((n_\varepsilon u_\varepsilon)^{m+1})_\parallel^{m+1}$.  
It is the discrete counterpart of~\eqref{DA4e5} but the link
with~\eqref{DA4e5} is not fully direct. Eq.~\eqref{DA5e14} is rather a
discretization of the following equation:  
\begin{equation}\label{DA4e5_0}
  \begin{aligned}
    &\varepsilon \partial_t \left(n_\varepsilon u_\varepsilon
    \right)_{\parallel} - T\, \nabla_{\parallel} \Big[ \int_{t^m}^t
      \nabla_{\parallel} \cdot \Big( \left(n_\varepsilon u_\varepsilon
      \right)_{\parallel} \Big) \, ds \Big] =  
    \\
    &\hspace{0.0cm}
    = \varepsilon \big[
      \left( \partial_t \bm b \right) \cdot \left( n_\varepsilon \bm
      u_\varepsilon \right) - \bm b \cdot \Big( \bm \nabla \left(
      n_\varepsilon \bm u_\varepsilon \otimes \bm u_\varepsilon \right) 
      \Big) \Big] 
    + T\,\nabla_{\parallel} \Big[ \int_{t^m}^t \bm \nabla 
      \cdot \left(n_\varepsilon \bm u_\varepsilon \right)_{\perp} \,ds
      \Big]\\  
    &\hspace{7.0cm} - \bm b \cdot \Big[ T\,\bm \nabla n_\varepsilon^m
    - n_\varepsilon \bm E \Big]\;, 
  \end{aligned}
\end{equation}
which is obtained through the reformulation process outlined in
section \ref{ss_DA4} when the mass conservation equation is used in
time-integrated form 
\begin{equation}\label{DA3e11}
  n_\varepsilon = n_\varepsilon^m - \int_{t^m}^t{\bm \nabla \cdot
  \left((n_\varepsilon \bm u_\varepsilon)_{\perp} 
  \right)\,ds} - \int_{t^m}^t{\nabla_{\parallel} \cdot
  \left((n_\varepsilon u_\varepsilon)_{\parallel} \right) \,ds}.  
\end{equation}    
That (\ref{DA4e5_0}) is equivalent to~\eqref{DA4e5} is easy and is
left to the reader.

Now, we investigate the limit $\varepsilon \to 0$ in~\eqref{DA5e8},
\eqref{DA5e14}, leaving $\Delta t$ unchanged. We get,       
\begin{eqnarray}
  &&\left(n \bm u\right)^{m+1}_{\perp} = - \frac1{B^{m+1}} \bm b^{m+1}
  \times \Big[-T\,\bm \nabla n^m + n^m \bm
    E^{m+1}\Big]\,,\label{DA5e18}\\ 
  &&    - T\,\Delta t \,\nabla_{\parallel}^{m+1}
  \Big(\nabla^{m+1}_{\parallel} \cdot \left( (n
  u)^{m+1} \right)^{m+1}_{\parallel} \Big)
  =T \,\Delta t\,\nabla_{\parallel}^{m+1}\Big(\bm \nabla \cdot 
  \left((n \bm u)^{m+1} \right)^{m+1}_{\perp} \Big)\nonumber \\   
  &&\hspace{6.0cm}
  -T\,\nabla_{\parallel}^{m+1} n^m +[ n^m \bm E^{m+1}
  ]_\parallel^{m+1} 
  \label{DA5e19}\,.
\end{eqnarray}
This is the discrete counterpart of the drift-fluid equations
~\eqref{DA3e5}, \eqref{DA3e13}. Therefore, the limit
$\varepsilon \to 0$ can be taken in the scheme (\ref{DA5e1}),
(\ref{DA5e8}), (\ref{DA5e14}) and the resulting scheme is consistent
with the drift-fluid equations. This shows that the time semi-discrete
scheme (\ref{DA5e1}), (\ref{DA5e8}), (\ref{DA5e14}) provides an
Asymptotic Preserving discretization of the Euler-Lorentz model in the
drift-fluid limit. This scheme enables us to compute the solution of
the isothermal Euler-Lorentz model for all regimes ranging from
$\varepsilon = O(1)$ to $\varepsilon \ll 1$ with the same time step
$\Delta t$. A conventional scheme would require to let $\Delta t \to
0$ simultaneously with $\varepsilon \to 0$. An AP scheme is free from
this constraint.  

As a comparison, let us investigate the conventional time
semi-discrete scheme for the Euler-Lorentz model which is the
following:  
\begin{eqnarray}
  &&\frac{n_\varepsilon^{m+1} - n_\varepsilon^m}{\Delta t} + \bm
  \nabla \cdot \left(n_\varepsilon \bm 
  u_\varepsilon \right)^m = 0\,,\label{DA5e26}\\
  &&\varepsilon \Big[\frac{\left(n_\varepsilon \bm u_\varepsilon
  \right)^{m+1}-\left(n_\varepsilon \bm u_\varepsilon
  \right)^m}{\Delta t} + \bm \nabla \cdot \left(n_\varepsilon \bm
  u_\varepsilon \otimes \bm u_\varepsilon \right)^m\Big] + T\,\bm
  \nabla n_\varepsilon^m \notag\\
  &&\hspace{2.0cm}= n_\varepsilon^{m+1} \bm E^{m+1} +
  \left(n_\varepsilon \bm u_\varepsilon \right)^{m+1} 
  \times \bm B^{m+1}\,.
  \label{DA5e27} 
\end{eqnarray}
The difference with our scheme is that the mass flux and the pressure
force are both evaluated explicitly. We note that the Lorentz force
at the right-hand side of~\eqref{DA5e27} is still implicit otherwise
some obvious instabilities arise in the discretization of the
cyclotron rotation (the $\bm u \times \bm B$ term in the Lorentz
force). 

In taking the formal limit $\varepsilon \to 0$ in this scheme, we find
that the parallel component of the velocity is now defined by the
constraint   
\begin{equation}\label{DA5e31}
  T\,\nabla_\parallel^{m+1} n^m - (n^{m+1} \bm
  E^{m+1})_\parallel^{m+1} = 0\,.
\end{equation}
We cannot reproduce the computations leading to the elliptic equation
for the parallel velocity (\ref{DA5e19}) (or an analogous equation),
because (\ref{DA5e31}) provides the value of $\nabla_\parallel^{m+1}
n^m$ instead of that of $\nabla_\parallel^{m+1} n^{m+1}$. So, we are
bound to using the discrete time derivative at the left hand-side of
(\ref{DA5e27}) to update the value of $(nu)_\parallel$ which requires
$\Delta t \leq \varepsilon$. Therefore, the conventional time
semi-discrete scheme is not Asymptotic-Preserving.  

We also see that in our AP scheme, we need to evaluate both the mass
flux and the parallel component of the pressure force to get an
elliptic equation for $(nu)_\parallel$. If the pressure force alone is
taken implicitely, this results in an equation for $(nu)_\parallel$
which is ill-posed.

%%%%%%%%%%%%%%%%%%%%%%%%%%%%%%%%%%%%%%%%%%%%%%%%%%%%%%%%%%%%%%%%
\subsection{Fully discrete  scheme}
\label{DA5_2}

\noindent
{\bf A two-dimensional case.} For the sake of simplicity, we restrict
ourselves to a 2-dimensional case with a constant in time, uniform in
space magnetic field lying in the computational plane. Therefore, we
assume that the magnetic field is directed along the $y$-axis and that
the plasma lies in $x,y$-plane, with translation invariance in the
$z$-direction. However, a possible non-zero plasma velocity is assumed
in the $z$ direction. In these conditions, the isothermal
Euler-Lorentz model in the drift-fluid asymptotics~\eqref{ec}, \eqref{em} is
written (we now omit the indices $\varepsilon$ for the sake of simplicity), 
\begin{equation}\label{DA5e32}
  \begin{aligned}
    &\partial_t{n} + \partial_x{\left(n u_x\right)} +
    \partial_y{\left(n u_y\right)} \hspace{3.3cm}= 0\,,\\ 
    &\varepsilon \Big[\partial_t{\left(nu_x\right)} +
    \partial_x{\left(n u_x^2\right)} + \partial_y{\left(n
    u_xu_y\right)}\Big] + T\,\partial_x{n} = nE_x - n u_zB_y\,,\\
    &\varepsilon \Big[\partial_t{\left(nu_y\right)} +
    \partial_x{\left(n u_xu_y\right)} + \partial_y{\left(n
    u_y^2\right)}\Big] + T\,\partial_y{n} = nE_y\,,\\
    &\varepsilon \Big[\partial_t{\left(nu_z\right)} +
      \partial_x{\left(n u_xu_z\right)} + \partial_y{\left(n
    u_yu_z\right)}\Big] \hspace{1.2cm} = nE_z + n u_x B_y\,,\\
  \end{aligned}
\end{equation}

Then the time semi-discrete AP scheme for the system~\eqref{DA5e32}
reads, 
\begin{eqnarray}
  &&\hspace{-1.0cm}\left(n u_x\right)^{m+1} -
  \dfrac{\varepsilon}{\Delta t}\dfrac1{B} \left(n
  u_z\right)^{m+1} = -\frac1{B} \left[\dfrac{\varepsilon}{\Delta
  t}\left(n u_z\right)^m - \varepsilon \Big(\partial_x{{\left(n u_x
  u_z\right)}^m} \right. \notag\\ 
  &&\hspace{6.0cm}\left. +\partial_y{{\left(n u_y
      u_z\right)}^m}\Big) + n^m E_z \right]\,,\label{DA5e33}\\
  &&\hspace{-1.0cm}\dfrac{\varepsilon}{\Delta t}\dfrac1{B}
  \left(n u_x\right)^{m+1} + \left(n u_z\right)^{m+1} = -\frac1{B}
  \left[-\dfrac{\varepsilon}{\Delta t}\left(n u_x\right)^m + 
    \varepsilon \Big(\partial_x{{\left(n u_x^2\right)}^m}
    \right.\notag\\ 
    &&\hspace{5.0cm}\left.+ \partial_y{{\left(n
	u_x u_y\right)}^m}\Big) + T\,\partial_x{n^m} - n^m E_x
    \right]\,,\label{DA5e34}\\ 
  &&\hspace{-1.0cm}\frac{\varepsilon}{\Delta t}\left(n
  u_y\right)^{m+1}-T\,\Delta t \,\partial_y\left(\partial_y\left(n
  u_y\right)^{m+1}\right) = T\, \Delta t \,
  \partial_y\left(\partial_x{\left(n u_x\right)^{m+1}}\right) \notag\\  
  &&\hspace{-0.1cm}+\Big[
    \dfrac{\varepsilon}{\Delta t}\left(n u_y\right)^m - \varepsilon 
    \Big(\partial_x{{\left(n u_x u_y\right)}^m} + \partial_y{{\left(n
	u_y^2\right)}^m}\Big) - T\,\partial_y{n^m} \Big]+n^m E_y\,,
  \label{DA5e35}\\
  &&\hspace{-1.0cm}\frac{n^{m+1} - n^m}{\Delta t} +
  \partial_x{{\left(n u_x\right)}^{m+1}} + \partial_y{{\left(n
  u_y\right)}^{m+1}} = 0\,, \label{DA5e36}
\end{eqnarray}
where $B=\bm b \cdot \bm B=B_y$. We now give the full space-time
discretization based on the above discussed AP scheme for
system~\eqref{DA5e32}.

\medskip
\noindent
{\bf Fully discrete scheme in the two-dimensional case.} For numerical
purpose, let us consider a Cartesian mesh of the calculation domain
$(x_{i-1/2},\;x_{i+1/2}) \times (y_{j-1/2},\;y_{j+1/2}), \;\;
i,j=1..N$. Then eqs.~\eqref{DA5e33}, \eqref{DA5e34}, \eqref{DA5e35}
and \eqref{DA5e36} are discretized according to, 
\begin{eqnarray}
&&\hspace{-1.0cm}\left(n u_x\right)^{m+1}_{ij} -
  \dfrac{\varepsilon}{\Delta t}\dfrac1{B} \left(n
  u_z\right)^{m+1}_{ij} = -\frac1{B}
  \left[\dfrac{\varepsilon}{\Delta t}\left(n u_z\right)^m_{ij}
  \right. \notag\\  
  &&\hspace{0.5cm}\left. - \dfrac{\varepsilon}{\Delta
  x}\left({\left(n u_x u_z\right)}^m_{i+1/2\,j} - {\left(n u_x
  u_z\right)}^m_{i-1/2\,j}\right) \right. \notag\\ 
  &&\hspace{1.0cm}\left. + \dfrac{\varepsilon}{\Delta
  y}\left({\left(n u_y u_z\right)}^m_{i\,j+1/2} - {\left(n u_y
  u_z\right)}^m_{i\,j-1/2}\right) 
    + \left(n^m E_z\right)_{ij} \right]\,,\label{DA5e37}\\
  &&\hspace{-1.0cm}\dfrac{\varepsilon}{\Delta t}\dfrac1{B}
  \left(n u_z\right)^{m+1}_{ij} + \left(n u_x\right)^{m+1}_{ij} -
  \frac1{B} \left[-\dfrac{\varepsilon}{\Delta t}\left(n
  u_x\right)^m_{ij} \right. \notag\\   
  &&\hspace{0.5cm}\left. + \dfrac{\varepsilon}{\Delta
  x}\left({\left(n u_x^2 + \dfrac{T}{\varepsilon}
  n\right)}^m_{i+1/2\,j} - {\left(n u_x^2 + \dfrac{T}{\varepsilon}
  n\right)}^m_{i-1/2\,j}\right) \right. \notag\\   
  &&\hspace{1.0cm}\left. + \dfrac{\varepsilon}{\Delta
  y}\left({\left(n u_x u_y\right)}^m_{i\,j+1/2} - {\left(n u_x
  u_y\right)}^m_{i\,j-1/2}\right) 
    - \left(n^m E_x\right)_{ij} \right]\,,\label{DA5e39}\\
  &&\hspace{-1.0cm}\frac{\varepsilon}{\Delta t}\left(n
  u_y\right)^{m+1}_{ij}-T\,\Delta t \, \left(\partial_y^2{{\left(n
  u_y\right)}^{m+1}}\right)_{ij} = T\,\Delta t \,
  \left(\partial_y{\left(\partial_x{{\left(n
  u_x\right)}^{m+1}}\right)}\right)_{ij} \notag\\
  &&\hspace{-0.2cm} + \Big[\frac{\varepsilon}{\Delta t}\left(n
  u_y\right)^m_{ij}  - \dfrac{\varepsilon}{\Delta x}\Big(\left(n u_x 
  u_y\right)^m_{i+1/2\,j} - \left(n u_x u_y\right)^m_{i-1/2\,j}\Big) 
  \notag\\
  &&\hspace{-0.2cm} 
  -\dfrac{\varepsilon}{\Delta
  y}\left(\left(nu_y^2+\dfrac{T}{\varepsilon}n\right)^m_{i\,j+1/2} -
  \left(nu_y^2+\dfrac{T}{\varepsilon}n\right)^m_{i\,j-1/2}\right)
  \Big] + \left(nE_y\right)^{m+1}_{ij}, \label{DA5e40}\\      
  &&\hspace{-1.0cm}\frac{n^{m+1}_{ij} - n^m_{ij}}{\Delta t} +
  \dfrac1{\Delta x}\left(\left(nu_x\right)^{m+1}_{i+1/2\,j} -
  \left(nu_x\right)^{m+1}_{i-1/2\,j}\right) \notag\\     
  &&\hspace{2.0cm}+\dfrac1{\Delta
  y}\left(\left(nu_y\right)^{m+1}_{i\,j+1/2} -
  \left(nu_y\right)^{m+1}_{i\,j-1/2}\right)= 0\,.   
  \label{DA5e41} 
\end{eqnarray}
Here $n^{m+1}_{ij}$ is the density in the cell
$(x_{i-1/2},\;x_{i+1/2})\times (y_{j-1/2},\;y_{j+1/2})$ at the time
$t^{m+1}$. The quantity $\left(\left(n
u_x\right)^{m+1}_{ij},\;\left(nu_z\right)^{m+1}_{ij}\right)^t$ is the
perpendicular part of the momentum while $\left(n
u_y\right)^{m+1}_{ij}$ is the parallel part of the momentum in the
cell $(x_{i-1/2},\;x_{i+1/2}) \times (y_{j-1/2},\;y_{j+1/2})$ at the
time $t^{m+1}$. The terms $\left(\cdot\right)_{i+1/2\,j}$ and
$\left(.\right)_{i\,j+1/2}$ denote the numerical fluxes at the
interfaces $x_{i+1/2}$ and $y_{j+1/2}$ of the corresponding
quantities, respectively. The second order terms
$\left(\partial_y^2{{\left(n u_y\right)}^{m+1}}\right)_{ij}$ and
$\left(\partial_y{\left(\partial_x{{\left(n
      u_x\right)}^{m+1}}\right)}\right)_{ij}$ will be discussed below.    

\bigskip
\noindent
{\bf Discretization of the hyperbolic part.}
To calculate the numerical fluxes at the interfaces
$\left(\cdot\right)_{i+1/2\,j}$ and $\left(.\right)_{i\,j+1/2}$, we
use the $P_0$ scheme~\cite{DegondPol}. To be more precise, let us
consider the interface $x_{i+1/2}$ separating data ${\cal
  U}_{ij}^m,\;{\cal U}_{i+1\,j}^m$ for the corresponding Riemann
problem at time $t^m$, where 
\begin{equation*}\label{DA5e42}
  {\cal U}_{ij}^m=\left(n_{ij}^m,\;(n \bm u)_{ij}^m
  \right)^t,\;{\cal U}_{i+1\,j}^m=\left(n_{i+1\,j}^m,\;(n \bm
  u)_{i+1\,j}^m\right)^t.
\end{equation*}
Let us denote by
$\widehat{\cal
  U}_{i+1/2\,j}^m=\left(\widehat{n}_{i+1/2\,j}^m,\;\widehat{n\bm
  u}_{i+1/2\,j}^m \right)^t$ the average state between the states
${\cal U}_{i\,j}^m$, ${\cal U}_{i+1\,j}^m$. The average state
$\widehat{\cal U}_{i+1/2\,j}^m$ is the Roe average state~\cite{Toro99} 
given here by following formula:
\begin{equation*}\label{DA5e43}
    \begin{array}{ccc}
      \widehat{n}_{i+1/2\,j}^m &= &\sqrt{n_{ij}^m\,n_{i+1\,j}^m}\,,\\
      \\
      \widehat{\bm u}_{i+1/2\,j}^m &= &\dfrac{\sqrt{n_{ij}^m}\,\bm
	u_{ij}^m+\sqrt{n_{i+1\,j}^m}\,\bm
	u_{i+1\,j}^m}{\sqrt{n_{ij}^m}+\sqrt{n_{i+1\,j}^m}}\,,
    \end{array}
\end{equation*}
and the momentum of the average state is reconstructed as
\begin{equation*}\label{DA5e44}
\widehat{n\bm u}_{i+1/2\,j}^m = \widehat{n}_{i+1/2\,j}^m \,
\widehat{\bm u}_{i+1/2\,j}^m\,. 
\end{equation*}
Then the numerical fluxes are given by,
\begin{eqnarray}
  &&\hspace{-1.0cm}{\left(n u_x u_z\right)}^m_{i+1/2\,j}=
  \frac{{\left(n u_x u_z\right)}^m_{i\,j} + {\left(n u_x 
  u_z\right)}^m_{i+1\,j}}2 \notag\\
  &&\hspace{5.0cm}- \frac{a_{i+1/2\,j}^m}2 \left({\left(n
  u_z\right)}^m_{i+1\,j} - {\left(n
  u_z\right)}^m_{ij}\right)\,,\label{DA5e45}  \\   
  &&\hspace{-1.0cm}{\left(n u_x^2 + \dfrac{T}{\varepsilon}
  n\right)}^m_{i+1/2\,j} 
  =\frac{{\left(n u_x^2 + \dfrac{T}{\varepsilon}
  n\right)}^m_{i\,j} + {\left(n u_x^2 + \dfrac{T}{\varepsilon}
  n\right)}^m_{i+1\,j}}2 \notag\\
  &&\hspace{5.0cm} - \frac{a_{i+1/2\,j}^m}2 \left({\left(n
  u_x\right)}^m_{i+1\,j} - {\left(n
  u_x\right)}^m_{ij}\right)\,,\label{DA5e46}\\   
  &&\hspace{-1.0cm}\left(n u_x u_y\right)^m_{i+1/2\,j} =
  \frac{{\left(n u_x u_y\right)}^m_{i\,j} + {\left(n u_x 
  u_y\right)}^m_{i+1\,j}}2 \notag\\
  &&\hspace{5.0cm} - \frac{a_{i+1/2\,j}^m}2 \left({\left(n
  u_y\right)}^m_{i+1\,j} - {\left(n
  u_y\right)}^m_{ij}\right)\,,\label{DA5e47}\\
  &&\hspace{-1.0cm}\left(nu_x\right)^{m+1}_{i+1/2\,j} = \frac{{\left(n 
  u_x\right)}^{m+1}_{i\,j} + {\left(n u_x\right)}^{m+1}_{i+1\,j}}2 - 
  \frac{a_{i+1/2\,j}^m}2 \left(n^m_{i+1\,j} -
  n^m_{ij}\right)\,,\label{DA5e48}\\
  &&\hspace{-1.0cm}\left(nu_y\right)^{m+1}_{i+1/2\,j} = \frac{{\left(n
  u_y\right)}^{m+1}_{i\,j} + {\left(n u_y\right)}^{m+1}_{i+1\,j}}2 - 
  \frac{a_{i+1/2\,j}^m}2 \left(n^m_{i+1\,j} -
  n^m_{ij}\right)\,.\label{DA5e49}
\end{eqnarray}
Here, the speed $a_{i+1/2\,j}^m$ is given by
\begin{equation}\label{DA5e50}
  a_{i+1/2\,j}^m = \max
  \left(|a_{i+1/2\,j}^{m,-}|,|a_{i+1/2\,j}^{m,+}|\right)\,. 
\end{equation}
with
\begin{equation}\label{DA5e51}
  \begin{aligned}
  &a_{i+1/2\,j}^{m,-} = \min
  \left({u_x}_{ij}^m-c^\varepsilon,\widehat{u}_{x_{i+1/2\,j}}^m-
  c^\varepsilon\right)\,,\\   
  &a_{i+1/2\,j}^{m,+} = \max \left(\widehat{u}_{x_{i+1/2\,j}}^m+
  c^\varepsilon, {u_x}_{i+1\,j}^m+c^\varepsilon\right)\,,  
  \end{aligned}
\end{equation}
and $c^\varepsilon=\sqrt{T/\varepsilon}$ standing for the sound
speed. The numerical fluxes across interfaces $y_{j+1/2}$ are computed
similiarly.  

\bigskip
\noindent
{\bf Discretization of the second order terms.}
The second order terms are computed with centered spatial
discretizations:
\begin{equation*}\label{DA5e52}
  \left(\partial_y^2{{\left(n u_y\right)}^{m+1}}\right)_{ij} =
      \dfrac{\left(n u_y\right)^{m+1}_{ij+1}-2\left(n
	u_y\right)^{m+1}_{ij}+\left(n
      u_y\right)^{m+1}_{i\,j-1}}{\Delta	y^2}\,,     
\end{equation*}   
and
\begin{equation*}\label{DA5e53}
 \left(\partial_y{\left(\partial_x{{\left(n
    u_x\right)}^{m+1}}\right)}\right)_{ij} = \dfrac1{\Delta y}\Big[ 
    \left(\partial_x \left(n u_x\right)^{m+1}\right)_{i\,j+1/2} -
    \left(\partial_x \left(n
    u_x\right)^{m+1}\right)_{i\,j-1/2}\Big]\,,  
\end{equation*} 
where   
\begin{equation*}\label{DA5e54}
  \begin{aligned}
 \hspace{-0.2cm}\Big(\partial_x \left(n
    u_y\right)^{m+1}\Big)_{i\,j+1/2} 
    & = \frac12 \, \left(\partial_x \left(n
    u_x\right)^{m+1}\right)_{i\,j+1} + \frac12 \, \left(\partial_x
    \left(n u_x\right)^{m+1}\right)_{i\,j}\\
    &= \frac12 \dfrac{\left(n u_x\right)^{m+1}_{i+1\,j+1} - \left(n
    u_x\right)^{m+1}_{i-1\,j+1}}{2\,\Delta x} \\
    &\hspace{0.2cm}
    + \frac12 \dfrac{\left(n
    u_x\right)^{m+1}_{i+1\,j} - \left(n
    u_x\right)^{m+1}_{i-1\,j}}{2\,\Delta x}\,.    
  \end{aligned}
\end{equation*}
To solve the elliptic equation for $\left(nu_y\right)$, suitable
boundary conditions need to be specified. In particular, using above
discretizations eq.~\eqref{DA5e40} can be recast in the form

\begin{equation}\label{DA5e40b}
  {\cal A} \;{\cal X}^{m+1} = {\cal RHS}^{m+1}\;,
\end{equation}
where ${\cal A} = {\cal A}\left( \varepsilon, \Delta t, \Delta y,
T\right)$ is a regular matrix,  
${\cal X}^{m+1} = (
\left(nu_y\right)^{m+1}_{i\,j})_{i,j=1..N}$ is the vector of the
parallel momenta in all the computational domain and 
${\cal RHS}^{m+1} = {\cal RHS}^{m+1} \left(\varepsilon, \Delta t,
\Delta x, \Delta y, T, \left(nu_x\right)^m,\left(nu_y\right)^m, E_y^m
\right)$ is the right-hand side, which is known. The linear system~\eqref{DA5e40b} is solved by a
  Gaussian elimination method with partial pivoting.      
 
\bigskip
\noindent
{\bf Choice of the time-step.} As usual, the time-step $\Delta t$ is
chosen such that the CFL condition is satisfied. In 2D geometry, this
condition takes the following form,  
\begin{equation}\label{DA5e55}
  \Delta t \, \left(\frac1{\Delta x}\max_{1\leq i,j \leq
    N}a_{i+1/2\,j}^m + \frac1{\Delta y}\max_{1\leq i,j \leq
    N}a_{i\,j+1/2}^m\right) = CFL\leq 1\,.    
\end{equation}
where $a_{i+1/2\,j}^m$ is defined by (\ref{DA5e50}). The
scheme~\eqref{DA5e37}--\eqref{DA5e41} with the time-step 
algorithm~\eqref{DA5e55} will be referred to as the
\textit{resolved AP scheme}.

When $\varepsilon$ tends to $0$, the sound speed
$c^\varepsilon=\sqrt{T/\varepsilon}$ takes large values and the
time-step calculated from \eqref{DA5e55} becomes very small. Due to
the AP character of the scheme, we do not need to constrain the time
step to stay of the order of $\varepsilon$. Therefore, the sound speed
can be removed from the definition of the velocities used to compute
the numerical viscosity in the interfacial fluxes. These new
velocities are defined as follows:       
\begin{equation}\label{DA5e59}
  {\tilde{a}_{i+1/2\,j}^m} = \max \left(|\tilde{a}_{i+1/2\,j}^{m,-}|,
  |\tilde{a}_{i+1/2\,j}^{m,+}|\right)\,,  
\end{equation}
with
\begin{equation}\label{DA5e60}
  \begin{aligned}
  &{\tilde{a}_{i+1/2\,j}^{m,-}} = \min
  \left({u_x}_{ij}^m,\widehat{u}_{x_{i+1/2\,j}}^m\right)\,, \quad
  {\tilde{a}_{i+1/2\,j}^{m,+}} = \max
  \left(\widehat{u}_{x_{i+1/2\,j}}^m, {u_x}_{i+1\,j}^m 
  \right)\,,    
  \end{aligned}
\end{equation}
and a new CFL condition is introduced: 
\begin{equation}\label{DA5e58}
  \Delta t \, \left(\frac1{\Delta x}\max_{1\leq i,j \leq
    N}{\tilde{a}_{i+1/2\,j}^m} + \frac1{\Delta y}\max_{1\leq i,j \leq
    N}{\tilde{a}_{i\,j+1/2}^m}\right) = CFL\leq 1\,.  
\end{equation}
It is important to notice that, with this new expression, the time
step can be chosen independent of $\varepsilon$.

The scheme~\eqref{DA5e37}--\eqref{DA5e41} where velocities
$a_{i+1/2\,j}^m$ are substituted by ${\tilde{a}_{i+1/2\,j}^m}$ given
by eq.~\eqref{DA5e60} and with the time-step algorithm~\eqref{DA5e58}
will be called the \textit{non-resolved AP scheme}. The numerical tests
will show that this choice gives rise to a correct solution.

%%%%%%%%%%%%%%%%%%%%%%%%%%%%%%%%%%%%%%%%%%%%%%%%%%%%%%%%%%%%%%%%
%%%%%%%%%%%%%%%%%%%%%%%%%%%%%%%%%%%%%%%%%%%%%%%%%%%%%%%%%%%%%%%%
%%%%%%%%%%%%%%%%%%%%%%%%%%%%%%%%%%%%%%%%%%%%%%%%%%%%%%%%%%%%%%%%
%%%%%%%%%%%%%%%%%%%%%%%%%%%%%%%%%%%%%%%%%%%%%%%%%%%%%%%%%%%%%%%%
%%%%%%%%%%%%%%%%%%%%%%%%%%%%%%%%%%%%%%%%%%%%%%%%%%%%%%%%%%%%%%%%
\setcounter{equation}{0}
\section{Numerical tests}
\label{DA6}

%%%%%%%%%%%%%%%%%%%%%%%%%%%%%%%%%%%%%%%%%%%%%%%%%%%%%%%%%%%%%%%%

\subsection{Geometry and test case}
\label{ss_testcase}

\begin{figure}[h!]
  \centering\epsfig{file= 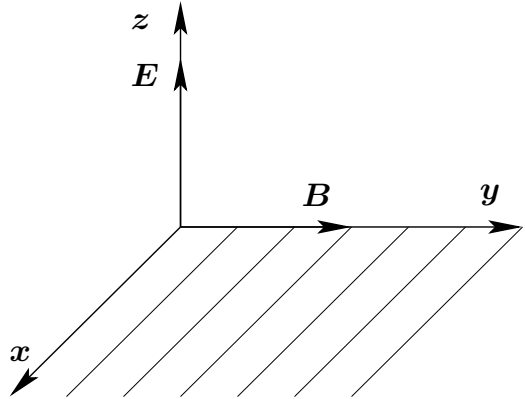,width=8cm,angle=0,clip}     
  \caption{The test-case geometry.}           
  \label{DAPf1} 
\end{figure}

Our test-case is two-dimensional: the physical quantities
depend only on the coordinates $x,\;y$. The magnetic field is assumed
to be uniform and directed along the $y$-axis \textit{i.e.} $\bm
B=(0,\;B_y,\;0)$ while the electric field is directed along the
$z$-axis, $\bm E=(0,\;0,\;E_z)$. The test-case geometry is
depicted in Fig.~\ref{DAPf1}. 

\begin{figure}[h!]
  \centering\epsfig{file= 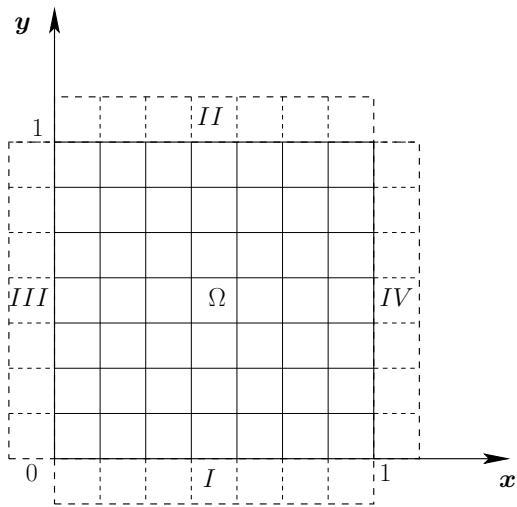,width=8cm,angle=0,clip}     
  \caption{Schematics of the computational domain.}           
  \label{DAPf12D} 
\end{figure}
The domain is a square of side $1$ as shown in Fig.~\ref{DAPf12D}
while the parameters are given in the table~\ref{t1}: the
  initial values of physical quantities are put in the $\Omega$ column
  while the boundary conditions are given in the columns $I$, $II$,
  $III$ and $IV$, accordingly.    

\begin{table}[h!] 
\caption{Simulation parameters for the test-case.} \label{t1}
\begin{center}
  \begin{tabular}{cccccccc}
    \hline\hline $$ & $\Omega$ & $I$ & $II$ & $III$ & $IV$\\  
    \hline $n$ & $1$ & $1+\varepsilon$ & $1$ & $1+\varepsilon$ & $1$ \\ 
    \hline $n u_x$ & $0$ & $-1$ & $-1$ & $-1+\varepsilon$
    & $-1+\varepsilon$ \\  
    \hline $n u_y$ & $0$ & $1$  & $1+\varepsilon$ &
    $1+\varepsilon$ & $1$ \\  
    \hline $n u_z$ & $0$ & $0$ & $\varepsilon$ & $0$ &
    $\varepsilon$ \\
    \hline $T$ & $1$ & $1$ & $1$ & $1$ & $1$ \\
    \hline $B_y$ & $1$ & $1$ & $1$ & $1$ & $1$ \\
    \hline $E_z$ & $1$ & $1$ & $1$ & $1$ & $1$ \\
    \hline\hline
  \end{tabular}
\end{center}
\end{table}

For the considered test case, the exact drift-fluid approximation 
is stationary and uniform in the whole domain, and given by 
\begin{equation}\label{DA6e1}
  n=1\,, nu_x=-1\,, nu_y=1\,, nu_z=0\,,T = 1
\end{equation}
We observe that the initial and boundary data are 'well prepared':
they are perturbations of order $\varepsilon$ of  the drift-fluid
limit. Indeed, if 'unprepared' initial and boundary data are used,
large (of order $1$) initial and boundary layers appear in which the
exact solution is significantly different from the drift-fluid
limit. In order to correctly capture these initial or boundary layers,
there is no other way than using a time and space resolved scheme in
which both the time and space steps are of order
$\varepsilon$. However, the goal of an AP scheme is {\bf not} to
capture the initial and boundary layers accurately, but to provide a
consistent approximation of the correct drift-fluid limit where it
applies, i.e. away from these layers. An accurate verification of this
property requires a test solution which is not polluted by the initial
and boundary layers and therefore, the need of well-prepared initial
and boundary conditions. In section \ref{DA6_3}, for the sake 
of completeness, we show some numerical results with unprepared initial and boundary conditions.

%%%%%%%%%%%%%%%%%%%%%%%%%%%%%%%%%%%%%%%%%%%%%%%%%%%%%%%%%%%%%%%%

\subsection{Simulations for $\varepsilon \ll 1$}
\label{DA6_1}   

Here we would like to demonstrate that the AP scheme is consistent
with the drift-fluid limit even for large time steps compared to 
$\varepsilon$. By contrast, the conventional scheme is shown to 
be unstable for time steps larger than $\varepsilon$.  
Three values of the parameter $\varepsilon$ are
used: $\varepsilon = 10^{-5}$, $\varepsilon = 10^{-6}$ and
$\varepsilon = \sqrt{\varepsilon_{machine}}=1.5\,10^{-8}$. For these
values, we observe the numerical solution at times 
$1$, $0.1$ and $0.01$, respectively. The reason for choosing smaller
observation times when $\varepsilon$ is smaller is due to the 
increase of computation time when the time
step resolves $\varepsilon$. The CFL number is taken equal to
$0.5$ for all simulations. A uniform mesh is used for both the $x$ and
$y$ direction with steps $\Delta x = \Delta y=0.01$.

\begin{figure}[h!]
  \vspace{-0.25cm}
     \begin{tabular}{c}
     \begin{tabular}{cc}
       \begin{minipage}{6cm}
	 \hspace{-0.25cm}
         \epsfig{file=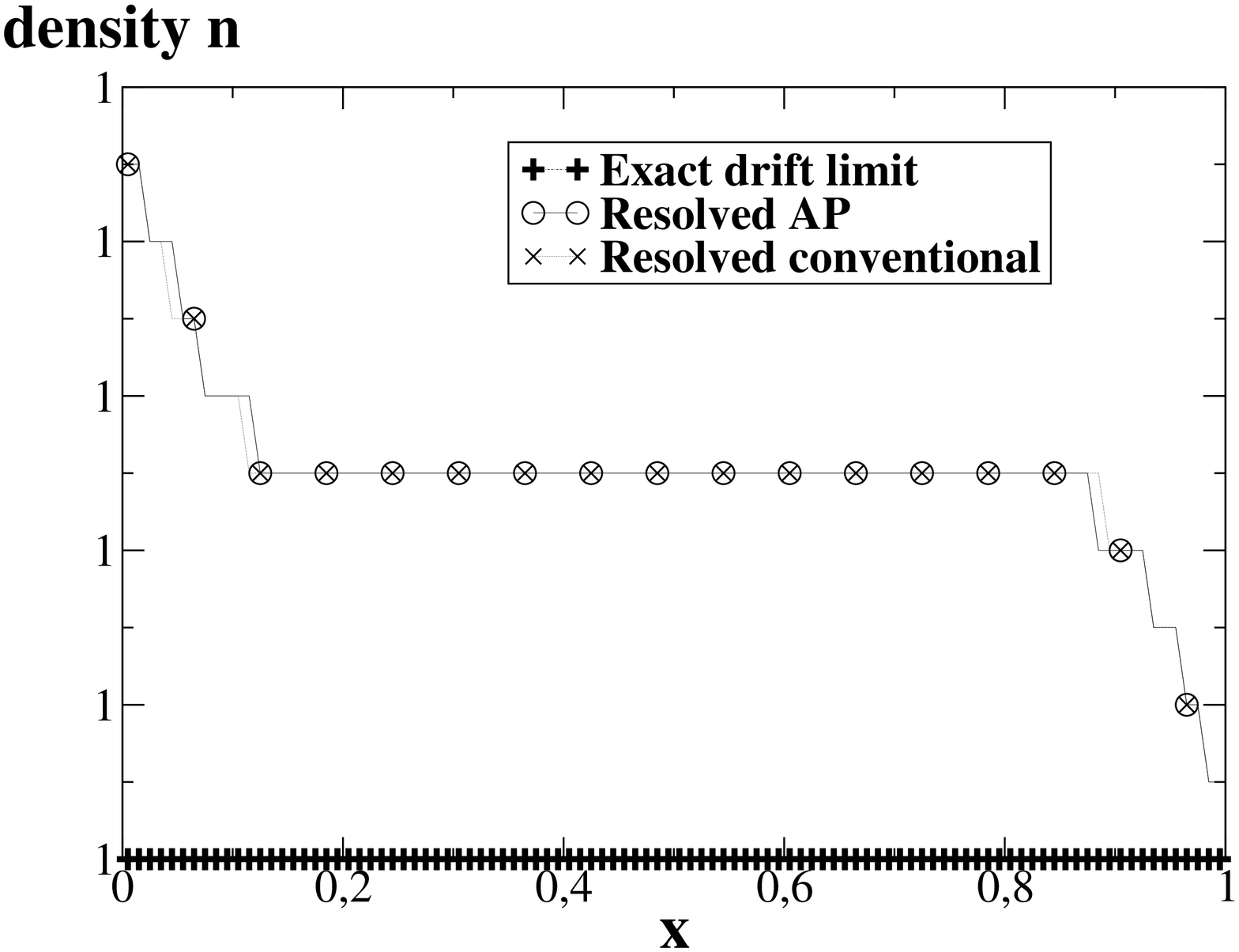,width=5cm,angle=-90,clip=} 
       \end{minipage} 
       &
       \begin{minipage}{6cm}
	 \hspace{-0.5cm}
         \epsfig{file=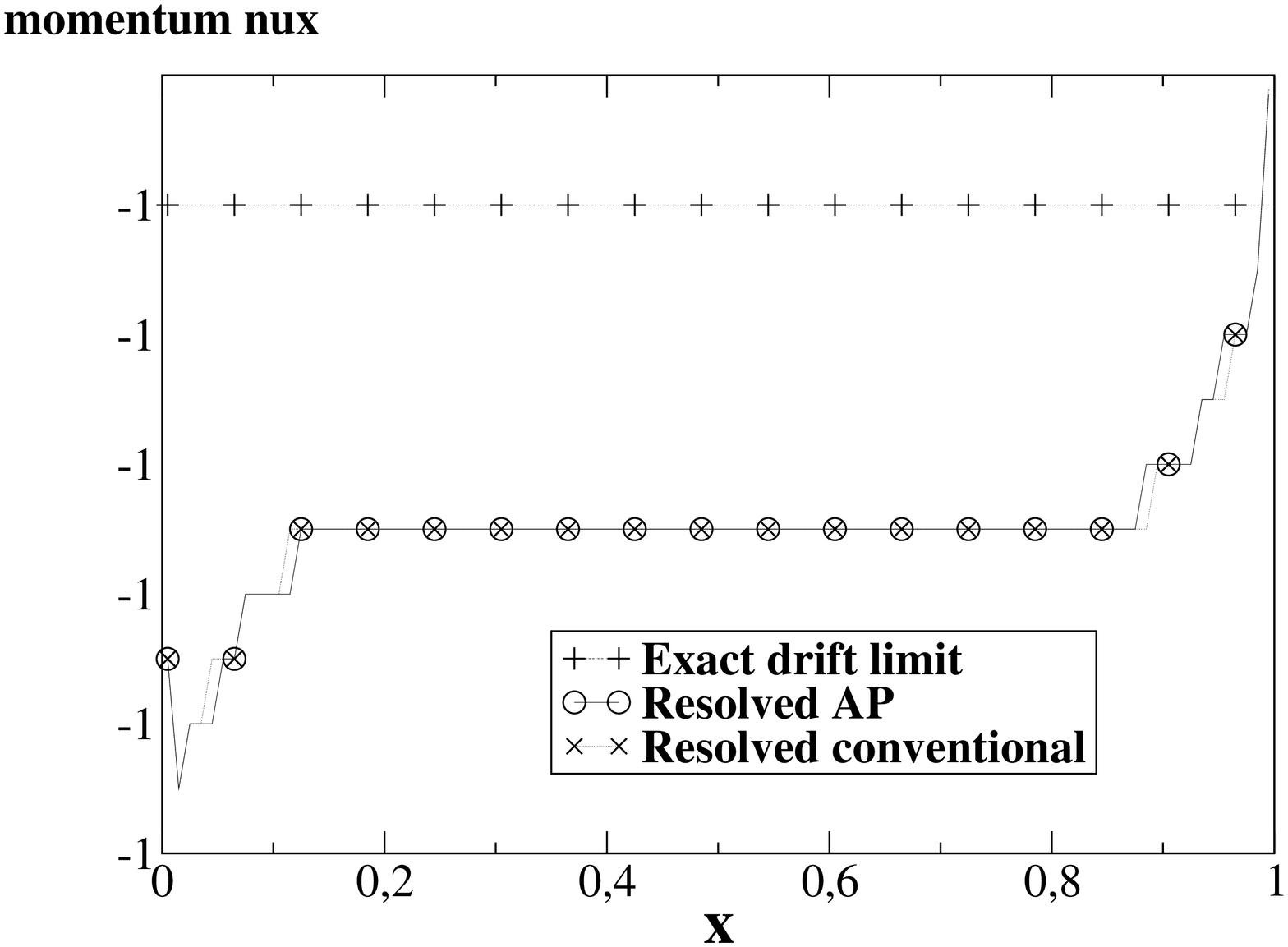,width=5cm,angle=-90,clip=}
       \end{minipage} 
     \end{tabular}
     \\
     \begin{tabular}{cc}
       \begin{minipage}{6cm}
	 \hspace{-0.5cm}
         \epsfig{file=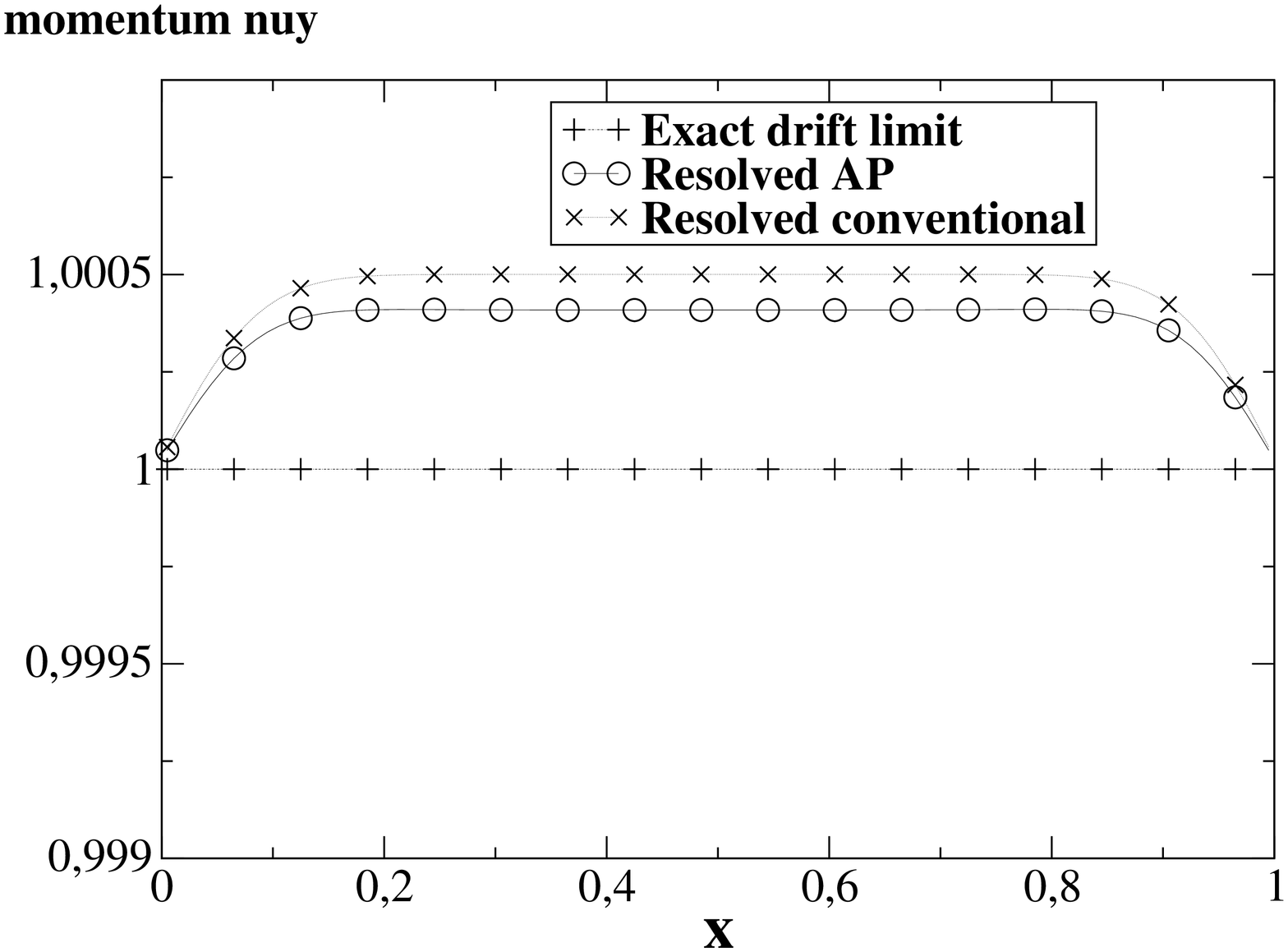,width=5cm,angle=-90,clip=} 
       \end{minipage} 
       &
       \begin{minipage}{6cm}
	 \hspace{-0.5cm}
         \epsfig{file=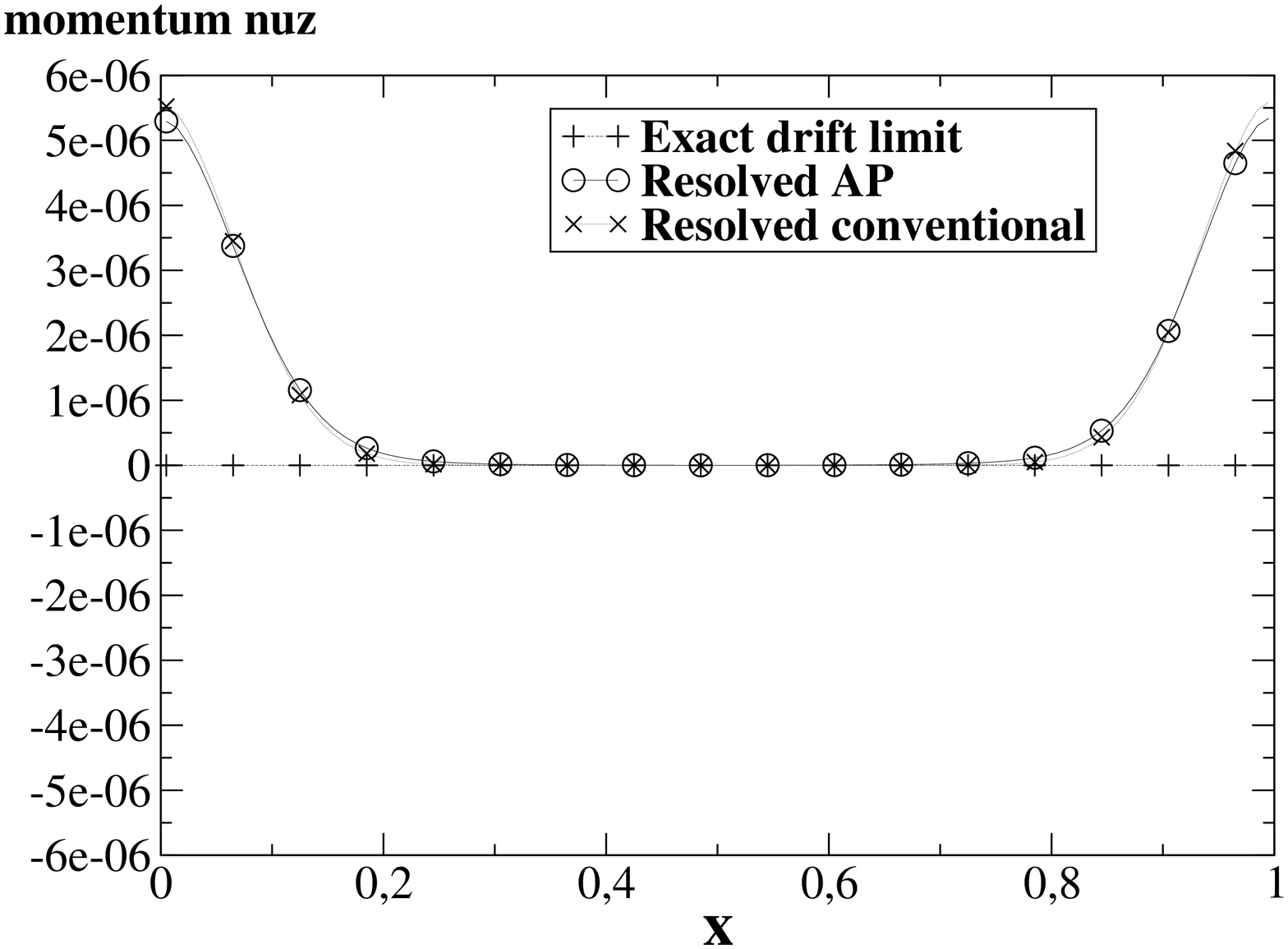,width=5cm,angle=-90,clip=} 
       \end{minipage} 
     \end{tabular}
   \end{tabular}
  \caption{Comparison of the resolved conventional scheme (crosses),
  the resolved AP scheme (circles) and the exact drift-fluid limit (vertical
  bars) at $t=0.1$ for $\varepsilon = 10^{-6}$ ; density $n$ (top
  left), $x$-component of the momentum $nu_x$ (top right),
  $y$-component of the momentum $nu_y$ (bottom left), $z$-component of
  the momentum $nu_z$ (bottom right).}
  \label{DAPfdm6ModifClasAPres}  
\end{figure}

\vspace{0.2cm}
\noindent 
{\bf Resolved case.} We first compare the conventional and AP 
schemes in the resolved case, i.e. when $\Delta t$ is smaller than
$\varepsilon$. The results given by both the conventional and AP
schemes are displayed in Figs.~\ref{DAPfdm6ModifClasAPres} for 
$\varepsilon = 10^{-6}$ and compared with the drift-fluid limit. 
Here, as well as in the forthcoming pictures, 
the various physical quantities (i.e. the density $n$ and the three 
components of the momentum $n \bm u$) are shown as functions of $x$
for a given value of $y=0.5$.  
The computed solutions are indistinguishable and very close to the 
drift-fluid limit.

\begin{figure}[h!]
   \begin{tabular}{c}
     \begin{tabular}{cc}
       \begin{minipage}{6cm}
	 \hspace{-0.25cm}
         \epsfig{file=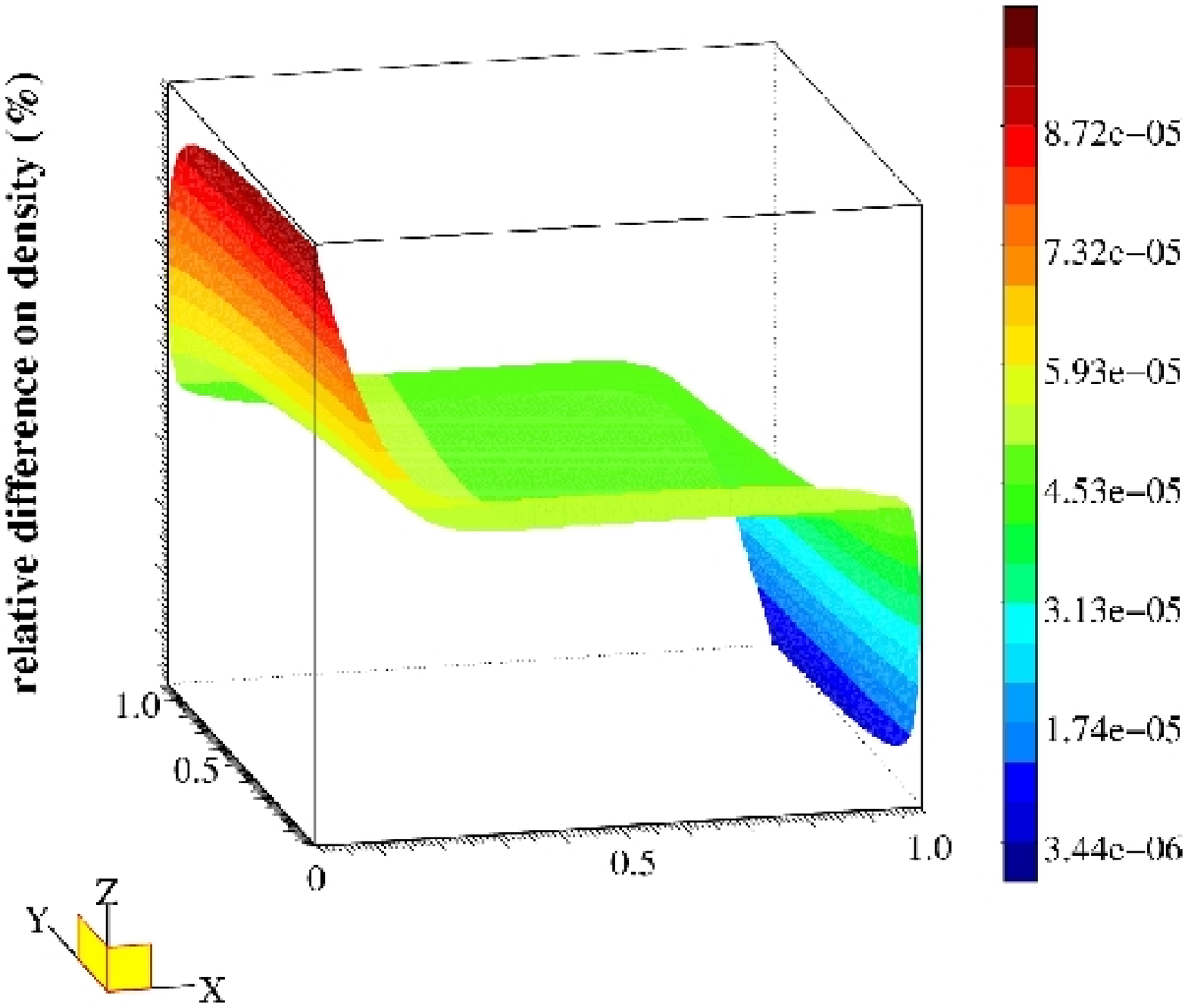,width=5cm,angle=0,clip=} 
       \end{minipage} 
       &
       \begin{minipage}{6cm}
	   \hspace{-0.5cm}
           \epsfig{file=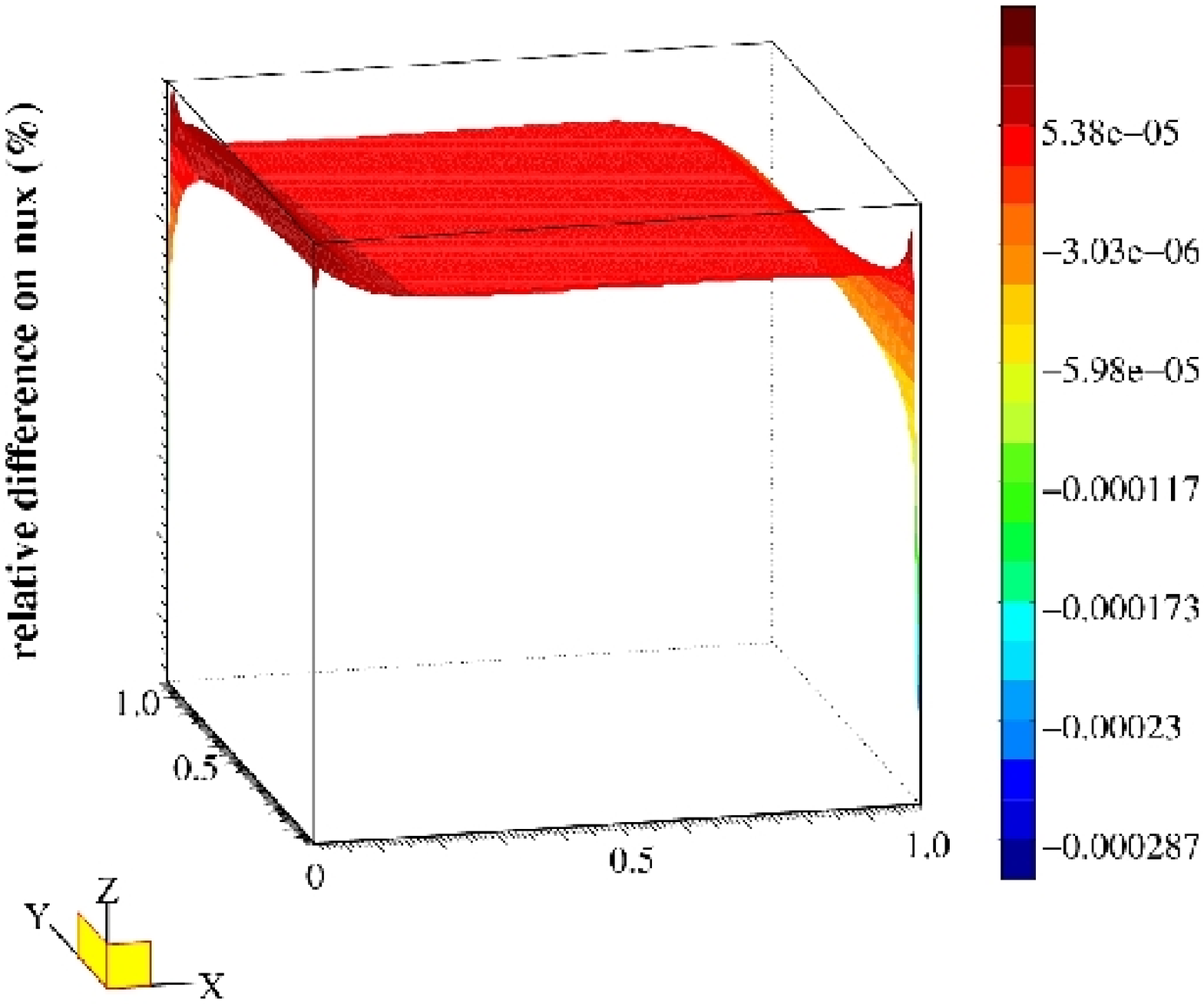,width=5cm,angle=0,clip=} 
       \end{minipage} 
     \end{tabular}
     \\
     \begin{tabular}{cc}
       \begin{minipage}{6cm}
	   \hspace{-0.5cm}
           \epsfig{file=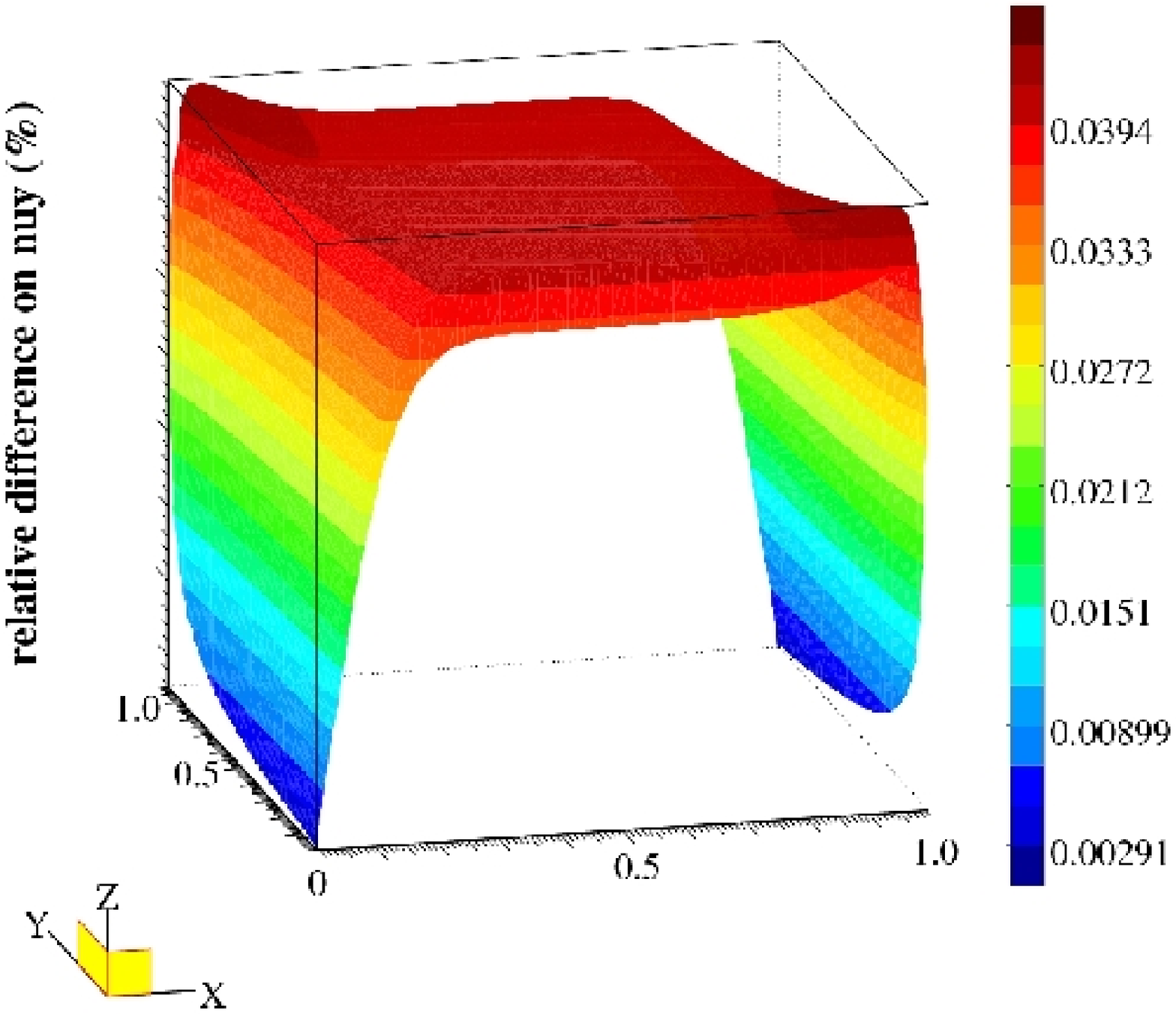,width=5cm,angle=0,clip=}
       \end{minipage} 
       &
       \begin{minipage}{6cm}
	   \hspace{-0.5cm}
          \epsfig{file=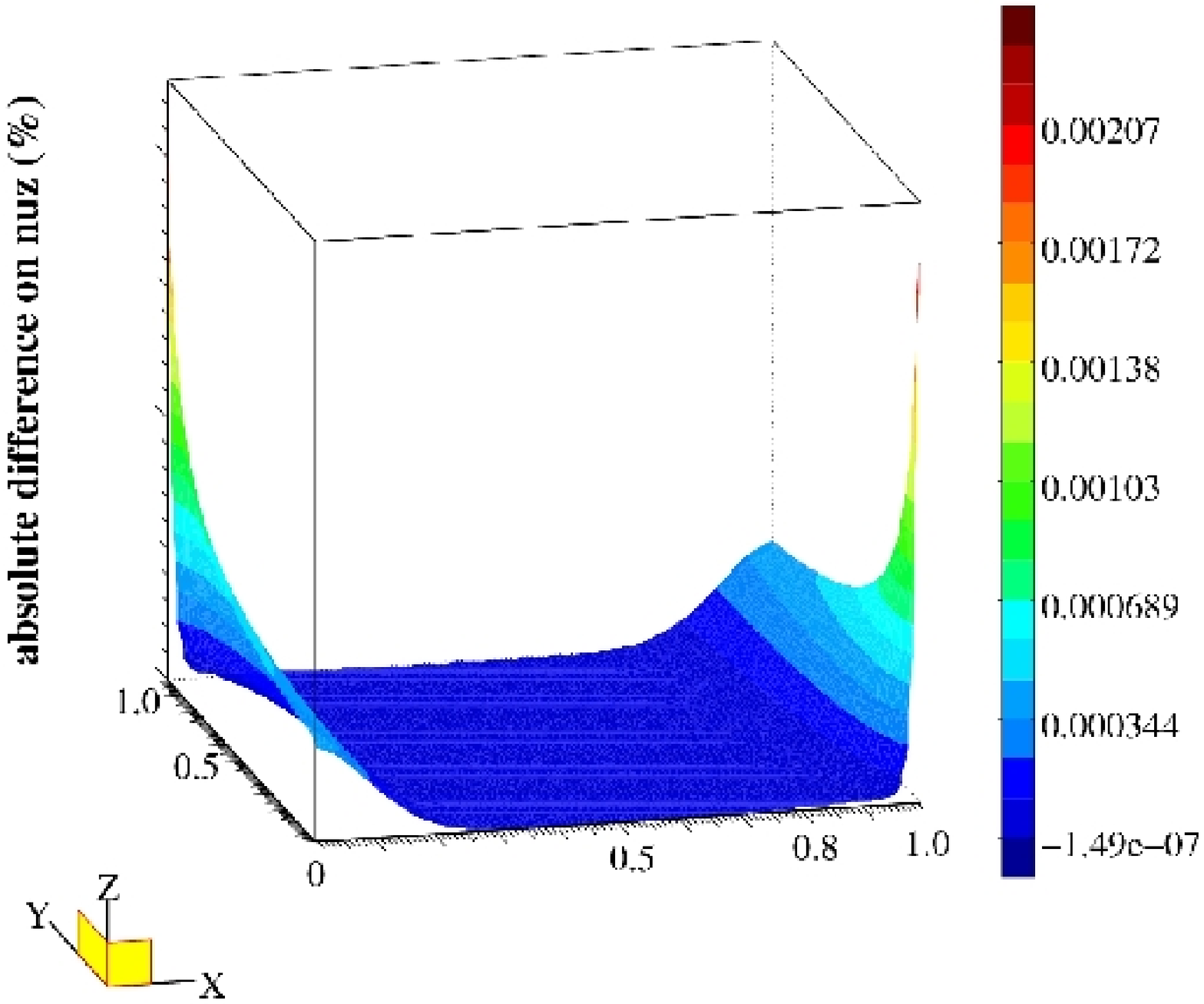,width=5cm,angle=0,clip=} 
       \end{minipage} 
     \end{tabular}
   \end{tabular}
   \caption{Relative difference between the computed
    solution and the exact drift-fluid limit of the resolved AP scheme
    at time $t=0.1$ for $\varepsilon =10^{-6}$; density $n$ (top
    left), $x$-component of the momentum $nu_x$ (top right),
    $y$-component of the momentum $nu_y$ (bottom left). Absolute
    difference between the computed solution and the exact drift-fluid
    limit on the $z$-component of the momentum $nu_z$ (bottom right).}
  \label{DAPfdm6erAPr}   
 \end{figure}

In this case $\varepsilon \ll 1$, we want to
test the consitency of the scheme with the drift-fluid limit.
To do so, we compute the difference 
between the numerical solution and the  
analytical solution (\ref{DA6e1}). This is not the numerical error in the 
conventional sense, since we do not compare the solution with 
the exact solution for the same value of $\varepsilon$ but rather
with the exact solution in the limit $\varepsilon \to 0$. For this
reason, we do not call this quantity, 'the error', but rather 'the
difference with the drift-fluid limit'. We normalize this difference by the exact value of the drift-fluid limit, except 
for $n u_z$ since this value is exactly
zero. We display these quantities  as  functions of $(x,y)$ for
the resolved AP scheme with the value $\varepsilon = 10^{-6}$ on Fig.~\ref{DAPfdm6erAPr}. The
picture is almost the same if we
replace the resolved AP scheme by the resolved conventional
scheme. For this reason, the latter is omitted.  

The maximal relative difference between the computed solution and the
exact drift-fluid limit on the density and momenta for three values of
$\varepsilon$ are given in table~\ref{t3} for the resolved conventional
scheme and in table~\ref{t4} for the resolved AP scheme. They both
show a very good agreement with the drift-fluid limit, which increases
as $\varepsilon$ decreases. This is an expected result since both the
resolved conventional and AP schemes use $\Delta t =
O(\varepsilon)$. For this range of time-steps the resolved
conventional scheme is only slightly more accurate than the resolved
AP scheme.

\begin{figure}[h!]
   \vspace{-0.25cm}
     \begin{tabular}{c}
     \begin{tabular}{cc}
       \begin{minipage}{6cm}
	   \hspace{-0.5cm}
           \epsfig{file=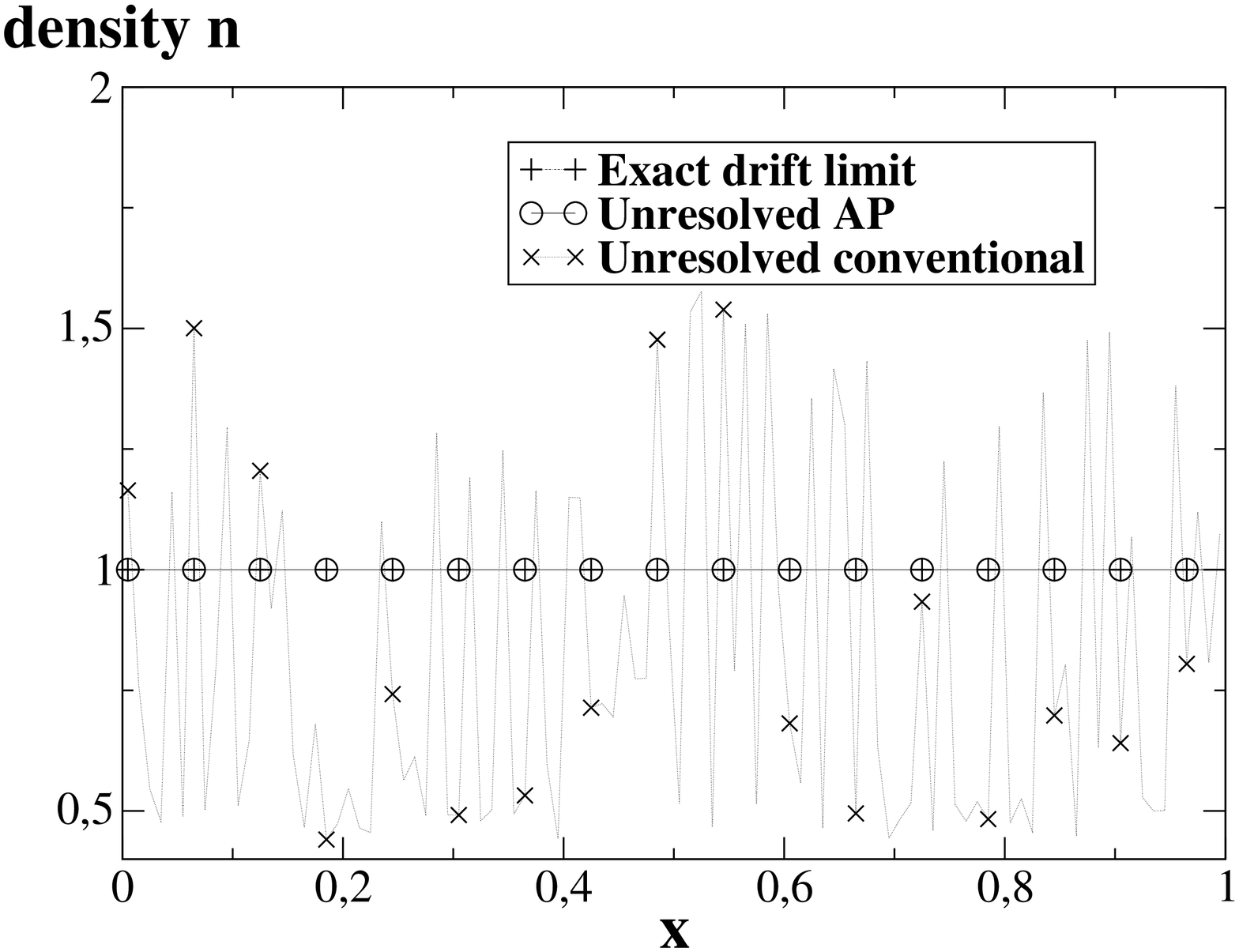,width=5cm,angle=-90,clip=} 
       \end{minipage} 
       &
       \begin{minipage}{6cm}
	   \hspace{-0.5cm}
           \epsfig{file=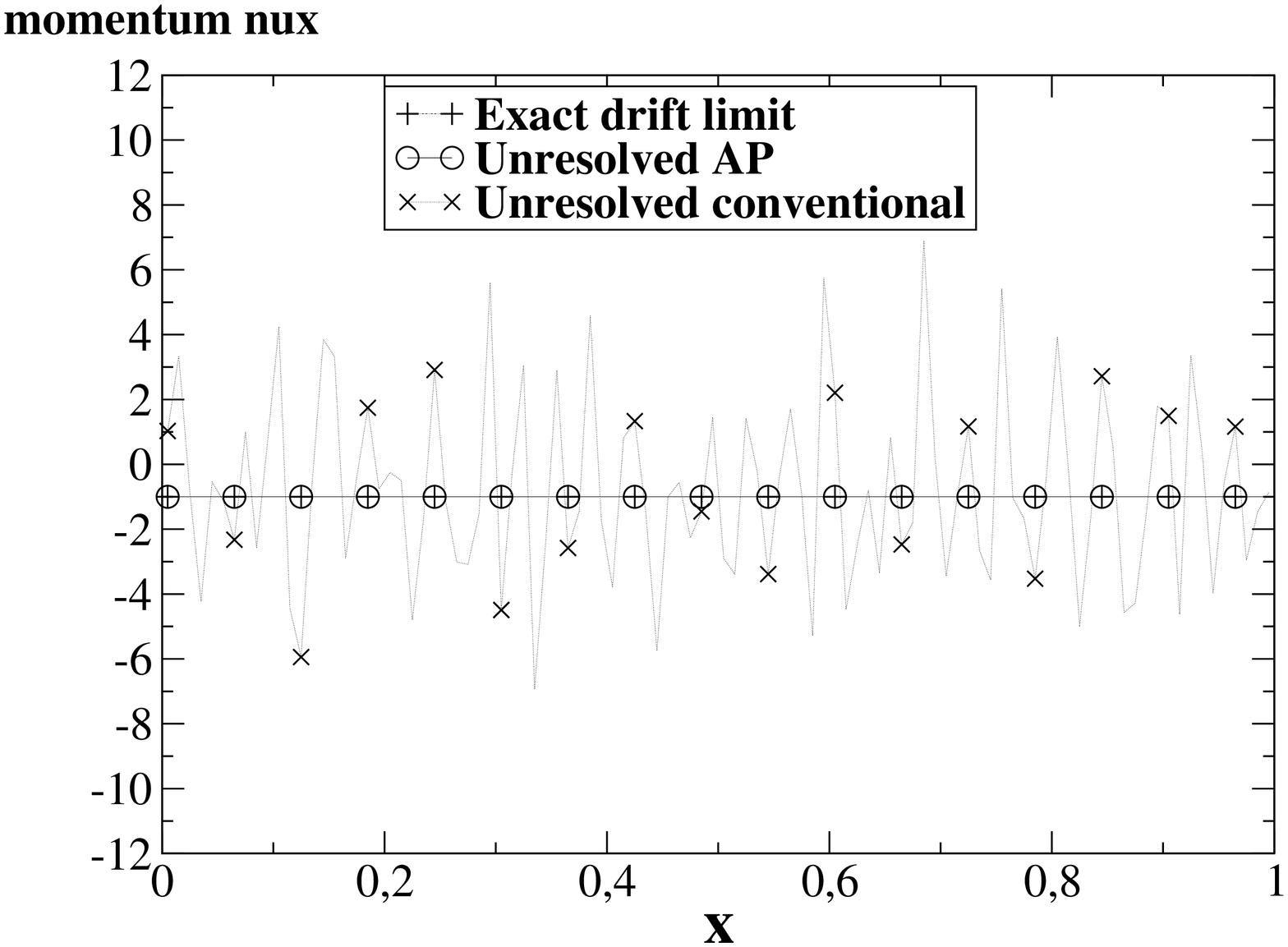,width=5cm,angle=-90,clip=}
       \end{minipage} 
     \end{tabular}
     \\
     \begin{tabular}{cc}
       \begin{minipage}{6cm}
	   \hspace{-0.5cm}
           \epsfig{file=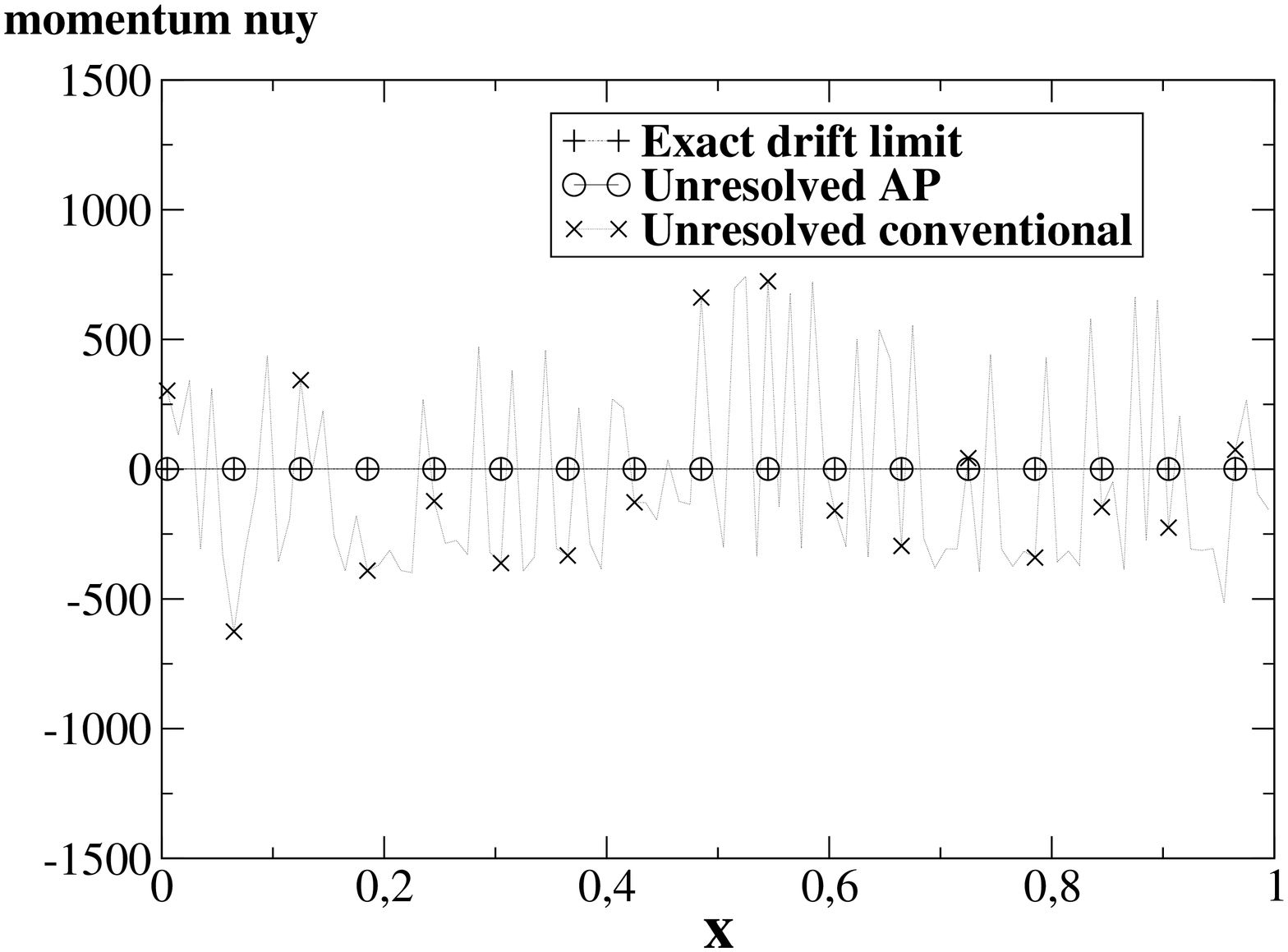,width=5cm,angle=-90,clip=} 
       \end{minipage} 
       &
       \begin{minipage}{6cm}
	   \hspace{-0.5cm}
           \epsfig{file=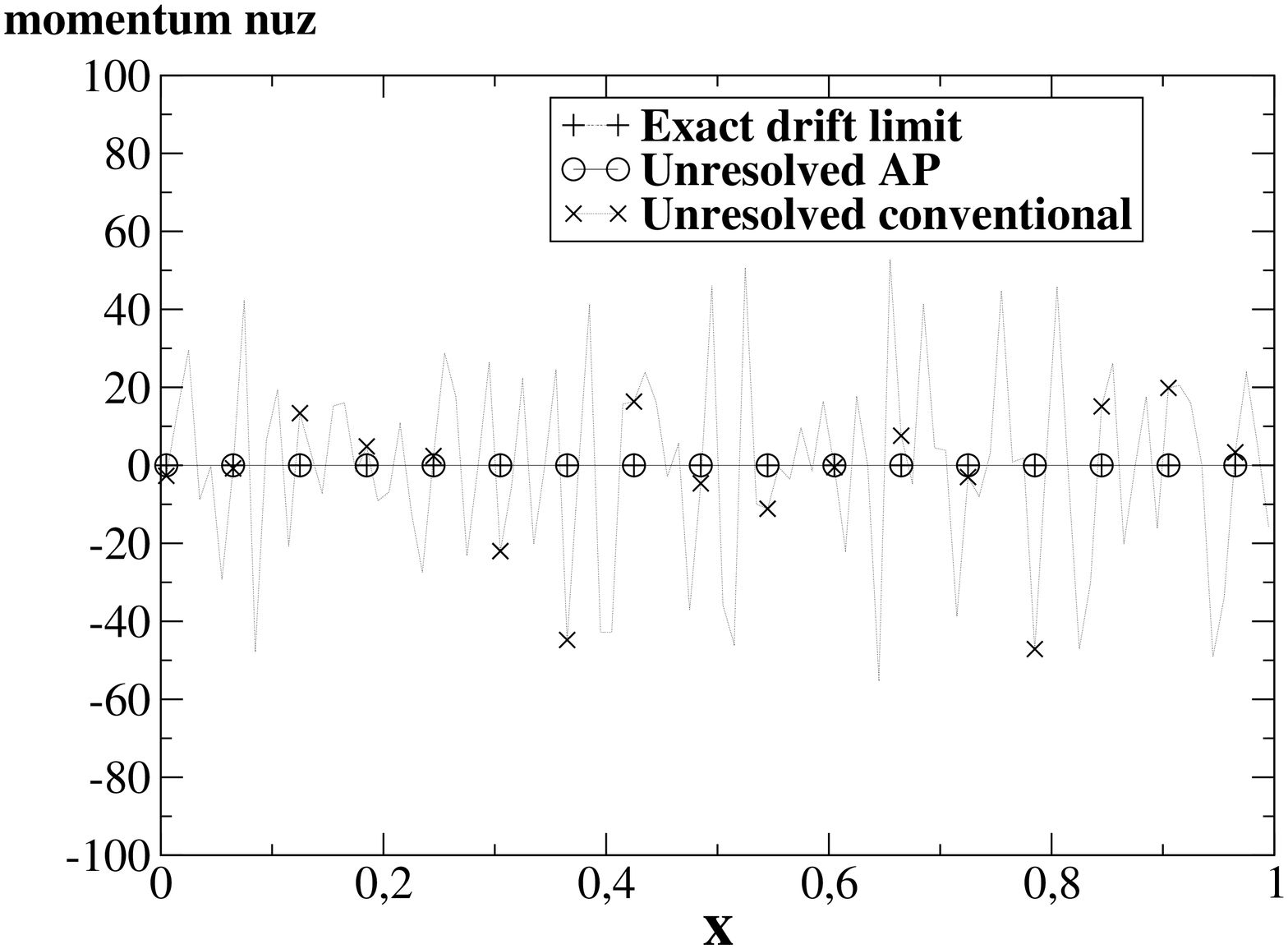,width=5cm,angle=-90,clip=} 
       \end{minipage} 
     \end{tabular}
   \end{tabular}
  \caption{Comparison of the non-resolved conventional scheme
  (crosses), the non-resolved AP scheme (circles) and the exact drift-fluid
  limit (vertical bars) at $t=0.1$ for $\varepsilon = 10^{-6}$ ;
  density $n$ (top left), $x$-component of the momentum $nu_x$ (top
  right), $y$-component of the momentum $nu_y$ (bottom left),
  $z$-component of the momentum $nu_z$ (bottom right).}               
  \label{DAPfdm6ModifClasAPnr} 
\end{figure}

\vspace{0.2cm}
\noindent 
{\bf Unresolved case.} We now examine the unresolved situation, where
$\varepsilon \ll \Delta t$. Results obtained by the non-resolved
conventional and AP schemes for $\varepsilon = 10^{-6}$ are displayed
in Fig.~\ref{DAPfdm6ModifClasAPnr}. We recall that ``non-resolved''
means that the viscosities are computed through (\ref{DA5e59}) and the
time-step through (\ref{DA5e58}) instead of (\ref{DA5e50}),
(\ref{DA5e55}) in the resolved case. 

\begin{figure}[h!]
   \begin{tabular}{c}
     \begin{tabular}{cc}
       \begin{minipage}{6cm}
	   \hspace{-0.5cm}
           \epsfig{file=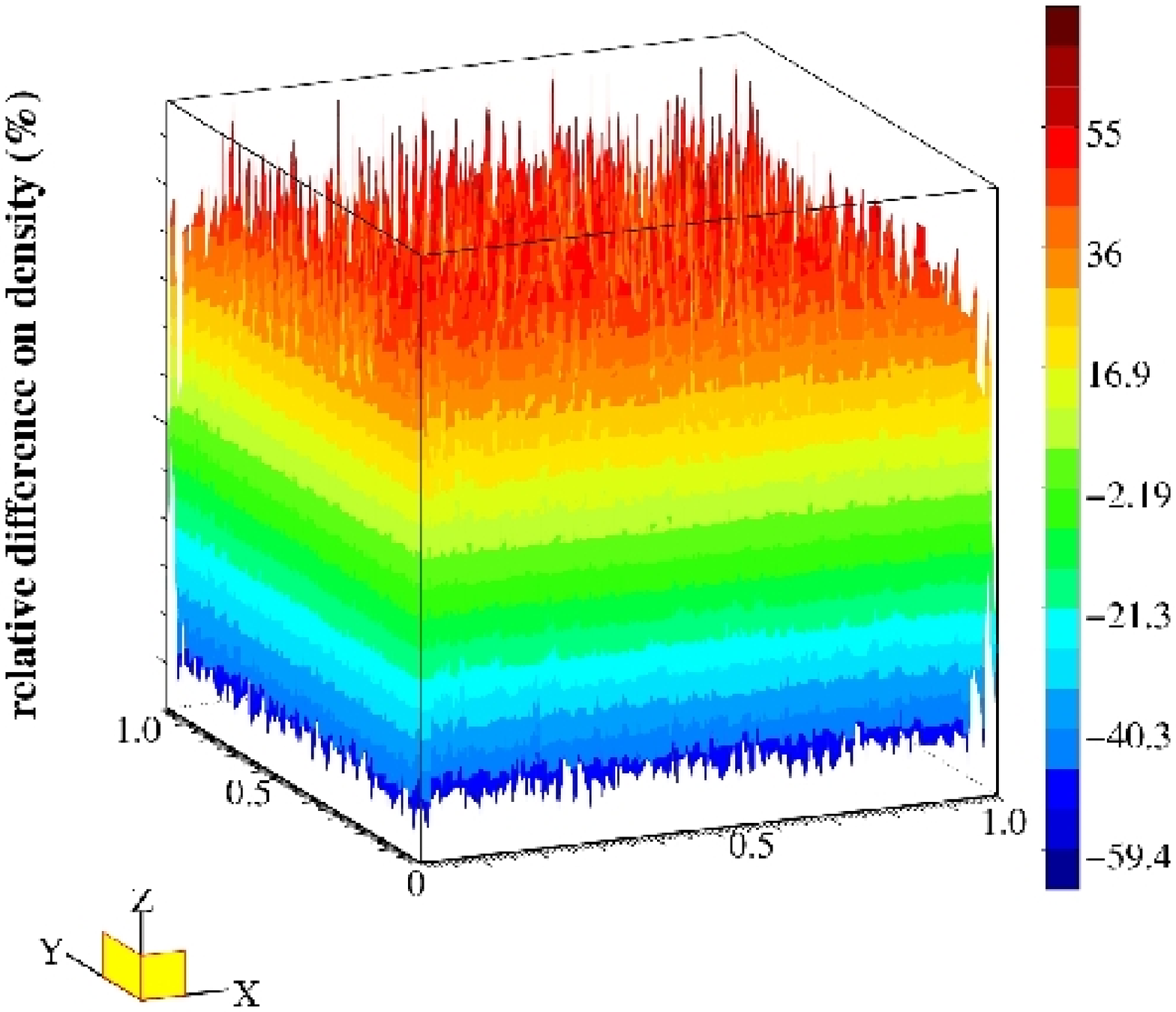,width=5cm,angle=0,clip=} 
       \end{minipage} 
       &
       \begin{minipage}{6cm}
	   \hspace{-0.5cm}
           \epsfig{file=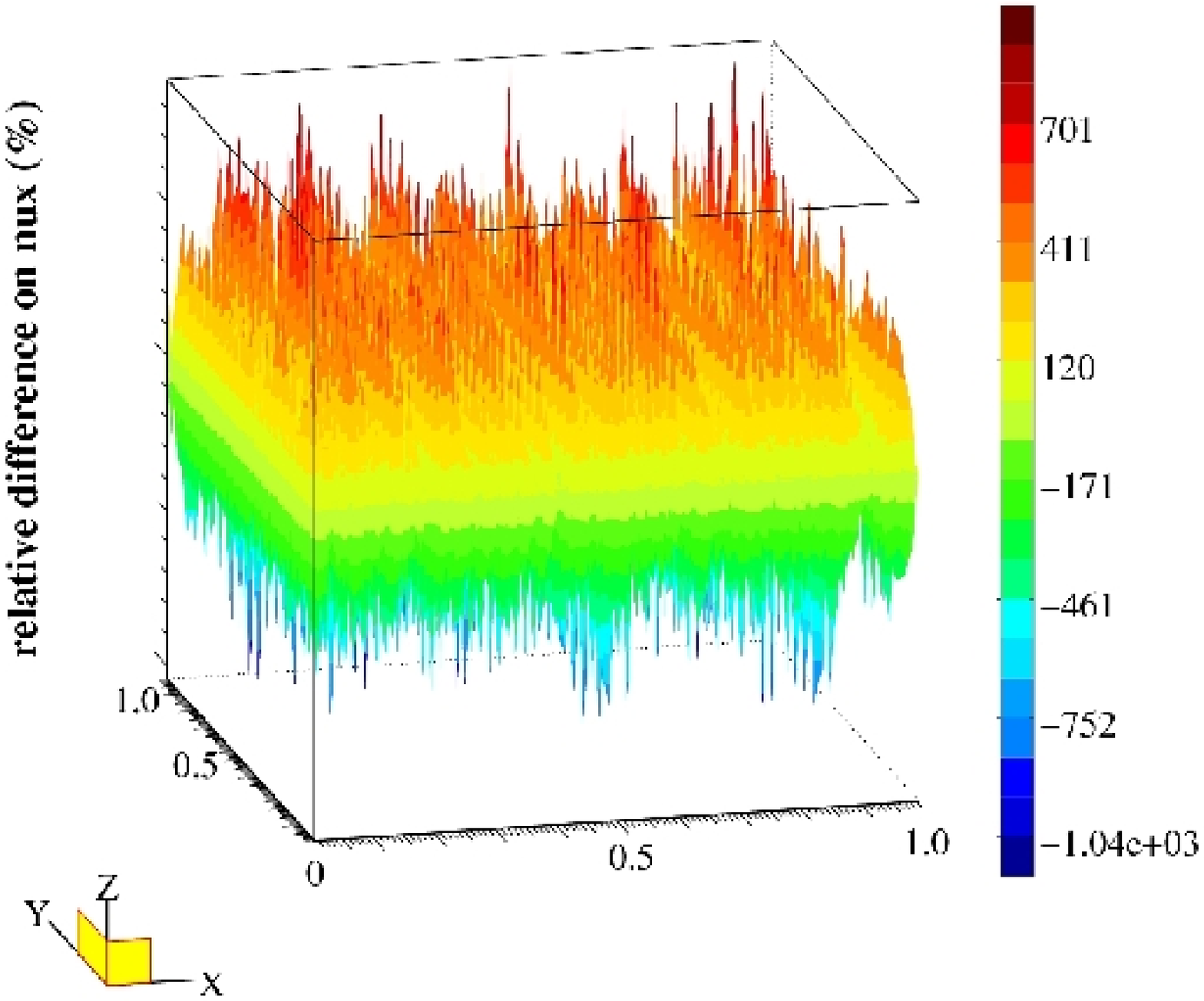,width=5cm,angle=0,clip=}
       \end{minipage} 
     \end{tabular}
     \\
     \begin{tabular}{cc}
       \begin{minipage}{6cm}
	   \hspace{-0.5cm}
           \epsfig{file=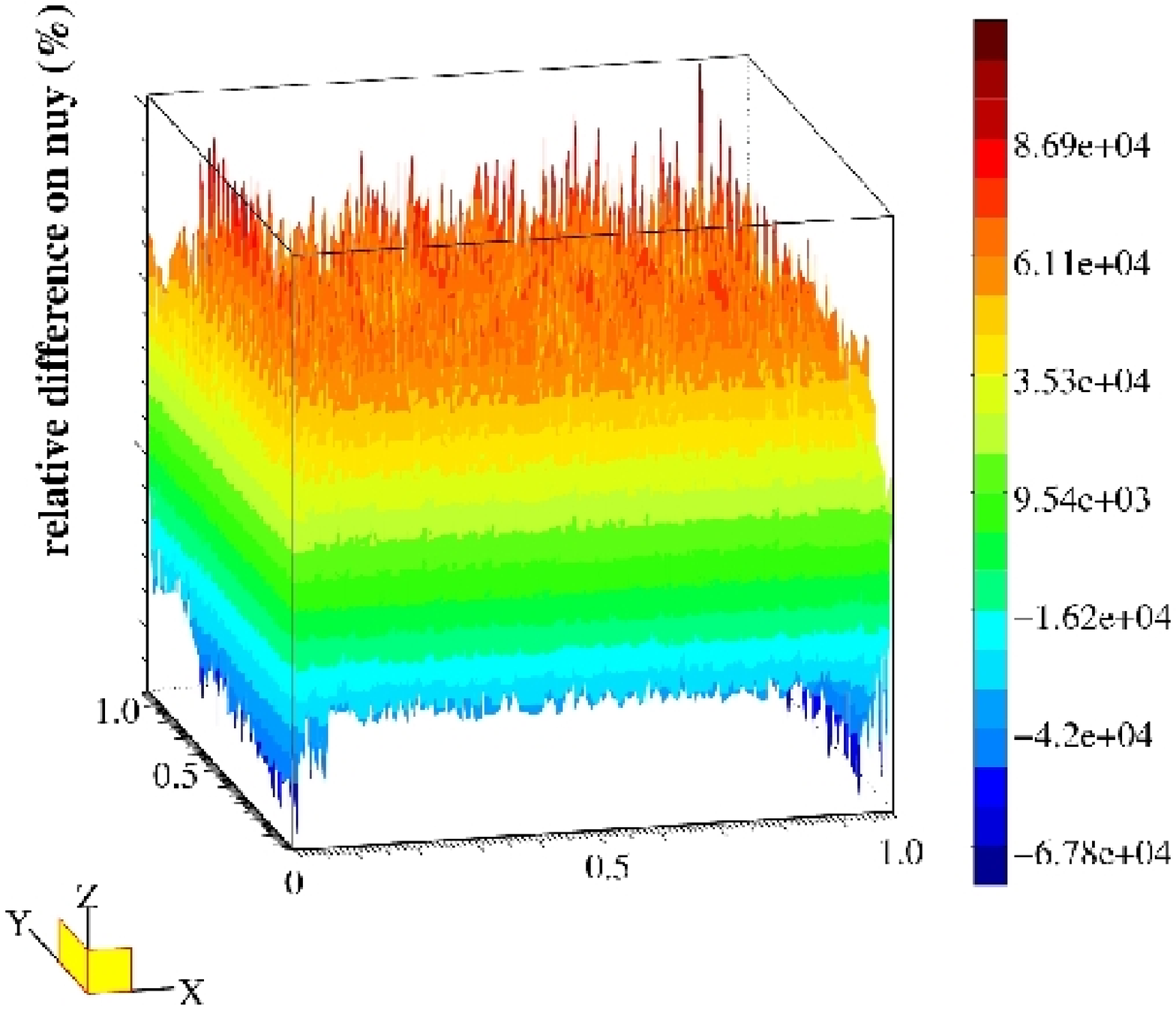,width=5cm,angle=0,clip=} 
       \end{minipage} 
       &
       \begin{minipage}{6cm}
	   \hspace{-0.5cm}
           \epsfig{file=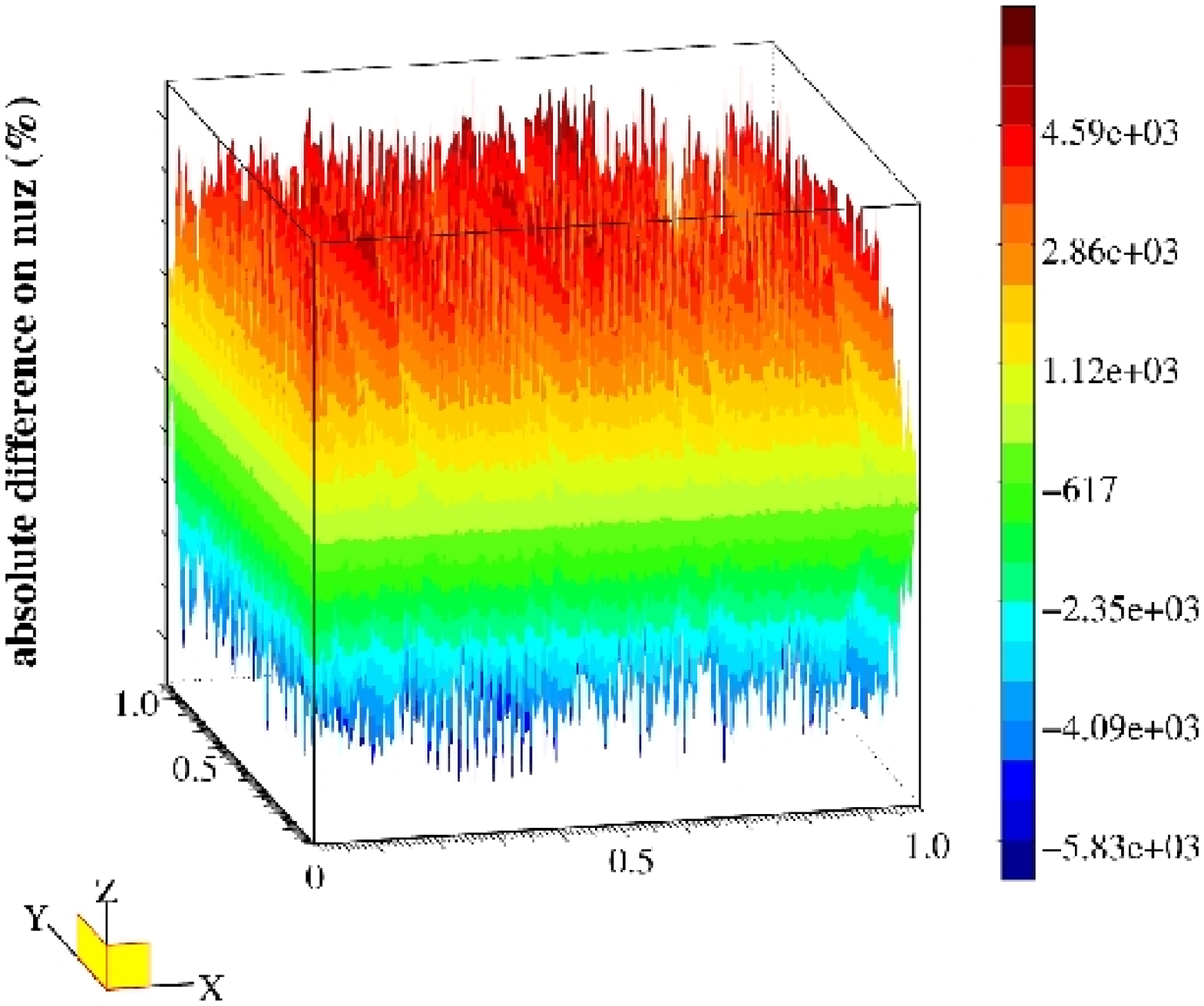,width=5cm,angle=0,clip=} 
       \end{minipage} 
     \end{tabular}
   \end{tabular}
   \caption{Relative difference between the computed solution and the
     exact drift-fluid limit of the non-resolved conventional scheme
     at time $t=0.1$ for $\varepsilon =10^{-6}$; density $n$ (top
     left), $x$-component of the momentum $nu_x$ (top right),
     $y$-component of the momentum $nu_y$ (bottom left). Absolute
     difference between the computed solution and the exact
     drift-fluid limit on the $z$-component of the momentum $nu_z$
     (bottom right).}             
  \label{DAPfdm6erMCnr} 
 \end{figure}

\begin{figure}[h!]
   \begin{tabular}{c}
     \begin{tabular}{cc}
       \begin{minipage}{6cm}
	   \hspace{-0.5cm}
           \epsfig{file=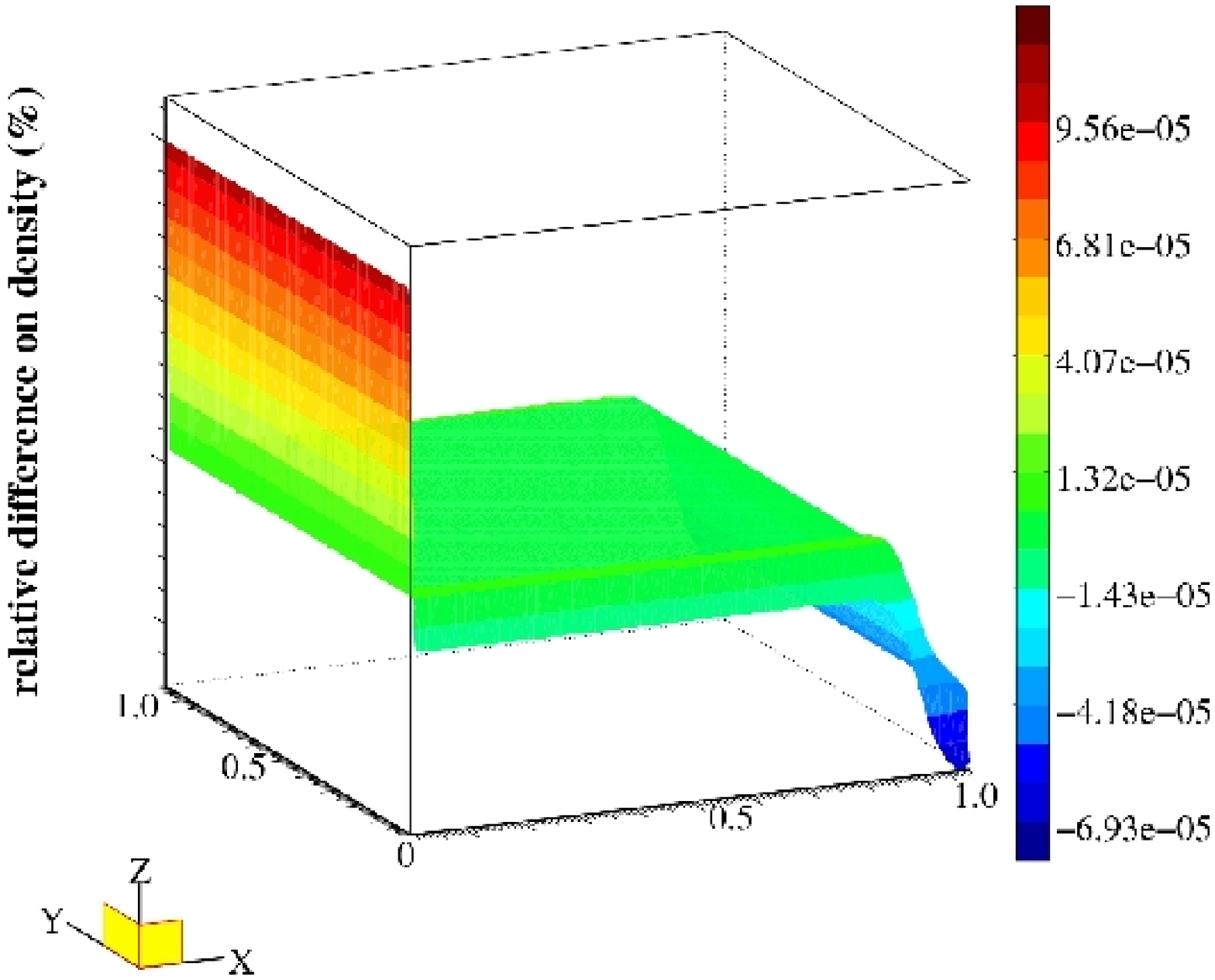,width=5cm,angle=0,clip=} 
       \end{minipage} 
       &
       \begin{minipage}{6cm}
	   \hspace{-0.5cm}
           \epsfig{file=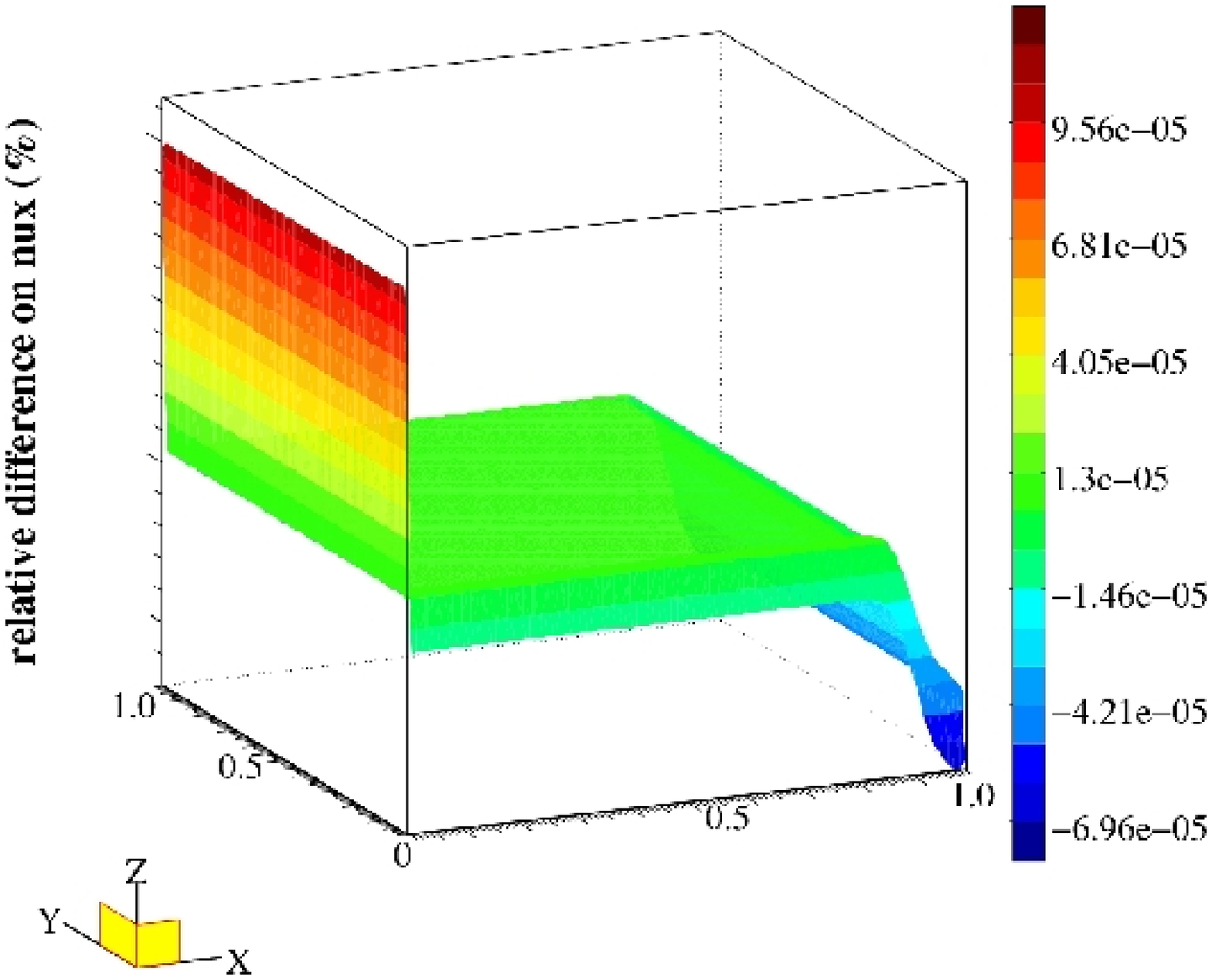,width=5cm,angle=0,clip=} 
       \end{minipage} 
     \end{tabular}
     \\
     \begin{tabular}{cc}
       \begin{minipage}{6cm}
	   \hspace{-0.5cm}
           \epsfig{file=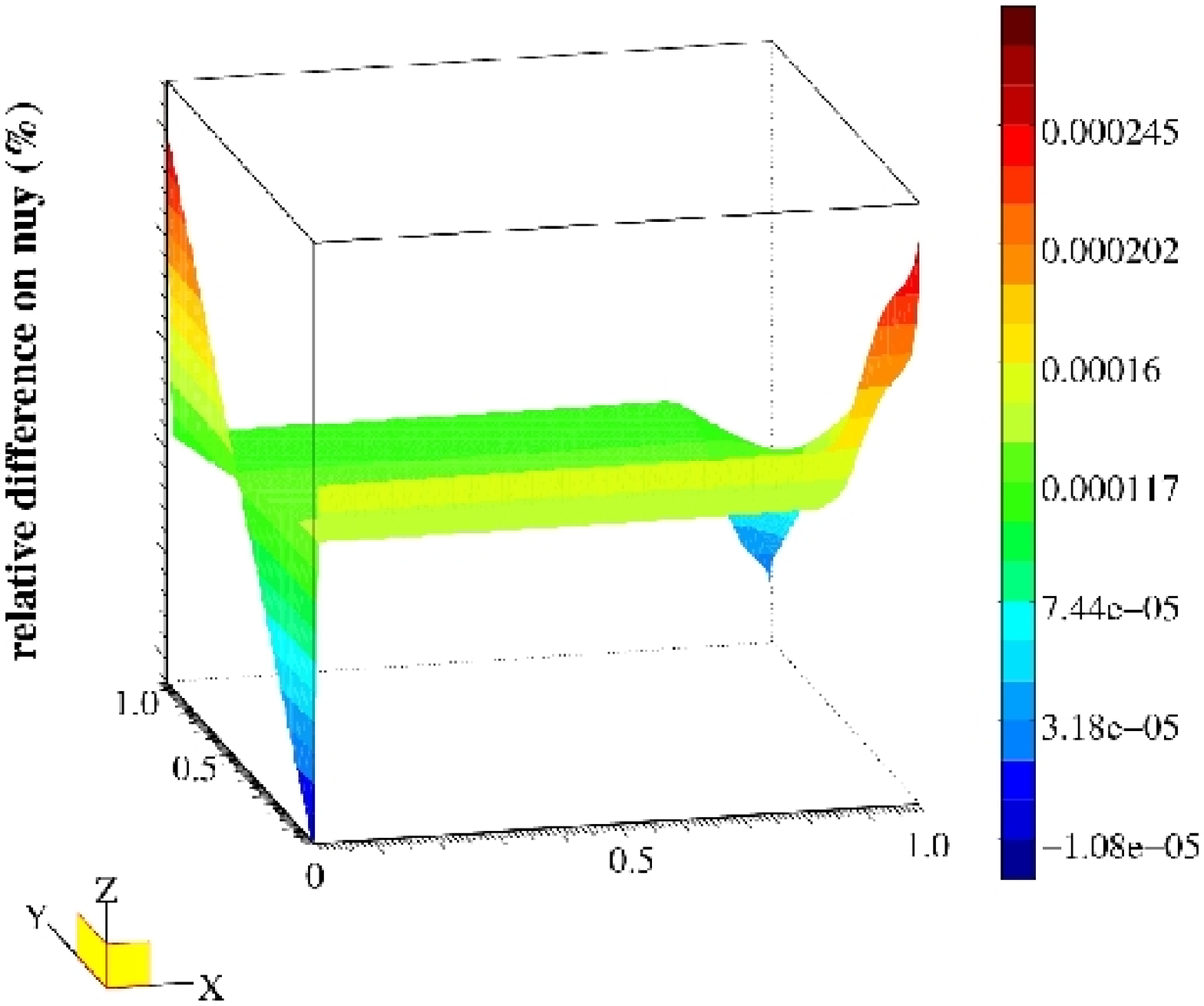,width=5cm,angle=0,clip=}
       \end{minipage} 
       &
       \begin{minipage}{6cm}
	   \hspace{-0.5cm}
           \epsfig{file=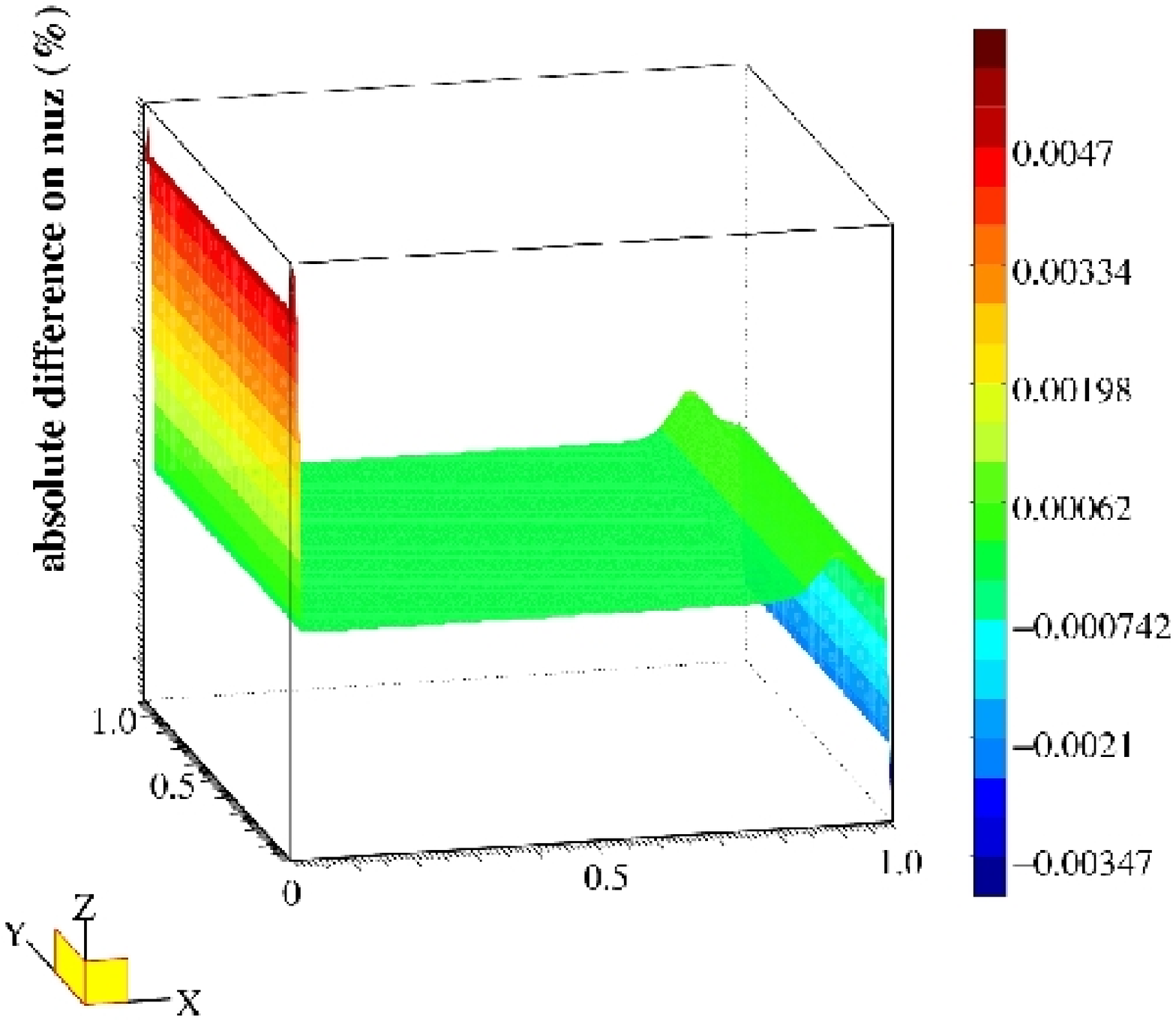,width=5cm,angle=0,clip=} 
       \end{minipage} 
     \end{tabular}
   \end{tabular}
   \caption{Relative difference between the computed solution and the
     exact drift-fluid limit of the non-resolved AP scheme at time
     $t=0.1$ for $\varepsilon =10^{-6}$; density $n$ (top left),
     $x$-component of the momentum $nu_x$ (top right), $y$-component
     of the momentum $nu_y$ (bottom left). Absolute difference between
     the computed solution and the exact drift-fluid limit on the
     $z$-component of the momentum $nu_z$ (bottom right).}            
  \label{DAPfdm6erAPnr} 
 \end{figure}

Clearly, the computed solutions with the
non-resolved conventional scheme are unstable, while those 
calculated with the non-resolved AP scheme remain stable and
consistent with the exact drift-fluid limit. The difference between the
computed solution and the exact drift-fluid limit on the density and
momenta (relative difference for $n$, $nu_x$ and $nu_y$ and absolute
difference for $nu_z$) of the solution for $\varepsilon = 10^{-6}$ are
given in Fig.~\ref{DAPfdm6erMCnr} for the non-resolved conventional
scheme and in Fig.~\ref{DAPfdm6erAPnr} for the non-resolved AP
scheme. They confirm the stability and accuracy of the non-resolved AP
scheme and the instability of the non-resolved conventional scheme.

In Table~\ref{t5}, the difference between the computed solution and
the exact drift-fluid limit for the non-resolved conventional scheme
for the three values of $\varepsilon$ are given, and similarly for the
non-resolved AP scheme in Table~\ref{t6}. Again, the consistency of
the non-resolved AP scheme with the drift-fluid limit on the one hand,
and the instability of the non-resolved conventional scheme on the
other hand, are confirmed.  

\begin{figure}[h!]
   \centering\epsfig{file=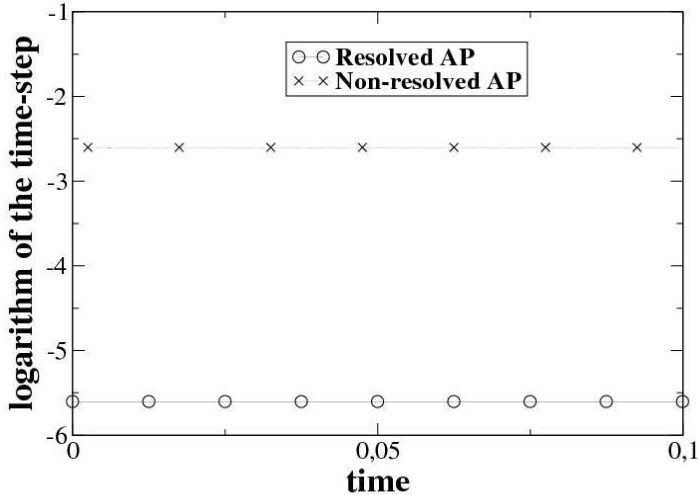,width=8cm,angle=-90,clip}      
  \caption{Time-step (log scale) as a function of time for the
    resolved and non-resolved AP schemes when $\varepsilon =
    10^{-6}$.}            
  \label{DAPfdm6time} 
\end{figure}

Fig. \ref{DAPfdm6time},
shows the evolution of the time step with
respect to time in the case $\varepsilon = 10^{-6}$
for both the unresolved and resolved AP schemes. 
We recall that the time step is not
fixed once for all, but is recomputed at each time step using the CFL
condition.   However, we can see that, in log scale, the time
step remains about constant. The time step for the unresolved AP
scheme is about 3 decades larger than for the resolved AP scheme. 
In Table~\ref{t7}, we compare the
time-step to the scaled gyro-period and we notice that it is of the
same order of magnitude for the resolved AP scheme, as it should, and
it is much larger (up to 4 decades !) for the unresolved AP
scheme. 

It is even more interesting to compare the CPU time between the
unresolved AP scheme and the conventional scheme. Indeed, 
the AP scheme involves more complex computations than
the conventional scheme, such as the inversion of linear systems.  
It is therefore legitimate to wonder whether this additional work 
does not completely balance the gain obtained through the use 
of larger time steps. To check this point, 
the CPU times for the conventional and non-resolved AP schemes
for three values of $\varepsilon$ are given in table~\ref{t8}.
We see that the gain in CPU time is up to almost 4 decades 
with the smallest value of $\varepsilon$. The gain scales
about like $\sqrt \varepsilon$ as it should.

These comparisons show that the proposed AP schemes are very powerful in
handling the numerical approximation of drift-fluid asymptotics.

%%%%%%%%%%%%%%%%%%%%%%%%%%%%%%%%%%%%Modif vs APr%%%%%%%%%%%%%% 

\begin{table}[h!] 
  \caption{Maximum of relative difference between the computed
    solution and the exact drift-fluid limit (\%) from the resolved
    conventional scheme on the density $n$, $x$-component of the
    momentum $nu_x$, $y$-component of the momentum $nu_y$ and absolute
    difference between the computed solution and the exact drift-fluid
    limit on the $z$-component of the momentum $nu_z$.} \label{t3}  
  \begin{center}
    \begin{tabular}{ccccc}
      \hline\hline $\varepsilon$ & $n$ & $nu_x$ & $nu_y$ & $nu_z$\\  
      \hline $10^{-5}$ & $0.0087$ & $0.00714$ & $0.145$ & $0.0174$\\ 
      \hline $10^{-6}$ & $8.68\,10^{-5}$ & $0.000274$ & $0.0455$ &
      $0.00204$\\   
      \hline $1.5\,10^{-8}$ & $1.29\,10^{-6}$ & $1.25\,10^{-6}$ &
      $0.00554$& $3.11\,10^{-5}$\\   
      \hline\hline
    \end{tabular}
  \end{center}
\end{table}

\begin{table}[h!] 
  \caption{Maximum of relative difference between the computed
    solution and the exact drift-fluid limit (\%) from the resolved AP
    scheme on the density $n$, $x$-component of the momentum $nu_x$,
    $y$-component of the momentum $nu_y$, and maximum of absolute
    difference between the computed solution and the exact drift-fluid
    limit  on the $z$-component of the momentum $nu_z$.} \label{t4}  
  \begin{center}
    \begin{tabular}{ccccc}
      \hline\hline $\varepsilon$ & $n$ & $nu_x$ & $nu_y$ & $nu_z$\\  
      \hline $10^{-5}$ & $0.00873$ & $0.0074$ & $0.126$ & $0.017$\\ 
      \hline $10^{-6}$ & $8.72\,10^{-5}$ & $0.000287$ & $0.0394$ &
      $0.00207$\\   
      \hline $1.5\,10^{-8}$ & $1.3\,10^{-6}$ & $1.25\,10^{-6}$ &
      $0.0048$& $3.16\,10^{-5}$\\   
      \hline\hline
    \end{tabular}
  \end{center}
\end{table}

%%%%%%%%%%%%%%%%%%%%%%%%%%%%%%%%%%%%Modif instable nr vs APnr%%%%%%%%%%%%%% 

\begin{table}[h!] 
  \caption{Maximum of relative difference between the computed
    solution and the exact drift-fluid limit (\% ) from the
    non-resolved conventional scheme on the density $n$, $x$-component
    of the momentum $nu_x$, $y$-component of the momentum $nu_y$ and
    maximum of absolute difference between the computed solution and
    the exact drift-fluid limit on the $z$-component of the momentum
    $nu_z$.} \label{t5}   
  \begin{center}
    \begin{tabular}{ccccc}
      \hline\hline $\varepsilon$ & $n$ & $nu_x$ & $nu_y$ & $nu_z$\\  
      \hline $10^{-5}$ & $75.2$ & $1.98\,10^3$ & $2.57\,10^3$ &
      $6.58\,10^3$\\ 
      \hline $10^{-6}$ & $59.4$ & $1.04\,10^3$ & $8.69\,10^4$ &
      $5.83\,10^3$\\   
      \hline $1.5\,10^{-8}$ & $58$ & $118$ & $6.44\,10^5$&
      $4.64\,10^3$\\   
      \hline\hline
    \end{tabular}
  \end{center}
\end{table}  

\begin{table}[h!] 
  \caption{Maximum of relative difference between the computed
    solution and the exact drift-fluid limit (\%) from the
    non-resolved AP scheme on the density $n$, $x$-component of the
    momentum $nu_x$, $y$-component of the momentum $nu_y$, and maximum
    of absolute difference between the computed solution and the exact 
    drift-fluid limit on the $z$-component of the momentum $nu_z$.} 
    \label{t6}  
  \begin{center}
    \begin{tabular}{ccccc}
      \hline\hline $\varepsilon$ & $n$ & $nu_x$ & $nu_y$ & $nu_z$\\  
      \hline $10^{-5}$ & $0.00104$ & $0.00104$ & $0.00255$ & $0.0447$\\ 
      \hline $10^{-6}$ & $9.56\,10^{-5}$ & $6.96\,10^{-5}$ &
      $0.000245$ & $0.0047$\\   
      \hline $1.5\,10^{-8}$ & $2.75\,10^{-6}$ & $7.12\,10^{-6}$ &
      $0.000554$& $0.00389$\\   
      \hline\hline
    \end{tabular}
  \end{center}
\end{table}

\begin{table}[h!] 
  \caption{Logarithms of the gyro-period $\tau$, maximum of
    time-steps used in the resolved AP scheme (AP) and non-resolved AP
    scheme (NAP)}\label{t7}
  \begin{center}
    \begin{tabular}{cccc}
      \hline\hline $\varepsilon$ & $\tau$ & $AP$ & $NAP$\\  
      \hline $10^{-5}$ & $-5$ & $-5.09$ & $-2.6$\\ 
      \hline $10^{-6}$ & $-6$ & $-5.6$ & $-2.6$\\  
      \hline $1.5\,10^{-8}$ & $-7.83$ & $-6.51$ & $-2.6$\\  
      \hline\hline
    \end{tabular}
  \end{center}
\end{table}

\begin{table}[h!] 
  \caption{ CPU time (in s) used in the resolved conventional
    scheme~(CONV) and non-resolved AP scheme~(NAP) for computing the
    Euler-Lorentz model at final time $t_{fin}$ (in s). Ratio of the CPU time of the conventional to the non-resolved AP schemes. }\label{t8}
  \begin{center}
    \begin{tabular}{ccccc}
      \hline\hline $\varepsilon$ & $t_{fin}$ & CONV & NAP & CONV/NAP  \\  
      \hline $10^{-5}$ & $1.00$ & $4940.32$ & $13.84$ & $357$\\ 
      \hline $10^{-6}$ & $0.1$  & $1584.21$ & $1.39$ & $1140$\\  
      \hline $1.5\,10^{-8}$ & $0.01$ & $1149.54$ & $0.17$ & $6762$\\  
      \hline\hline
    \end{tabular}
  \end{center}
\end{table}

%%%%%%%%%%%%%%%%%%%%%%%%%%%%%%%%%%%%%%%%%%%%%%%%%%%%%%%%%%%%%%%%

\subsection{Simulations for $\varepsilon=1$}
\label{DA6_2}

When $\varepsilon = O(1)$, the resolved and unresolved AP schemes 
are almost similar and we want to show that they give similar
results as the resolved conventional scheme. In this way, we show
that the AP scheme is as good as the conventional scheme 
when $\varepsilon = O(1)$. We have already shown in the previous
section that the former is much better than the latter when
$\varepsilon \ll 1$.

\begin{figure}[h!]
   \vspace{-0.25cm}
     \begin{tabular}{c}
     \begin{tabular}{cc}
       \begin{minipage}{6cm}
	   \hspace{-0.5cm}
           \epsfig{file=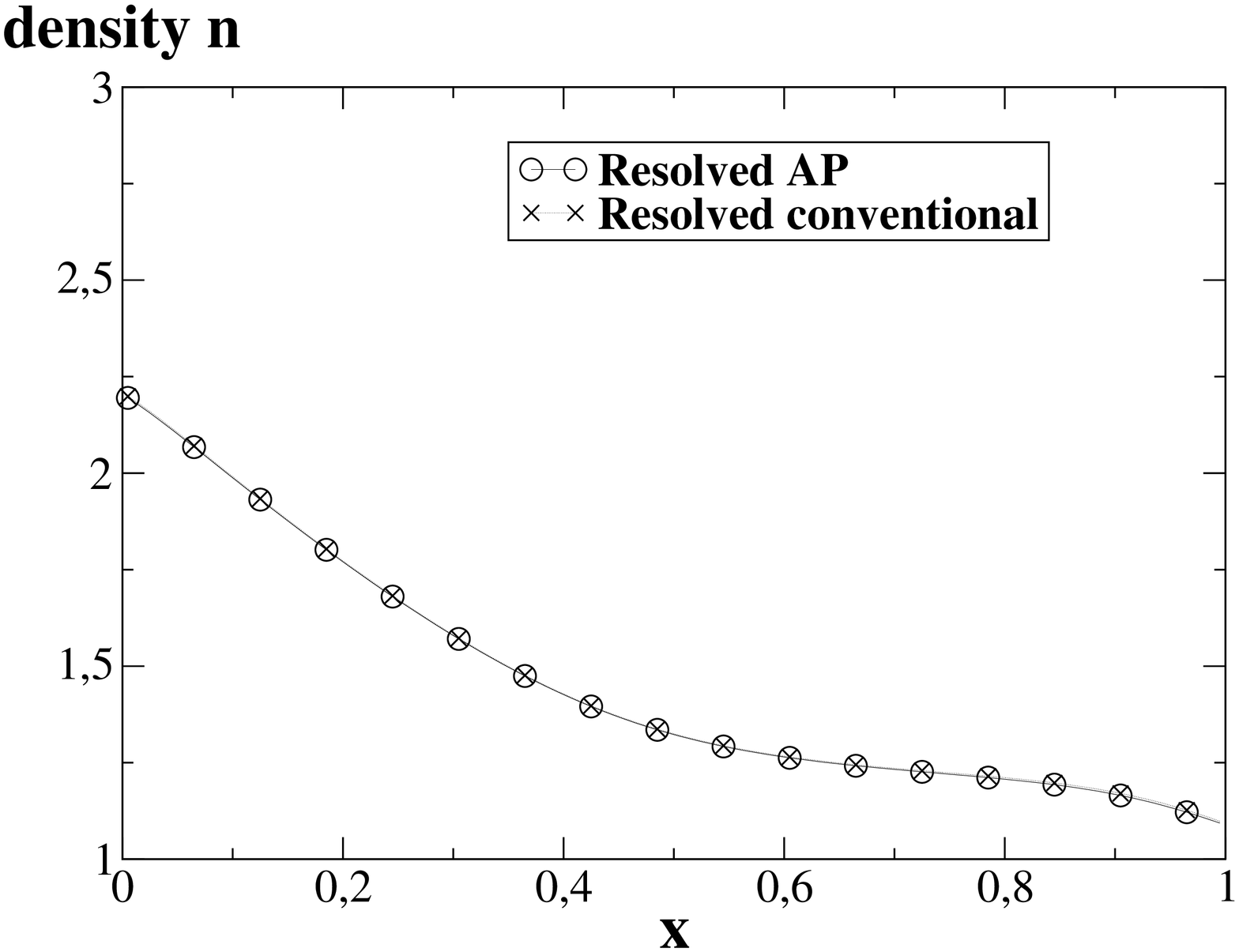,width=5cm,angle=-90,clip=} 
       \end{minipage} 
       &
       \begin{minipage}{6cm}
	   \hspace{-0.5cm}
           \epsfig{file=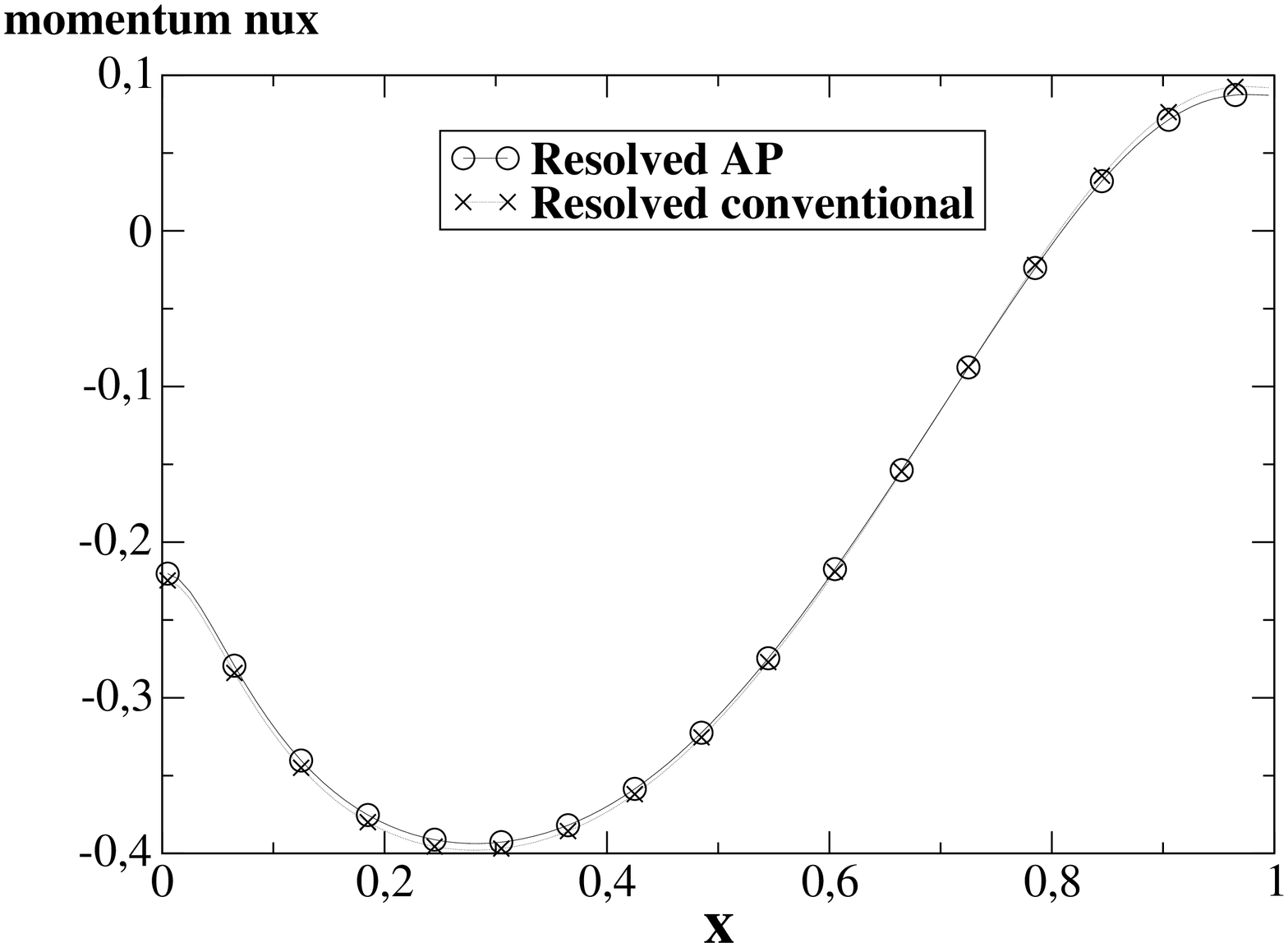,width=5cm,angle=-90,clip=}
       \end{minipage} 
     \end{tabular}
     \\
     \begin{tabular}{cc}
       \begin{minipage}{6cm}
	   \hspace{-0.5cm}
           \epsfig{file=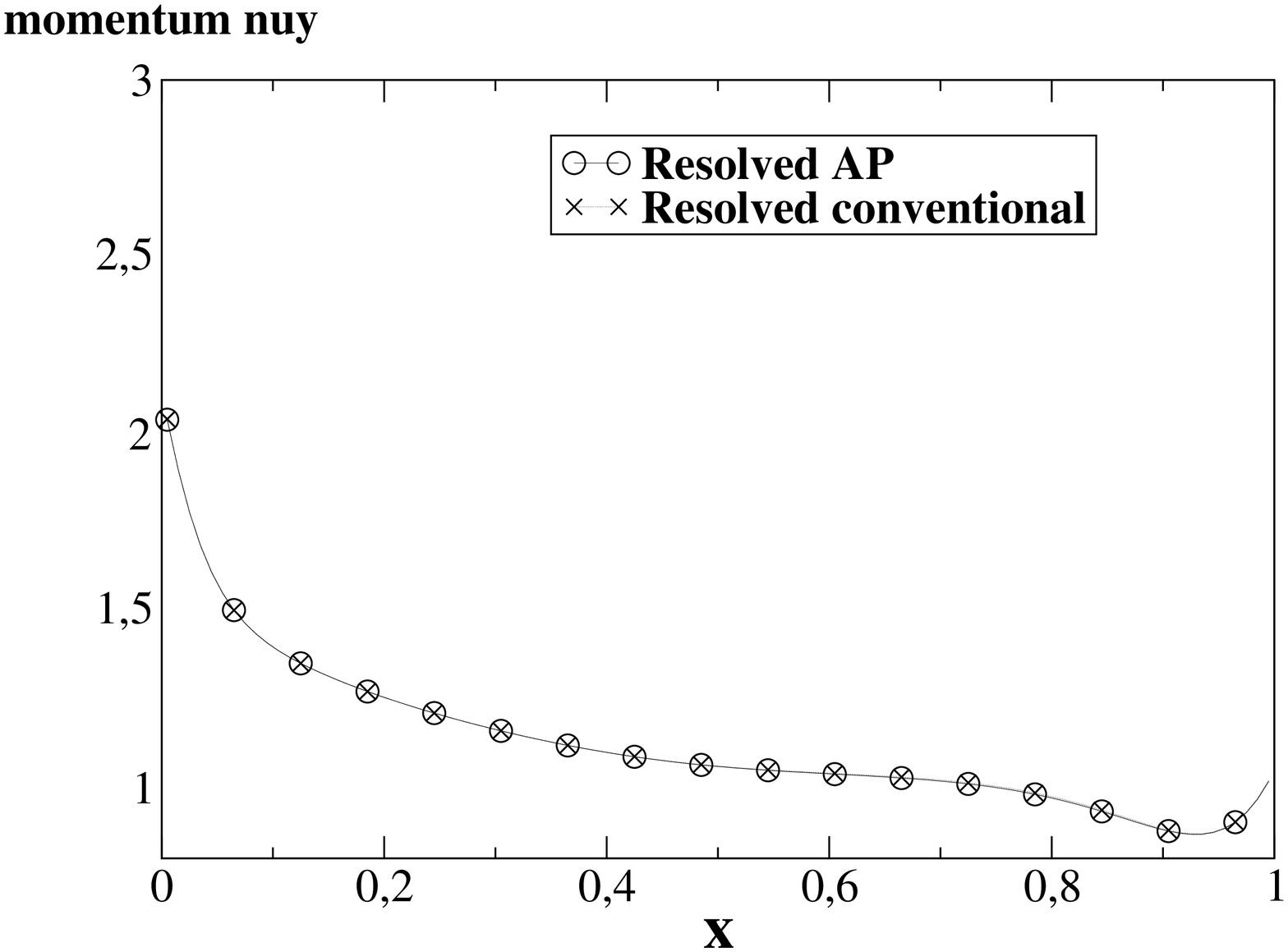,width=5cm,angle=-90,clip=} 
       \end{minipage} 
       &
       \begin{minipage}{6cm}
	   \hspace{-0.5cm}
           \epsfig{file=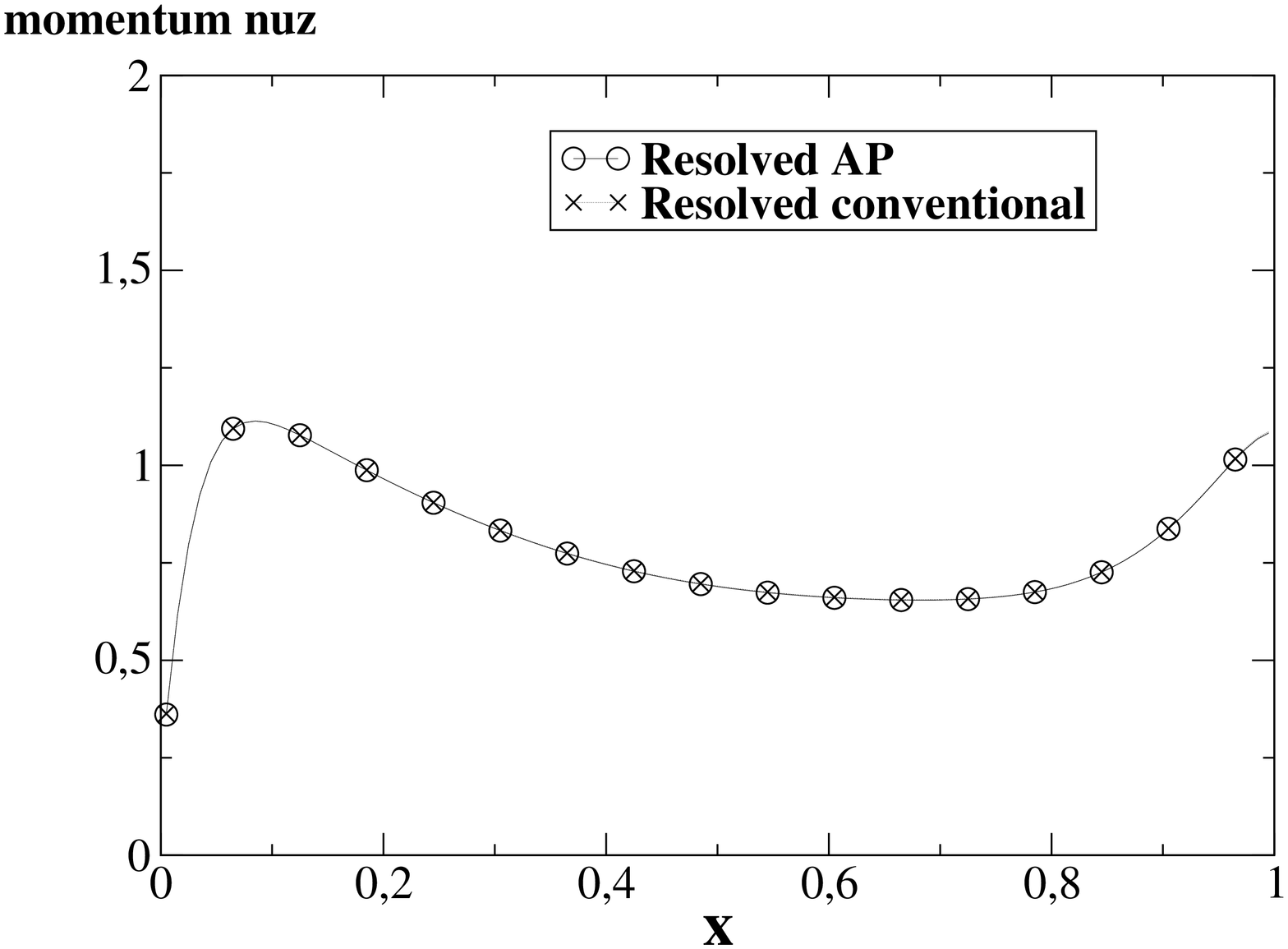,width=5cm,angle=-90,clip=} 
       \end{minipage} 
     \end{tabular}
   \end{tabular}
  \caption{Comparison of the  conventional scheme (crosses), and
  resolved AP scheme (circles)  at $t=1$ for $\varepsilon = 1$ ;
  density $n$ (top left), $x$-component of the momentum $nu_x$ (top
  right), $y$-component of the momentum $nu_y$ (bottom left),
  $z$-component of the momentum $nu_z$ (bottom right).} 
  \label{DAPf1ModifClasAPr} 
\end{figure}

When $\varepsilon=1$, the exact solution is no longer the drift-fluid limit
solution. So, we restict ourselves to a comparison between the 
resolved AP scheme and the resolved conventional scheme. This 
comparison is shown in Fig.~\ref{DAPf1ModifClasAPr}. We notice that the
calculated solutions with the two schemes are
indistinguishable. Therefore for $\varepsilon$ of order of 1, both the
resolved AP scheme and resolved conventional scheme are
comparable.

%%%%%%%%%%%%%%%%%%%%%%%%%%%%%%%%%%%%%%%%%%%%%%%%%%%%%%%%%%%%%%%%

\subsection{Simulations for unprepared conditions}
\label{DA6_3}

For the sake of completeness, we show some numerical results
obtained with unprepared boundary conditions. For this purpose, 
we introduce a second parameter $\varepsilon' = 10^{-2}$ and
use initial and boundary conditions as given by table \ref{t1} 
with $\varepsilon$ replaced by $\varepsilon'$. On the other hand, 
$\varepsilon$ is kept at the value $\varepsilon = 10^{-6}$ in the 
model (\ref{ec}), (\ref{em}) and in the scheme (\ref{DA5e1}), 
(\ref{DA5e2}). 

\begin{figure}[h!]
   \vspace{-0.25cm}
     \begin{tabular}{c}
     \begin{tabular}{cc}
       \begin{minipage}{6cm}
	   \hspace{-0.5cm}
           \epsfig{file=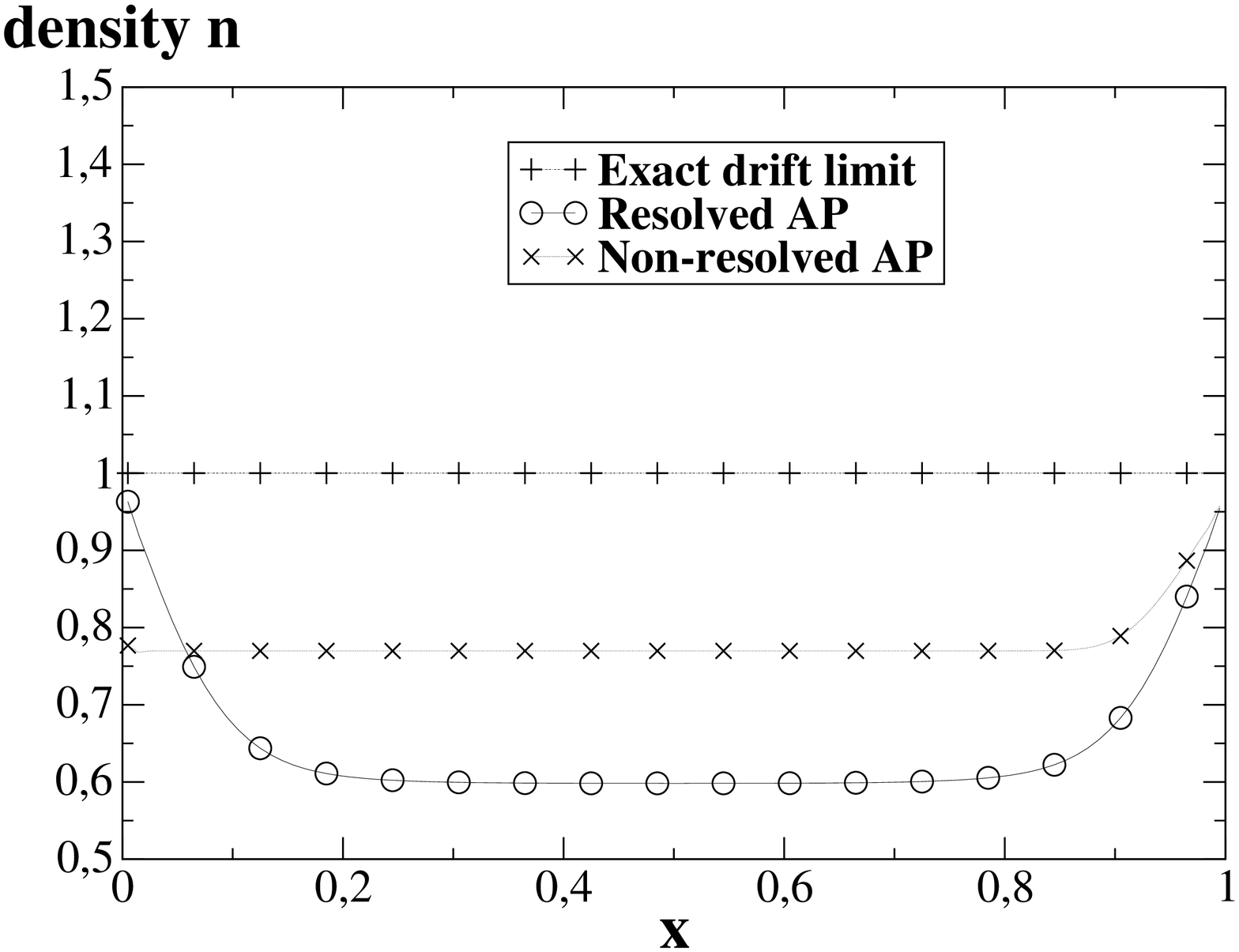,width=5cm,angle=-90,clip=} 
       \end{minipage} 
       &
       \begin{minipage}{6cm}
	   \hspace{-0.5cm}
           \epsfig{file=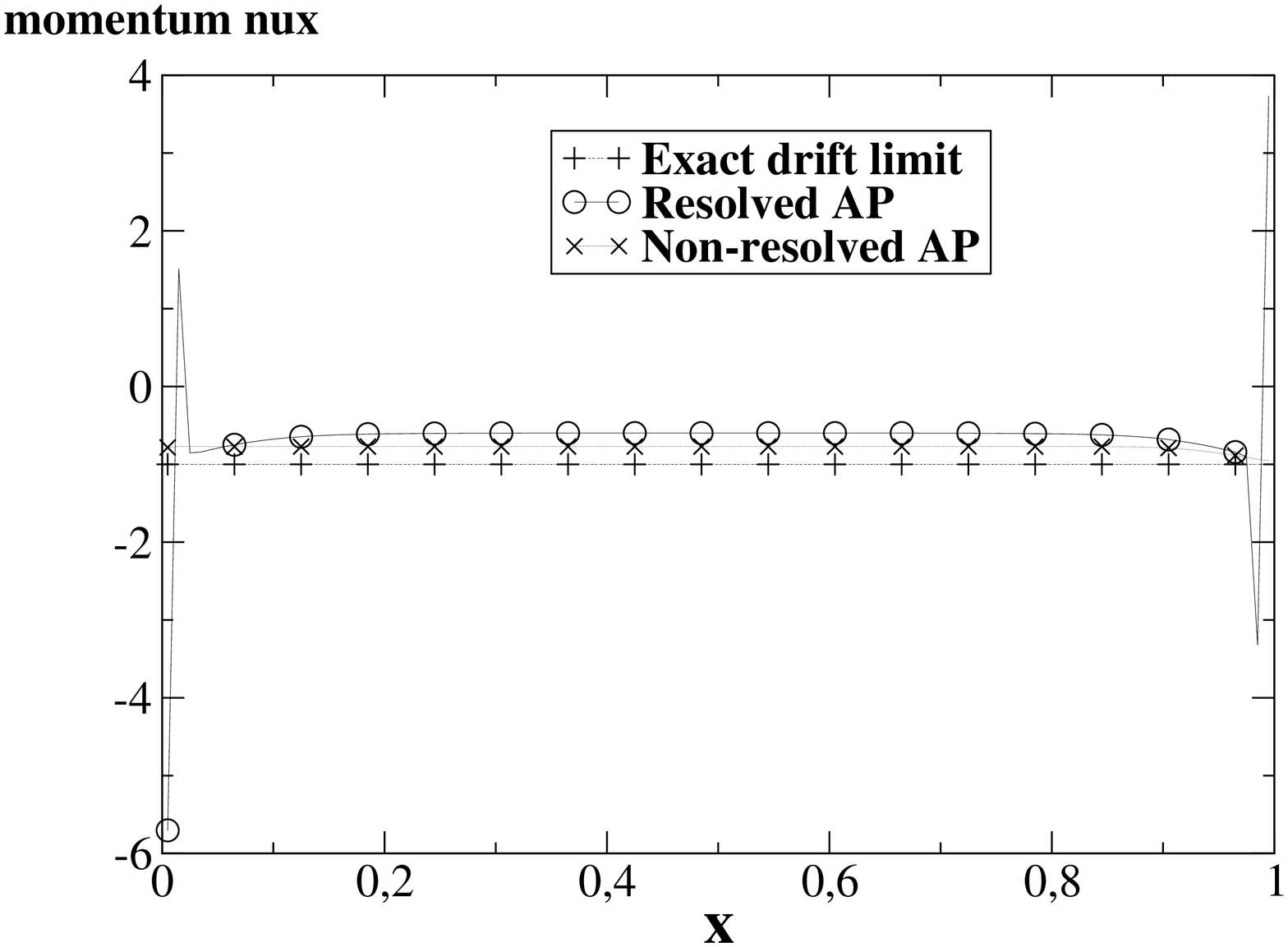,width=5cm,angle=-90,clip=}
       \end{minipage} 
     \end{tabular}
     \\
     \begin{tabular}{cc}
       \begin{minipage}{6cm}
	   \hspace{-0.5cm}
           \epsfig{file=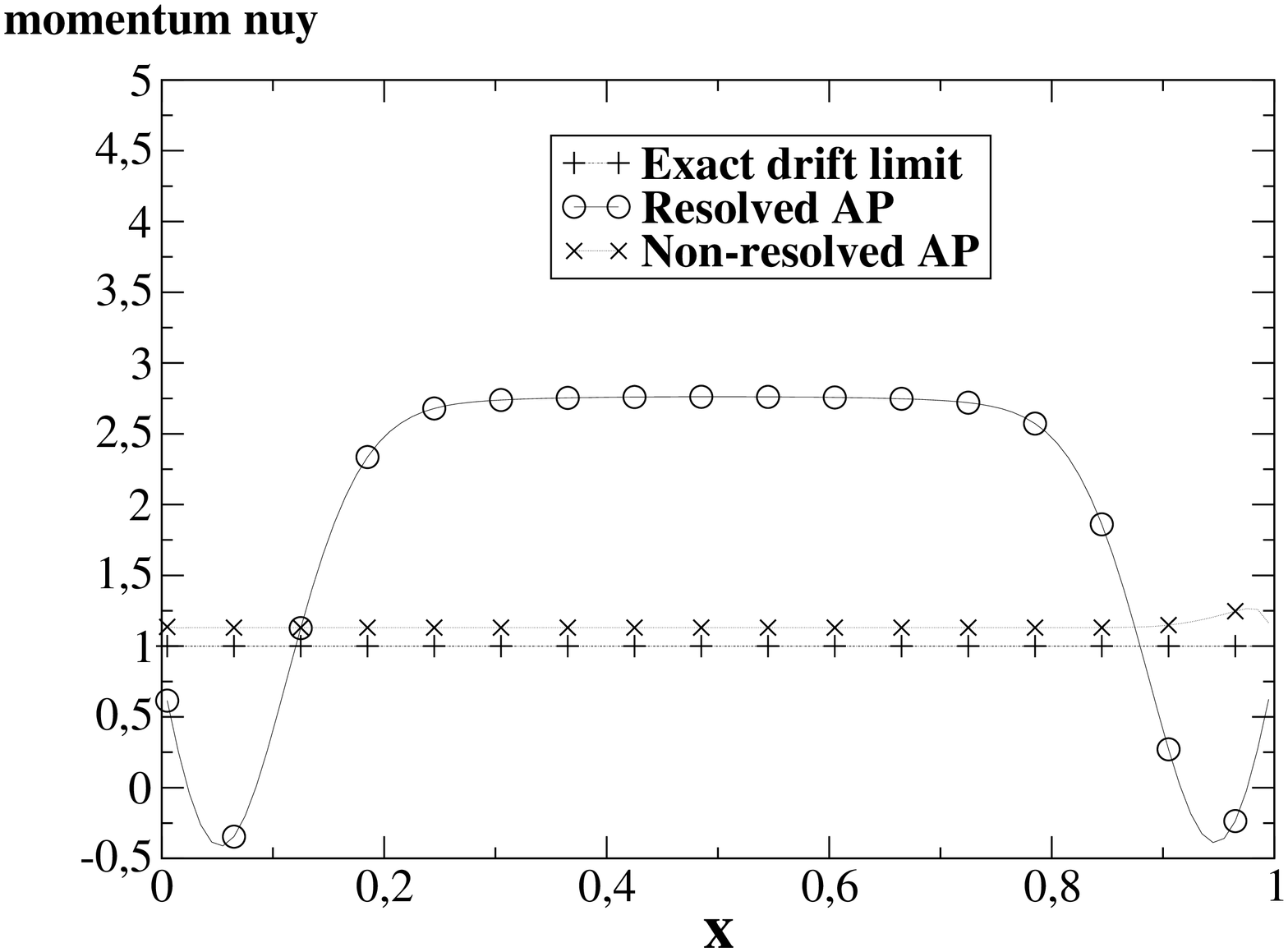,width=5cm,angle=-90,clip=} 
       \end{minipage} 
       &
       \begin{minipage}{6cm}
	   \hspace{-0.5cm}
           \epsfig{file=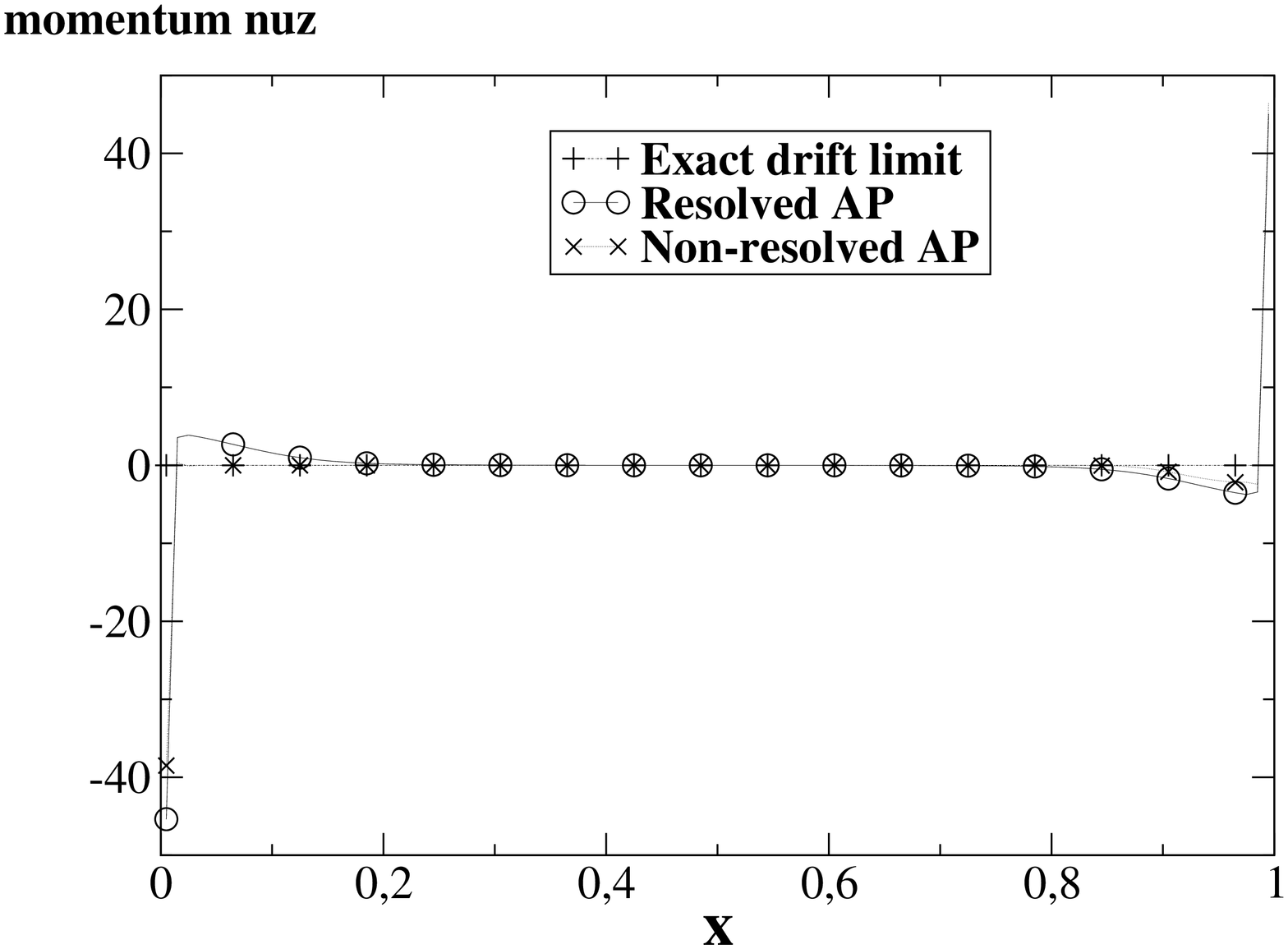,width=5cm,angle=-90,clip=} 
       \end{minipage} 
     \end{tabular}
   \end{tabular}
  \caption{Comparison of the resolved AP scheme (crosses), and
  unresolved AP scheme (circles)  at $t=0.1$ for unprepared initial
  and boundary conditions $\varepsilon = 10^{-6}$ and
  $\varepsilon^{\prime} = 10^{-2}$ ; density $n$ (top left),
  $x$-component of the momentum $nu_x$ (top right), $y$-component of
  the momentum $nu_y$ (bottom left), $z$-component of the momentum
  $nu_z$ (bottom right). The computed solutions are shown for the
  section at middle $y=0.5$ of the calculation domain $\Omega$ along
  the $x$-direction.}  
  \label{DAPf1APrnrx} 
\end{figure}

\begin{figure}[h!]
   \vspace{-0.25cm}
     \begin{tabular}{c}
     \begin{tabular}{cc}
       \begin{minipage}{6cm}
	   \hspace{-0.5cm}
           \epsfig{file=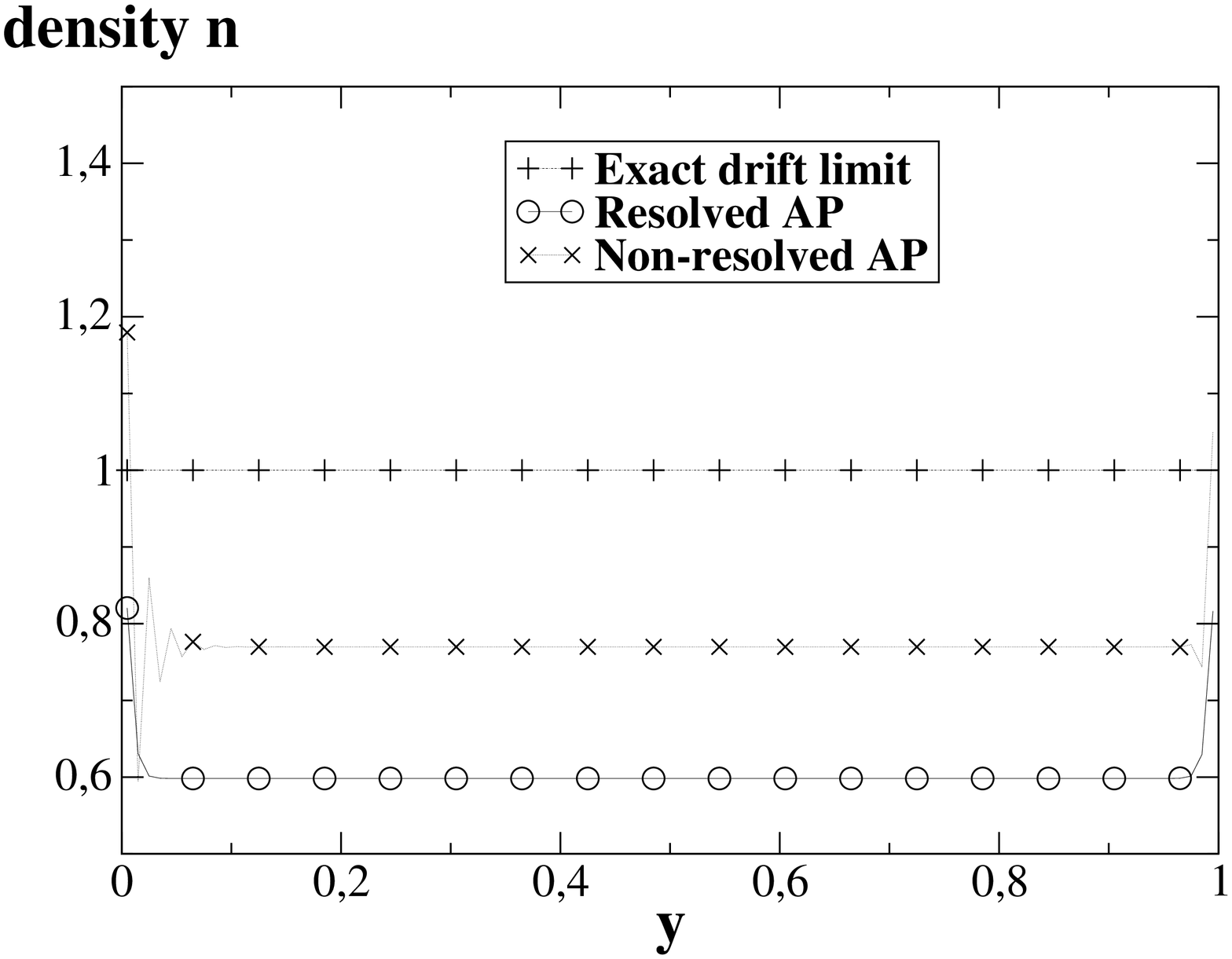,width=5cm,angle=-90,clip=} 
       \end{minipage} 
       &
       \begin{minipage}{6cm}
	   \hspace{-0.5cm}
           \epsfig{file=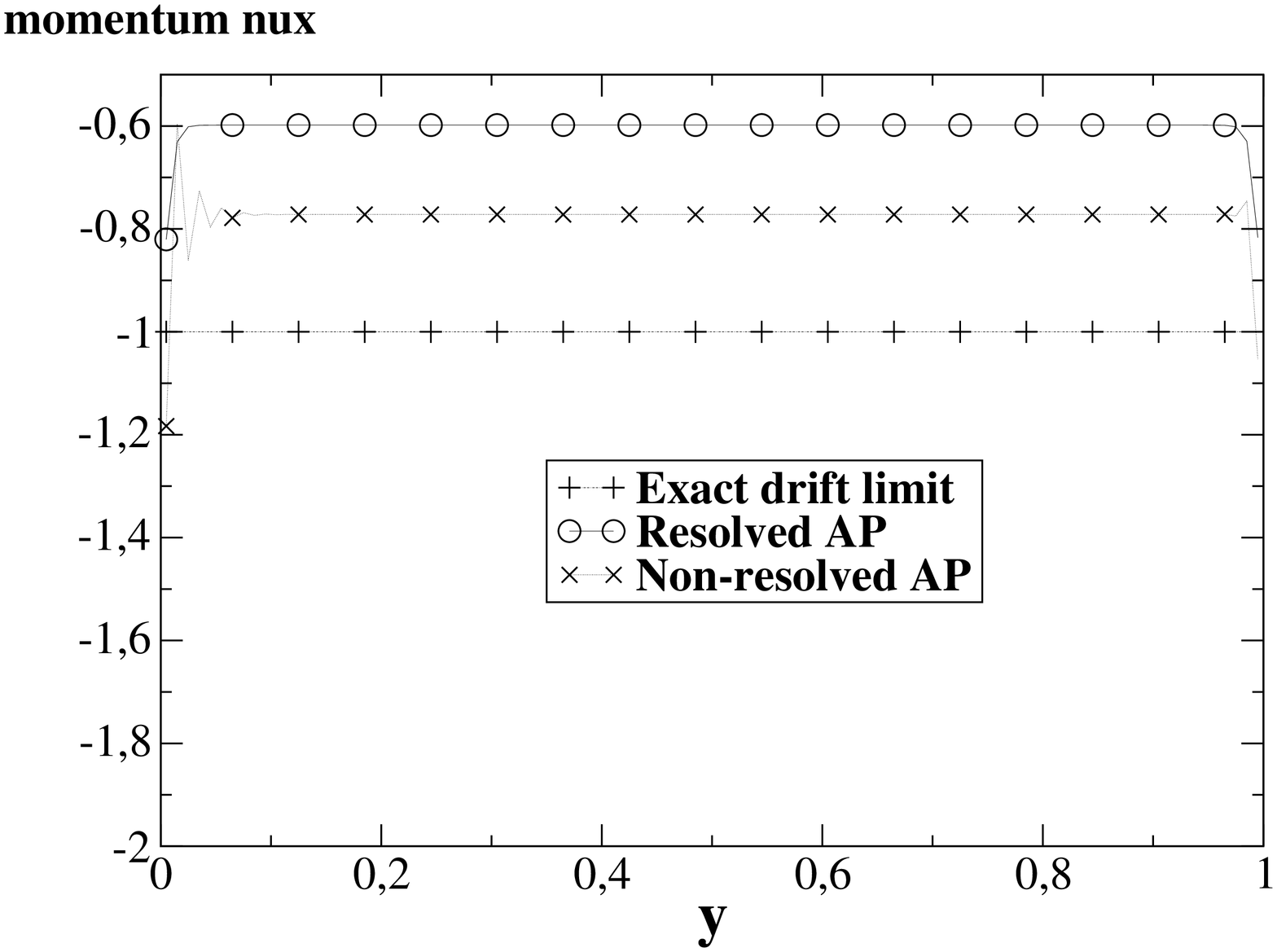,width=5cm,angle=-90,clip=}
       \end{minipage} 
     \end{tabular}
     \\
     \begin{tabular}{cc}
       \begin{minipage}{6cm}
	   \hspace{-0.5cm}
           \epsfig{file=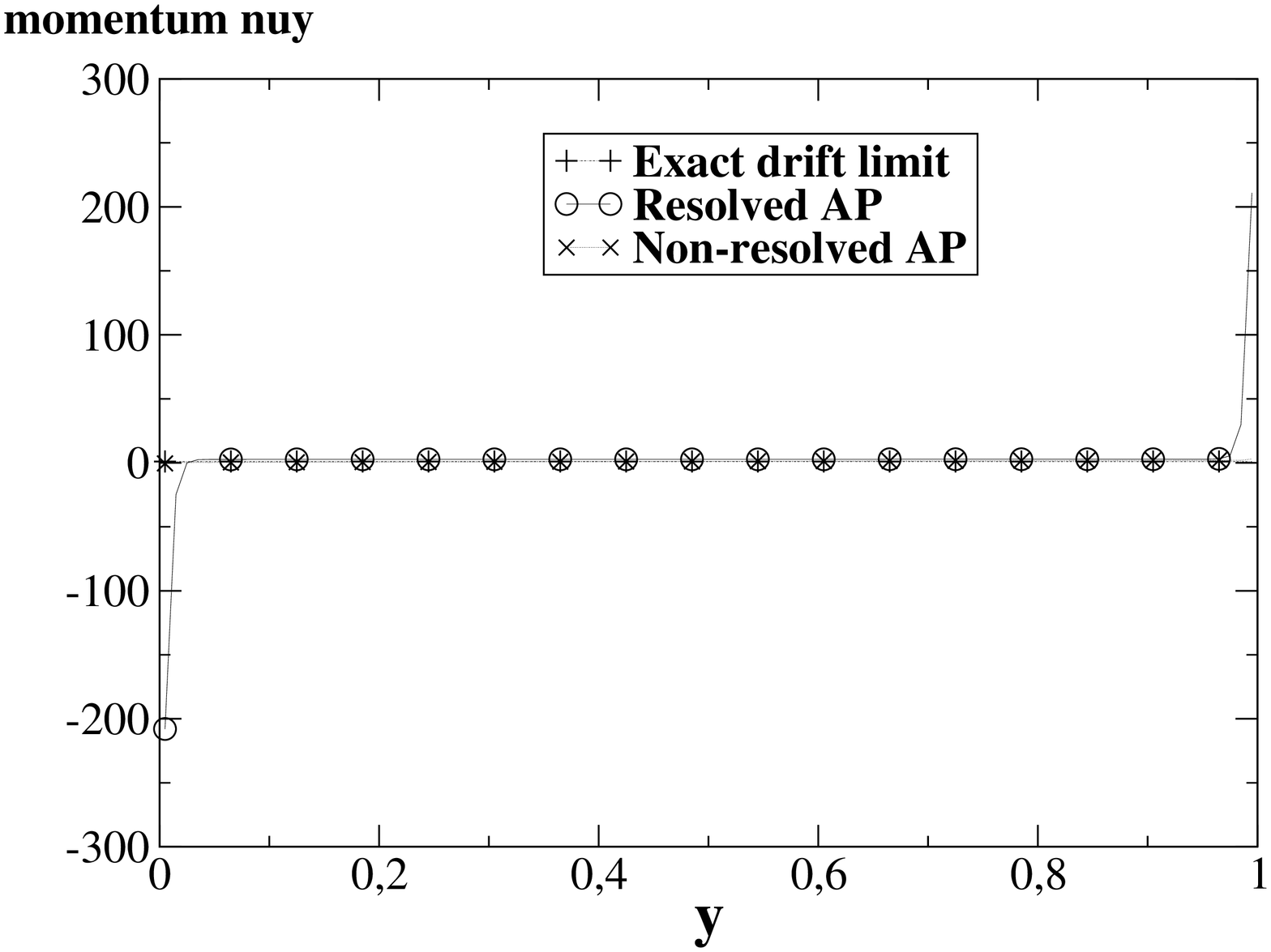,width=5cm,angle=-90,clip=} 
       \end{minipage} 
       &
       \begin{minipage}{6cm}
	   \hspace{-0.5cm}
           \epsfig{file=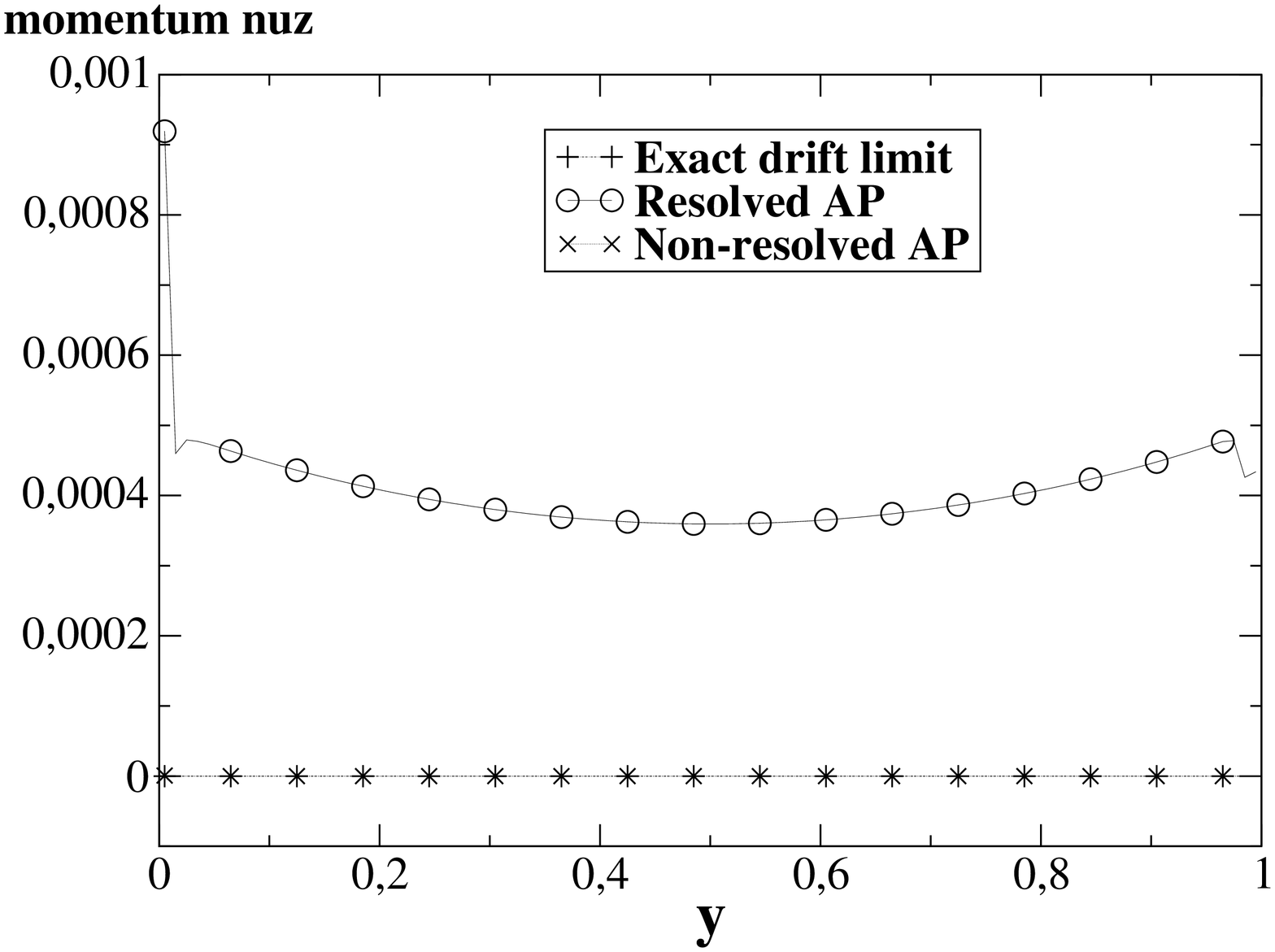,width=5cm,angle=-90,clip=} 
       \end{minipage} 
     \end{tabular}
   \end{tabular}
  \caption{Comparison of the resolved AP scheme (crosses), and
  unresolved AP scheme (circles) at $t=0.1$ for unprepared initial and
  boundary conditions $\varepsilon = 10^{-6}$ and
  $\varepsilon^{\prime} = 10^{-2}$ ; density $n$ (top left),
  $x$-component of the momentum $nu_x$ (top right), $y$-component of
  the momentum $nu_y$ (bottom left), $z$-component of the momentum
  $nu_z$ (bottom right). The computed solutions are shown for the
  section at middle $x=0.5$ of the calculation domain $\Omega$ along
  the $y$-direction.}  
  \label{DAPf1APrnry}
\end{figure}

\begin{figure}[h!]
   \begin{tabular}{c}
     \begin{tabular}{cc}
       \begin{minipage}{6cm}
	   \hspace{-0.5cm}
           \epsfig{file=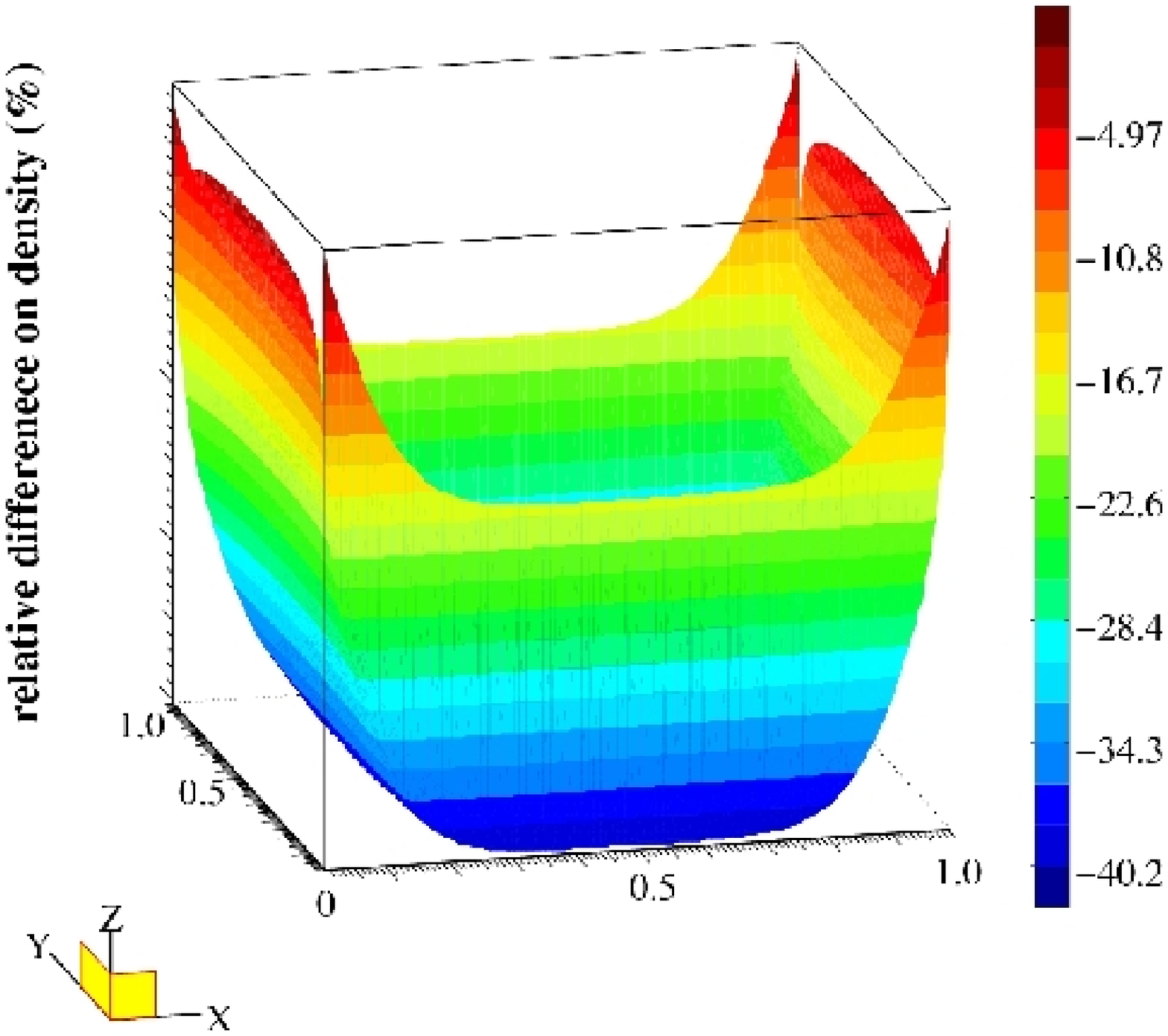,width=5cm,angle=0,clip=} 
       \end{minipage} 
       &
       \begin{minipage}{6cm}
	   \hspace{-0.5cm}
           \epsfig{file=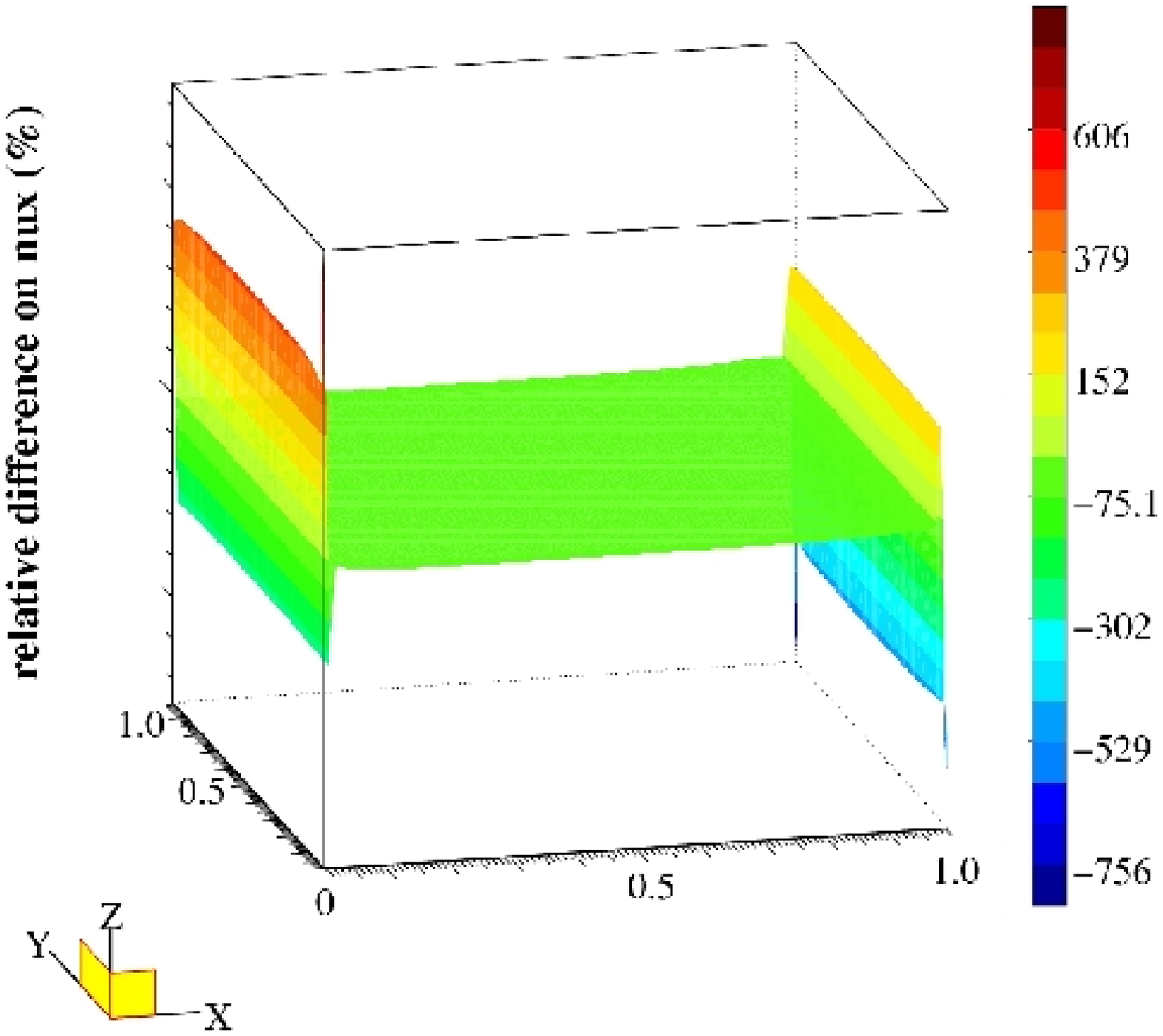,width=5cm,angle=0,clip=}
       \end{minipage} 
     \end{tabular}
     \\
     \begin{tabular}{cc}
       \begin{minipage}{6cm}
	   \hspace{-0.5cm}
           \epsfig{file=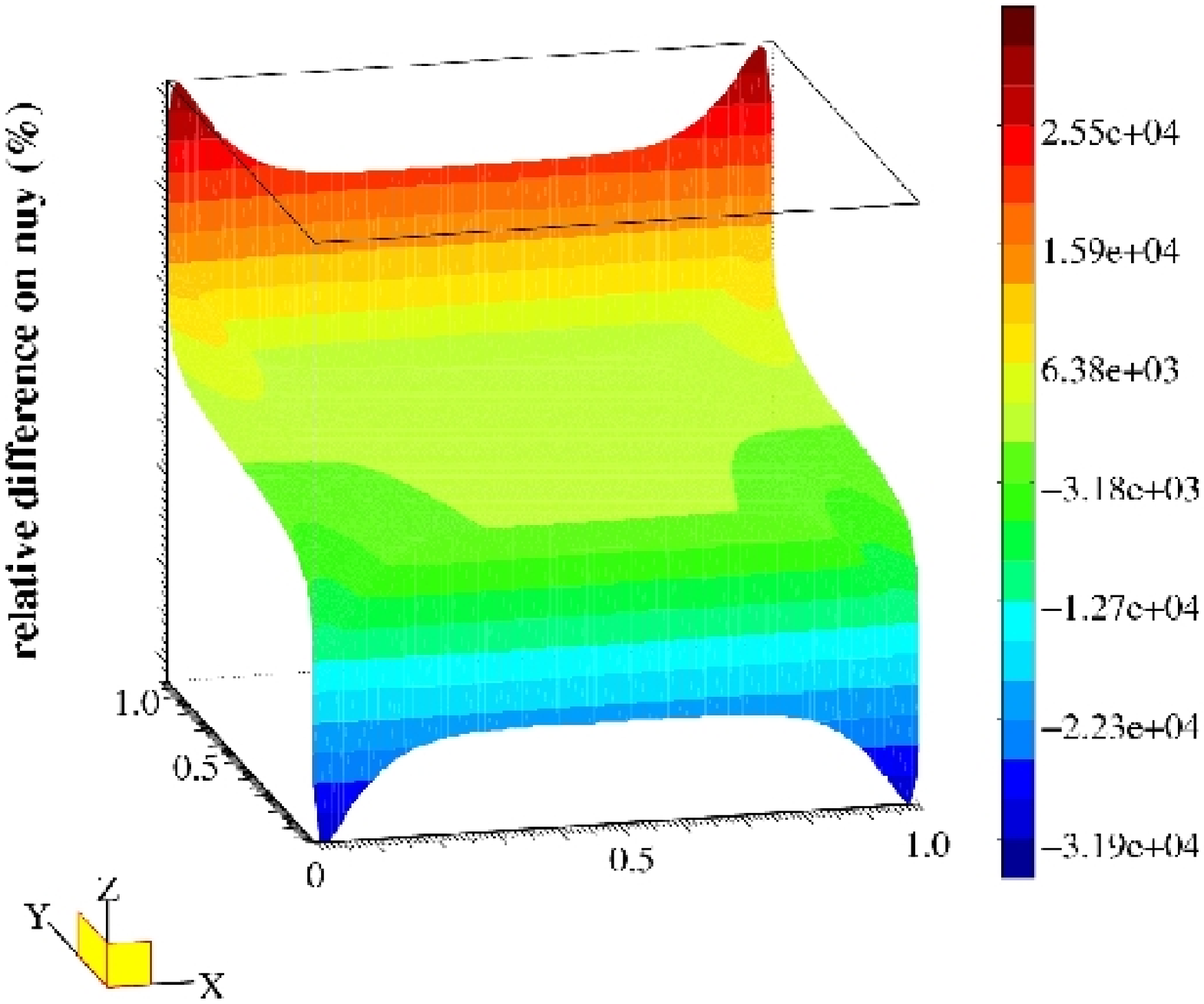,width=5cm,angle=0,clip=} 
       \end{minipage} 
       &
       \begin{minipage}{6cm}
	   \hspace{-0.5cm}
           \epsfig{file=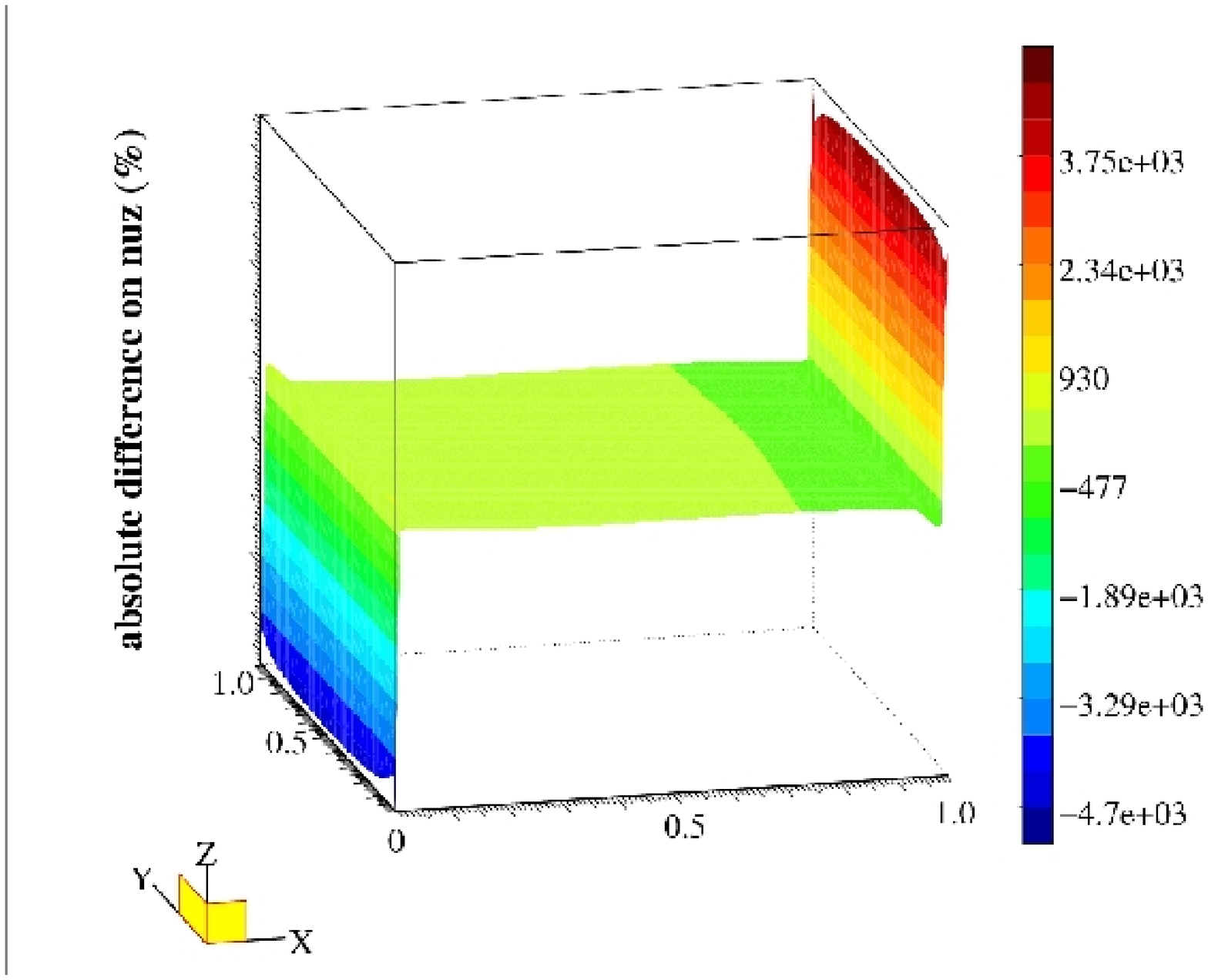,width=5cm,angle=0,clip=} 
       \end{minipage} 
     \end{tabular}
   \end{tabular}
   \caption{Relative difference between the computed solution and the
     exact drift-fluid limit of the resolved AP scheme at time $t=0.1$
     for unprepared initial and boundary conditions $\varepsilon =
     10^{-6}$ and $\varepsilon^{\prime} = 10^{-2}$ ; density $n$ (top
     left), $x$-component of the momentum $nu_x$ (top right), 
     $y$-component of the momentum $nu_y$ (bottom left). Absolute
     difference between the computed solution and the exact
     drift-fluid limit on the $z$-component of the momentum $nu_z$
     (bottom right).}             
  \label{DAPf1APrerr}
 \end{figure}

\begin{figure}[h!]
   \begin{tabular}{c}
     \begin{tabular}{cc}
       \begin{minipage}{6cm}
	   \hspace{-0.5cm}
           \epsfig{file=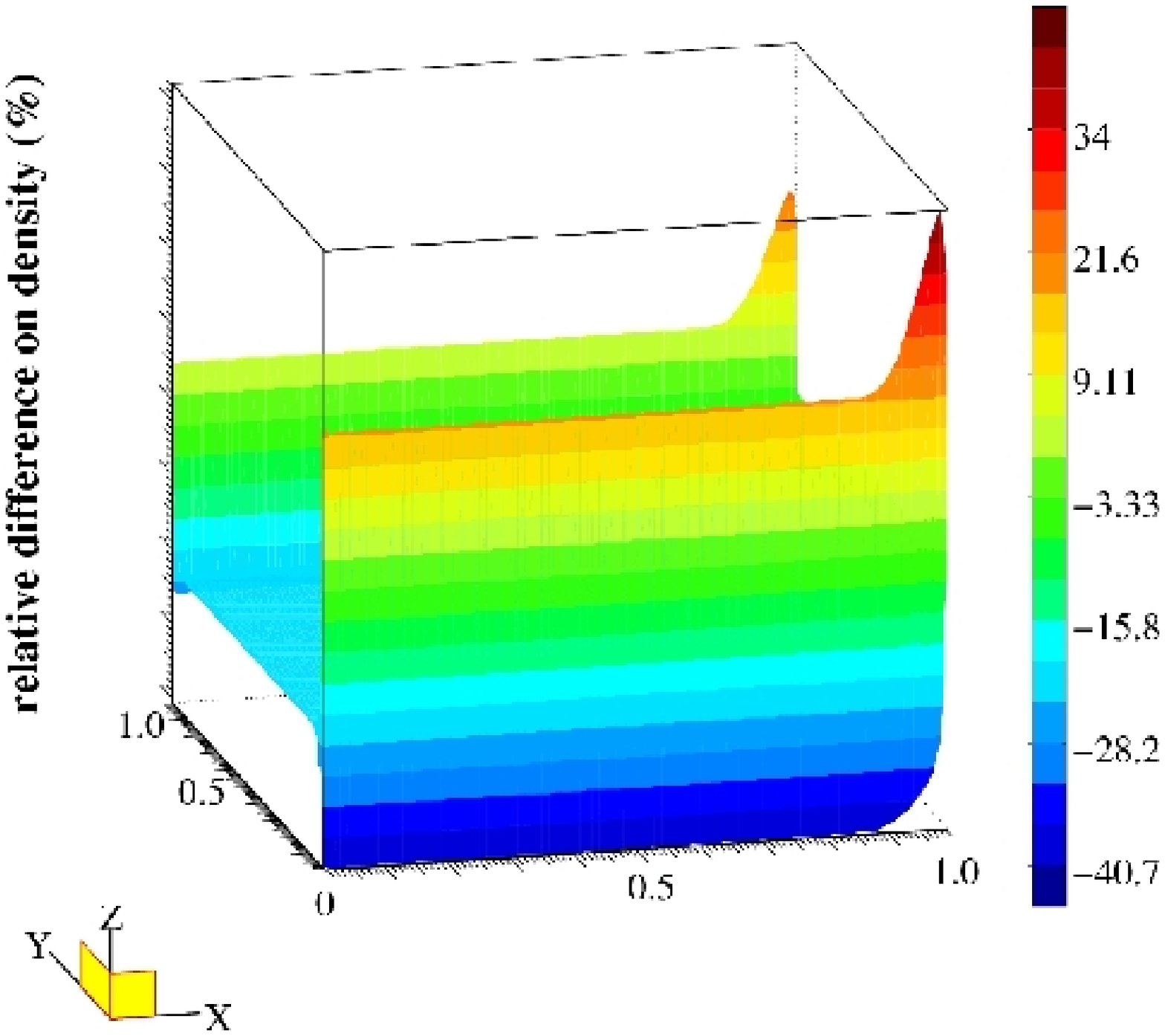,width=5cm,angle=0,clip=} 
       \end{minipage} 
       &
       \begin{minipage}{6cm}
	   \hspace{-0.5cm}
           \epsfig{file=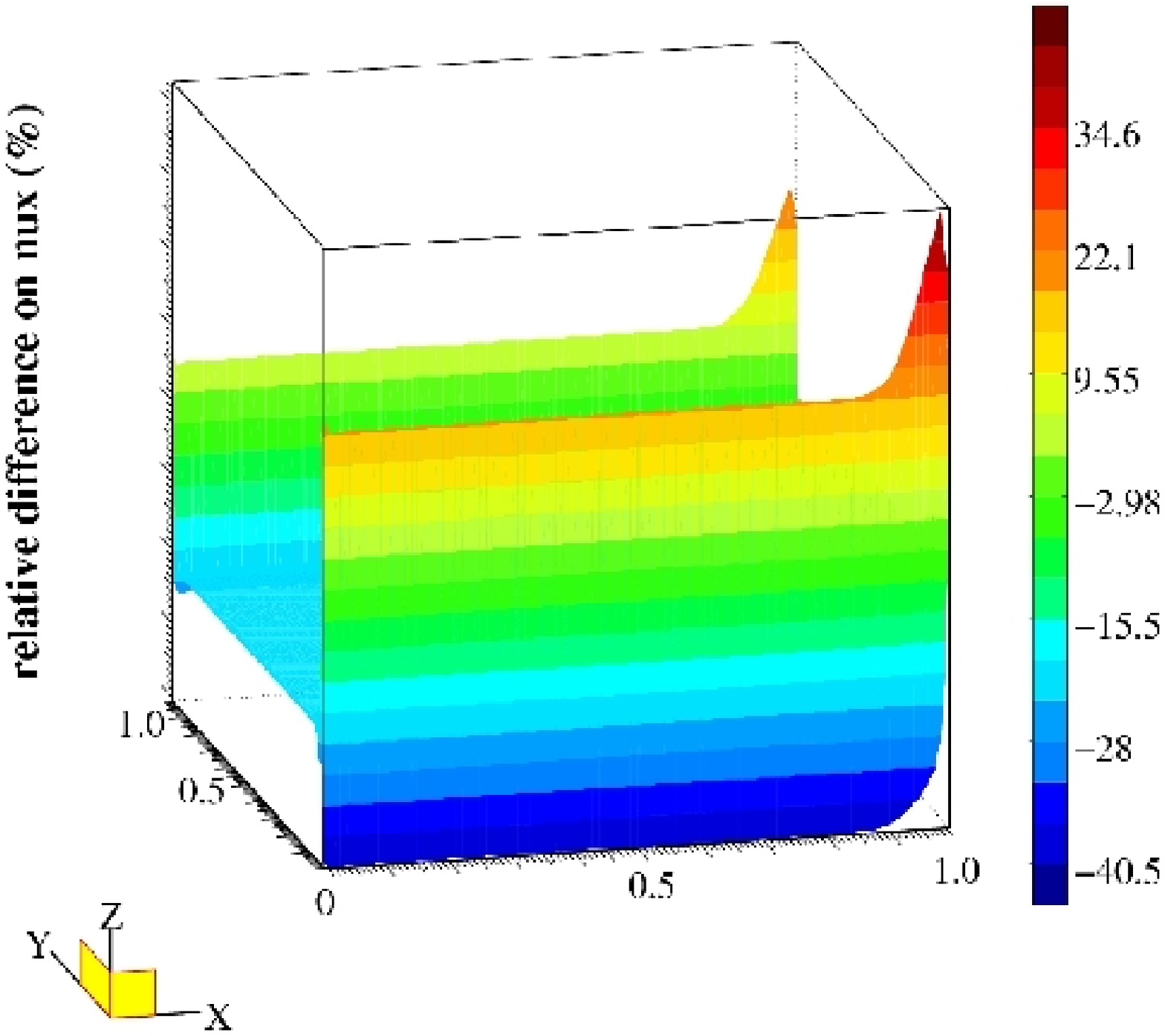,width=5cm,angle=0,clip=} 
       \end{minipage} 
     \end{tabular}
     \\
     \begin{tabular}{cc}
       \begin{minipage}{6cm}
	   \hspace{-0.5cm}
           \epsfig{file=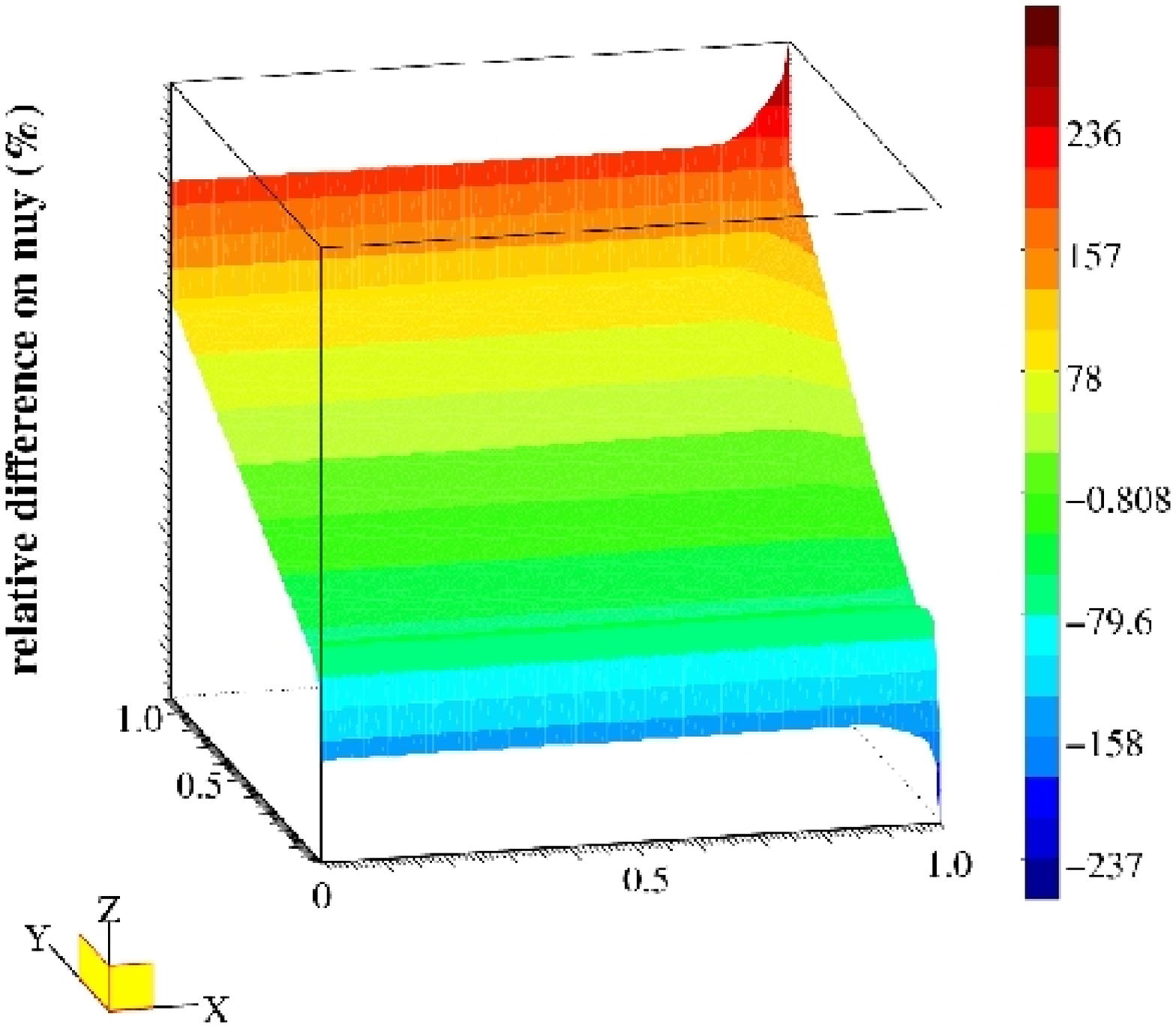,width=5cm,angle=0,clip=}
       \end{minipage} 
       &
       \begin{minipage}{6cm}
	   \hspace{-0.5cm}
           \epsfig{file=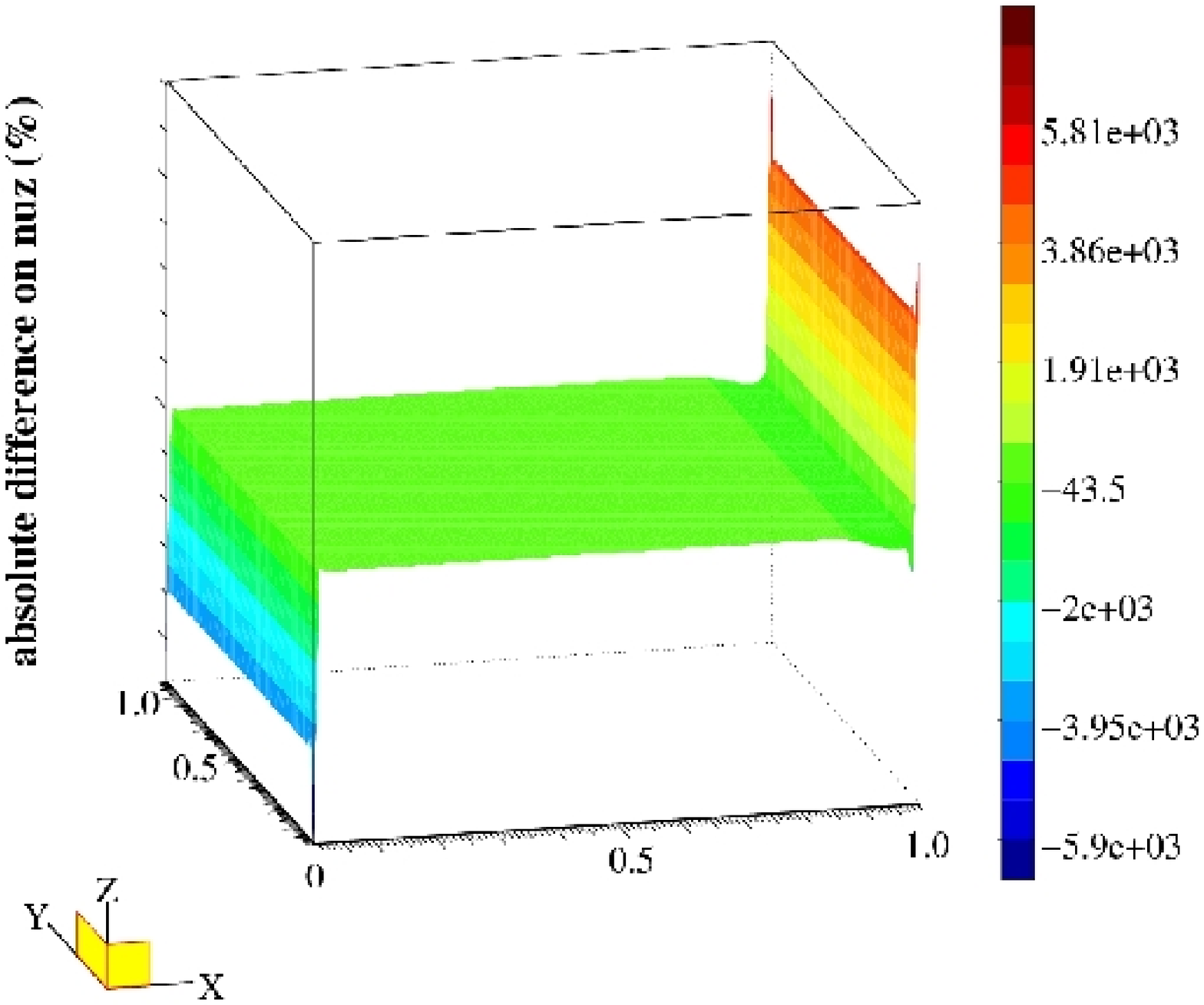,width=5cm,angle=0,clip=} 
       \end{minipage} 
     \end{tabular}
   \end{tabular}
   \caption{Relative difference between the computed solution and the
     exact drift-fluid limit of the non-resolved AP scheme at time
     $t=0.1$ for unprepared initial and boundary conditions
     $\varepsilon = 10^{-6}$ and $\varepsilon^{\prime} = 10^{-2}$;
     density $n$ (top left), $x$-component of the momentum $nu_x$ (top
     right), $y$-component of the momentum $nu_y$ (bottom
     left). Absolute difference between the computed solution and the
     exact drift-fluid limit on the $z$-component of the momentum
     $nu_z$ (bottom right).}             
  \label{DAPf1APnrerr}
 \end{figure}

On Fig. \ref{DAPf1APrnrx} and \ref{DAPf1APrnry}, we display the
values of the density and the three components of the momentum 
as functions of $x$ (resp. $y$) along the line $y=0.5$ 
(resp. $x=0.5$) for the resolved and non-resolved AP schemes and 
for the drift-fluid limit. The relative differences (absolute
difference in the case of $nu_z$) of the solution 
with the drift-fluid limit are given on Fig. \ref{DAPf1APrerr} for
the resolved AP scheme and on Fig. \ref{DAPf1APnrerr} for the 
non-resolved one.

Both the resolved and non-resolved AP schemes exhibit a significant
discrepancy with the drift-fluid limit. This discrepancy originates
in the appearance of boundary layers which pollute the accuracy 
of the solution inside the domain. However, the discrepancy is
much larger for the resolved AP scheme than for the non-resolved one. 
In many instances, the non-resolved AP scheme provides a fairly correct
solution and its oscillations inside the boundary layers are less 
pronounced. This can be attributed to the larger time steps which provide
a stronger relaxation rate towards the drift-fluid limit as well as a bigger 
amount of numerical diffusion than the small time-steps
used in the resolved-AP schemes. 

These results show that the use of the non-resolved AP scheme in the case of
non well-prepared boundary data at least provides a stable if not
accurate solution. Additionally, it is expected that a suitable 
boundary layer analysis will permit to derive corrected boundary 
conditions which will take into account the influence of the 
boundary layer. The search for adequate boundary layer correctors
will be the subject of future work.

%%%%%%%%%%%%%%%%%%%%%%%%%%%%%%%%%%%%%%%%%%%%%%%%%%%%%%%%%%%%%%%%
%%%%%%%%%%%%%%%%%%%%%%%%%%%%%%%%%%%%%%%%%%%%%%%%%%%%%%%%%%%%%%%%
%%%%%%%%%%%%%%%%%%%%%%%%%%%%%%%%%%%%%%%%%%%%%%%%%%%%%%%%%%%%%%%%
%%%%%%%%%%%%%%%%%%%%%%%%%%%%%%%%%%%%%%%%%%%%%%%%%%%%%%%%%%%%%%%%
%%%%%%%%%%%%%%%%%%%%%%%%%%%%%%%%%%%%%%%%%%%%%%%%%%%%%%%%%%%%%%%%

\setcounter{equation}{0}
\section{Conclusion}
\label{DA7}

The Euler-Lorentz model in the drift-fluid scaling for ion flow has been
investigated. First the drift-fluid limit has been studied and it has
been shown that the parallel fluid velocity to the magnetic field is a
solution of an elliptic equation. Then, the full Euler-Lorentz model
in the drift-fluid scaling has been investigated. A reformulation of the
model has been provided, which shows that the parallel velocity is the
solution of a wave equation with wave velocities tending to infinity
as the scaling parameter $\varepsilon$ goes to zero. This
reformulation allows to derive an Asymptotic Preserving scheme for the 
Euler-Lorentz model in the drift-fluid scaling, i.e. a scheme which is
consistent with the full Euler-Lorentz model when $\varepsilon = O(1)$
and which is consistent with its drift-fluid limit when $\varepsilon
\to 0$. The scheme allows to compute the solution of the Euler-Lorentz
model when $\varepsilon \ll 1$ with a time step independent of
$\varepsilon$. This property has been demonstrated numerically on a
test example. It has been shown that the scheme is as good as the
conventional scheme when $\varepsilon = O(1)$ and that it provides a stable
if not accurate solution in the case of non well-prepared boundary
data.  

Forthcoming work will be devoted to the application of the AP scheme
in the case of time varying and inhomogeneous magnetic fields, the
coupling of the ion flow to the electron flow and to the electric and
magnetic fields as well as to more rigorous stability analyses of the 
proposed schemes.

%%%%%%%%%%%%%%%%%%%%%%%%%%%%%%%%%%%%%%%%%%%%%%%%%%%%%%%%%%%%%%%%
%%%%%%%%%%%%%%%%%%%%%%%%%%%%%%%%%%%%%%%%%%%%%%%%%%%%%%%%%%%%%%%%
%%%%%%%%%%%%%%%%%%%%%%%%%%%%%%%%%%%%%%%%%%%%%%%%%%%%%%%%%%%%%%%%
%%%%%%%%%%%%%%%%%%%%%%%%%%%%%%%%%%%%%%%%%%%%%%%%%%%%%%%%%%%%%%%%
%%%%%%%%%%%%%%%%%%%%%%%%%%%%%%%%%%%%%%%%%%%%%%%%%%%%%%%%%%%%%%%%


\begin{thebibliography}{bDA}   

\bibitem{Beer_CowleyHammett} M.A. Beer, S.C. Cowley, G.W. Hammett,
  Field-aligned coordinates for nonlinear simulations of tokamak
  turbulence, Phys. Plasmas 2 (1995) 2687.

\bibitem{Beer_Hammett} M.A. Beer, G.W. Hammett, Toroidal Gyrofluid
  Equations for Simulations of Tokamak Turbulence, 
  Phys. Plasmas 3 (1996) 4046. 

\bibitem{BCDS} R. Belaouar, N. Crouseilles, P. Degond,
  E. Sonnendr\"ucker, An asymptotically stable semi-lagrangian scheme
  in the quasi-neutral limit, submitted. 

\bibitem{Buet_Cordier_Lucquin_Mancini} C. Buet, S. Cordier,
  B. Lucquin-Desreux and S. Mancini, Diffusion limit of the Lorentz
  model: asymptotic preserving schemes, Mathematical Modelling and
  Numerical Analysis 36 (2002) 631. 
  
\bibitem{CJR} R. Caflisch, S. Jin and G. Russo,
  Uniformly Accurate Schemes for Hyperbolic Systems with
  Relaxations, SIAM J. Num. Anal. 34 (1997) 246.
  
\bibitem{CDV_CRAS_05} P. Crispel, P. Degond, M-H. Vignal,
  An asymptotically stable discretization for the Euler-Poisson system
  in the quasineutral limit, C. R. Acad. Sci. Paris Ser. I 341 (2005)
  341.

\bibitem{CDV} P. Crispel, P. Degond  M.-H. Vignal, An asymptotic
  preserving scheme for the two-fluid Euler-Poisson model in the
  quasineutral limit, J. Comput. Phys. 223 (2007) 208.
  
 
\bibitem{DDN_CRAS_06} P. Degond, F. Deluzet and L. Navoret, An
  asymptotically stable Particle-in-Cell (PIC) scheme for
  collisionless plasma simuations near quasineutrality,
  C. R. Acad. Sci. Paris, Ser I, 343 (2006) 613. 

  
\bibitem{DJL} P. Degond, S. Jin and J.-G. Liu, Mach-number uniform
  asymptotic-preserving Gauge schemes for compressible flows,
  Bulletin of the Institute of Mathematics, Academia Sinica 
  (New Series) 2 (2007) 851. 

\bibitem{DJT} P. Degond, S. Jin and M. Tang, On the time splitting
  spectral method for the complex Ginzburg-Landau equation in the
  large time and space scale limit, SIAM J. Sci. Comput. 30 (2008)
  2466.  
  
\bibitem{DLV} P. Degond, J.-G. Liu and M.-H. Vignal,
  Analysis of an asymptotic preserving scheme for the Euler-Poisson
  system in the quasineutral limit, SIAM J. Num.
  Anal. 46 (2008) 1298.  
  
\bibitem{DegondPol} P. Degond, P.-F. Peyrard, G. Russo, P. Villedieu,
  Polynomial Upwind Schemes for  Hyperbolic Systems,
  C. R. Acad. Sci. Paris Ser. I  328 (1999) 479.  

\bibitem{Dimits} A.M. Dimits, Fluid simulations of tokamak turbulence
  in quasiballooning coordinates, Phys. Rev. E
  48 (1993) 4070. 

\bibitem{Dorland_Hammett} W. Dorland, G.W. Hammett, Gyrofluid
  Turbulence Models with Kinetic Effects, Phys. Fluids B-Plasmas 
  5 (1993) 812. 

\bibitem{Falchetto_Ottaviani_PRL04} G.L. Falchetto and M. Ottaviani,
  Effect of Collisional Zonal-Flow Damping on Flux-Driven Turbulent
  Transport, Phys. Rev. Lett. 92 (2004) 025002.  

\bibitem{Garbet_PhPl01} X. Garbet, C. Bourdelle, G.T. Hoang, P. Maget,
  S. Benkadda, P. Beyer, C. Figarella, I. Voitsekovitch, O. Agullo,
  N. Bian, Global simulations of ion turbulence with magnetic shear
  reversal, Phys. Plasmas 8 (2001) 2793.  

\bibitem{Gosse_Toscani} L. Gosse and G. Toscani, Asymptotic preserving
  and well-balanced schemes for radiative transfer and the Rosseland
  approximation, Numer. Math. 98 (2004) 223.

\bibitem{Gosse_Toscani_2} L. Gosse and G. Toscani,  An asymptotic
  preserving well-balanced scheme for the hyperbolic heat equation,
  C. R. Acad. Sci. Paris Ser I, 334 (2002) 1. 

\bibitem{Hallatschek_PhPl00} K. Hallatschek, A. Zeiler, Nonlocal
  simulation of the transition from ballooning to ion temperature
  gradient mode turbulence in the tokamak edge, Phys. Plasmas 7 (2000)
  2554.  


\bibitem{Hammett_Dorland} G.W. Hammett, M.A. Beer, W. Dorland,
  S.C. Cowley, S. A. Smith, Developments in the Gyrofluid Approach
  to Tokamak Turbulence Simulations, Plasma Phys. Contr. F. 35 (1993)
  973.   
  
\bibitem{Hazeltine_Meiss} R.D. Hazeltine, J.D. Meiss, Plasma
  Confinement, Dover Publications, Mineola, New York (2003). 
  
\bibitem{Jin} S. Jin, Efficient Asymptotic-Preserving (AP) Schemes for
  Some Multiscale Kinetic Equations, SIAM J. Sci. Comp., 21 (1999)
  441.
  
\bibitem{Jin-JCP} S. Jin,
  Runge-Kutta Methods for Hyperbolic Conservation Laws with Stiff
  Relaxation Terms, J. Comp. Phys.  122 (1995) 51.
  
  
\bibitem{JL} S. Jin, C.D. Levermore,
  Numerical Schemes for Hyperbolic Conservation Laws with Stiff
  Relaxation Terms, J. Comp.  Phys.  126 (1996) 449.
  
\bibitem{JPT1} S. Jin, L. Pareschi, G. Toscani,
  Diffusive Relaxation Schemes for Discrete-Velocity Kinetic
  Equations,   SIAM J. Num. Anal., 35 (1998) 2405.
  
\bibitem{JPT2} S. Jin, L. Pareschi, G. Toscani, Uniformly Accurate
  Diffusive Relaxation Schemes for Multiscale Transport Equations,
  SIAM J. Num. Anal. 38 (2000) 913. 

\bibitem{Lifshitz} E.M. Lifshitz and L.P. Pitaevskii, Physical
  Kinetics, Course in theoretical physics, Volume 10, 
  Butterworth-Heinemann, Oxford (2006).

\bibitem{Naulin_PhPl03} V. Naulin, Electromagnetic transport
  components and sheared flows in drift-Alfv\`en turbulence,
  Phys. Plasmas 10 (2003) 4016.   

\bibitem{Naulin_PhPl05} V. Naulin, A. Kendl, O.E. Garcia,
  A.H. Nielsen, J. Juul Rasmussen, Shear flow generation and
  energetics in electromagnetic turbulence, Phys. Plasmas 12 (2005)
  052515. 

\bibitem{Ottaviani_Manfredi_PhPl99} M. Ottaviani, G. Manfredi, The
  gyro-radius scaling of ion thermal transport from global numerical
  simulations of ion temperature gradient driven turbulence,
  Phys. Plasmas 6 (1999) 3267.  

\bibitem{Pareschi_Russo} L.Pareschi, G.Russo, Asymptotic preserving
  Monte Carlo methods for the Boltzmann equation, Transp. Theory
  Stat. Phys.  29 (2000) 415. 

\bibitem{Scott_PlPhCF97} B.D. Scott, Three-dimensional computation of
  drift Alfv\`en turbulence Plasma Phys. Control. F. 39 (1997) 1635. 

\bibitem{Scott_PhPl05} B.D. Scott, Free-energy conservation in local
  gyrofluid models, Phys. Plasmas 12 (2005) 102307.

\bibitem{Toro99} E.F. Toro, Riemann  Solvers and Numerical
  Methods for Fluids dynamics, A Practical Introduction, Second
  edition, Springer, Berlin (1999).

\bibitem{Xu_PhPl00} X.Q. Xu, R.H. Cohen, T.D. Rognlien, J.R. Myra,
  Low-to-high confinement transition simulations in divertor geometry,
  Phys. Plasmas~7 (2000) 1951.  

\end{thebibliography}
\end{document}